\documentclass[numberedappendix,twocolappendix,iop,apj]{emulateapj}
\usepackage{psfig,amsfonts,amsmath,graphicx,natbib,nicefrac,url,hyperref,longtable}

\usepackage{booktabs}
\hypersetup{colorlinks=true, urlcolor=blue, citecolor=blue}

\newcommand{\Swift}{\textit{Swift}}
\newcommand{\Fermi}{\textit{Fermi}}

\newcommand{\EK}{\ensuremath{E_{\rm K}}}
\newcommand{\EKiso}{\ensuremath{E_{\rm K,iso}}}
\newcommand{\Egamma}{\ensuremath{E_{\gamma}}}
\newcommand{\Egammaiso}{\ensuremath{E_{\gamma,\rm iso}}}	     
\newcommand{\epse}{\ensuremath{\epsilon_{\rm e}}}
\newcommand{\epsb}{\ensuremath{\epsilon_{\rm B}}}
\newcommand{\dens}{\ensuremath{n_{0}}}
\newcommand{\Astar}{\ensuremath{A_{*}}}
\newcommand{\tdec}{\ensuremath{t_{\rm dec}}}
\newcommand{\tjet}{\ensuremath{t_{\rm jet}}}
\newcommand{\thetajet}{\ensuremath{\theta_{\rm jet}}}

\newcommand{\AV}{\ensuremath{A_{\rm V}}}

\newcommand{\pcmsq}{\ensuremath{{\rm cm}^{-2}}}

\newcommand{\me}{GRB~140304A}

\newcommand{\nua}{\ensuremath{\nu_{\rm a}}}
\newcommand{\nusa}{\ensuremath{\nu_{\rm sa}}}
\newcommand{\nuac}{\ensuremath{\nu_{\rm ac}}}
\newcommand{\numax}{\ensuremath{\nu_{\rm m}}}
\newcommand{\nuc}{\ensuremath{\nu_{\rm c}}}

\newcommand{\nuar}{\ensuremath{\nu_{\rm a,r}}}
\newcommand{\numr}{\ensuremath{\nu_{\rm m,r}}}
\newcommand{\nucr}{\ensuremath{\nu_{\rm c,r}}}
\newcommand{\fnumr}{\ensuremath{F_{\nu,\rm m,r}}}

\newcommand{\nunir}{\ensuremath{\nu_{\rm NIR}}}

\newcommand{\nux}{\ensuremath{\nu_{\rm X}}}

\newcommand{\nupk}{\ensuremath{\nu_{\rm pk}}}
\newcommand{\fnupk}{\ensuremath{F_{\nu,\rm pk}}}
\newcommand{\fnumax}{\ensuremath{F_{\nu,\rm m}}}
\newcommand{\td}{\ensuremath{t_{\rm d}}}
\newcommand{\tcol}{\ensuremath{t_{\rm col}}}

\shorttitle{GRB~140304A}
\shortauthors{Laskar et al.}

\def\nrao{1}
\def\ucb{2}
\def\cfa{3}
\def\northwestern{4}
\def\ash{5}
\def\einsteinfellow{6}
\def\ariz{7}
\def\huji{8}

\begin{document} 

\title{A VLA Study of High-redshift GRBs II --
The Complex Radio Afterglow of GRB\,140304A: Shell 
Collisions and Two Reverse Shocks}
\author{Tanmoy Laskar\altaffilmark{\nrao,\ucb},
        Edo Berger\altaffilmark{\cfa},
        Raffaella Margutti\altaffilmark{\northwestern}
        Ashley Zauderer\altaffilmark{\ash},
        Peter K.~G. Williams\altaffilmark{\cfa},\\
        Wen-fai Fong\altaffilmark{\einsteinfellow,\ariz},
        Re'em Sari\altaffilmark{\huji},
        Kate D.~Alexander\altaffilmark{\cfa},        
        and Atish Kamble\altaffilmark{\cfa}}
\altaffiltext{\nrao}{National Radio Astronomy Observatory,
520 Edgemont Road, Charlottesville, VA 22903, USA} 
\altaffiltext{\ucb}{Department of Astronomy, University of California, 501 Campbell Hall, 
Berkeley, CA 94720-3411, USA} 
\altaffiltext{\cfa}{Harvard-Smithsonian Center for Astrophysics, 60 Garden St, 
Cambridge, MA 02138, USA}
\altaffiltext{\northwestern}{Center for Interdisciplinary Exploration and Research in Astrophysics 
(CIERA) and Department of Physics and Astrophysics,Northwestern University, Evanston, IL 60208, USA
}
\altaffiltext{\ash}{Center for Cosmology and Particle Physics, New York 
University, 4 Washington Place, New York, NY 10003, USA}
\altaffiltext{\einsteinfellow}{Einstein Fellow}
\altaffiltext{\ariz}{Steward Observatory, University of Arizona, 933 N. Cherry Ave, Tucson, AZ 
85721, USA}
\altaffiltext{\huji}{Racah Institute of Physics, The Hebrew University, 
Jerusalem 91904, Israel}

\begin{abstract}
We present detailed multi-frequency, multi-epoch radio observations of 
GRB~140304A at $z=5.283$ from 1 to 86\,GHz and 0.45\,d to 89\,d. 
The radio and mm data exhibit unusual multiple spectral components, 
which cannot be simply explained by standard forward and reverse shock scenarios. 
Through detailed multi-wavelength analysis spanning radio to X-rays, 
we constrain the forward shock parameters to 
$\EKiso\approx4.9\times10^{54}$\,erg,
$\Astar \approx 2.6\times10^{-2}$, 
$\epse\approx2.5\times10^{-2}$,
$\epsb\approx5.9\times10^{-2}$,
$p\approx2.6$, 
and $\thetajet\approx1.1^{\circ}$, 
yielding a beaming corrected $\gamma$-ray and kinetic energy,
$\Egamma\approx2.3\times10^{49}$\,erg and $\EK\approx9.5\times10^{50}$\,erg, 
respectively.
We model the excess radio emission as due to a combination of 
a late-time reverse shock (RS) launched by a shell collision, which also
produces a re-brightening in the X-rays at $\approx0.26$\,d,
and either a standard RS or diffractive interstellar scintillation.
Under the standard RS interpretation, we invoke
consistency arguments between the forward and reverse shocks to derive 
a deceleration time, $\tdec\approx100$\,s, 
the ejecta Lorentz factor, 
$\Gamma(\tdec)\approx300$, and a low RS magnetization, 
$R_{\rm B}\approx0.6$. Our observations highlight both the power
of radio observations in capturing RS emission and thus constraining
the properties of GRB ejecta and central engines, and the challenge
presented by interstellar scintillation in conclusively identifying 
RS emission in GRB radio afterglows.
\end{abstract}

\section{Introduction}
The mechanism producing the relativistic jets responsible for long-duration
$\gamma$-ray bursts (GRBs) is understood to involve a compact
central engine such as a magnetar or accreting black hole, formed during 
core collapse of a massive star \citep{wb06,pir05,mgt+11}.
A crucial clue to uncovering the nature of this mechanism and of the engine 
is provided by studies of GRB jets, requiring detailed observations and 
theoretical modeling of both the prompt $\gamma$-ray radiation from 
magnetic reconnection or shell collisions within the jet itself, and the 
afterglow generated when the jet is decelerated by the circumburst environment 
\citep{spn98}.

Facilitated by data from \Swift\ and \Fermi, such studies have revealed 
complex spectral and temporal features in both the prompt emission and the afterglow, 
suggesting that GRB jets are 
episodic and variable;
theoretical studies suggest the variability may be an intrinsic 
feature of the jet acceleration mechanism \citep{frw99,aimu02,mlb10,llm16}.
While internal shocks within the ejecta arising from the collision of material 
moving with different Lorentz factors are believed to be responsible for the 
production of the prompt $\gamma$-ray radiation \citep{kps97,dm98}, 
long-lasting central engine activity is a leading model for flares observed in 
their X-ray afterglows \citep{fw05,fbl+06,cmr+07,cmm+10,mgc+10,mbbd+11,mcg+11,bmc+11}. 

The observed $\gamma$-ray variability and late-time X-ray and optical flaring activity 
has encouraged a range of theoretical models predicting ejecta stratification, 
including fragmentation of the accretion disk due to viscous instabilities,
two-stage collapse, fall-back accretion, variability in the accretion rate,
and shell collisions
\citep{kog+05,paz06,pz06,gngc09,vvemk11,gwhy13,gdf+15,ywh+15,dptm17}.
The resulting structured ejecta profiles are expected to have a long-term impact on the afterglow,
producing a long-lasting energy injection phase as slower shells catch up with the 
forward shock \cite[FS;][]{sm00}. The injection is expected to flatten the afterglow decay, and,
if it occurs rapidly enough, to cause an achromatic re-brightening in the 
afterglow light curves \citep{kp00a,zm02,gnp03,bm17}. 
Such simultaneous optical and X-ray re-brightenings have been seen in a few instances 
\citep{mhm+07,mgg+10,hdpm+12,llt+12,gkn+13,pvw13,nef+14,dpko+15}.

The injection process is expected to be accompanied by a reverse shock (RS) 
if the collision between shells is violent, i.e., at 
large relative Lorentz factor \citep{zm02}. Identification and characterization of this RS
may lead to deeper insight into the jet production mechanisms, and by extension,
the accretion process. 
Whereas multi-wavelength modeling 
of the observed X-ray to radio light curves suggests that complex ejecta profiles
may be ubiquitous in GRB afterglows, 
these studies did not find evidence for the injection RS,
possibly due to physical effects such as the shell collision process being gentle, or due to 
observational constraints, such as limited wavelength coverage and temporal sampling of the 
spectral energy distribution (SED) of the afterglow
\citep{lbm+15}.
In particular, the RS produced by shell collisions is expected to peak in the mm-band,
where observational facilities have been scarce
\citep{duplm+12}. 

In Paper I of this series, we introduced our VLA study of radio afterglows of GRBs at $z\gtrsim5$  
(Laskar et al.~2017, submitted). Here, we present radio through X-ray observations of GRB~140304A 
at $z=5.283$, 
together with detailed multi-band modeling using physical afterglow models.
The radio and mm observations exhibit multiple components, which cannot be explained
as a standard FS and RS combination. We interpret the data in the context of a model 
requiring a RS initiated by a shell collision, and show that the resultant 
SEDs and light curves are consistent with the signatures 
of energy injection visible in the X-ray and optical observations of this event. 
The model suggests expansion into a wind-like medium, and identification
of one of the radio components as the standard RS yields a measurement of the 
ejecta Lorentz factor at the deceleration time. We employ a 
standard cosmology of 
$\Omega_{m}=0.31$, $\Omega_{\Lambda}=0.69$, and $H_0=68$\,km\,s$^{-1}$\,Mpc$^{-1}$,
all magnitudes are in the AB system, all uncertainties are at $1$\,sigma,
and all times refer to the observer frame, unless otherwise specified.

\section{GRB Properties and Observations}
\label{text:GRB_Properties_and_Observations}
GRB~140304A was discovered by the \Swift\ Burst Alert Telescope \citep[BAT,][]{bbc+05} on 2014 
March 4 at 13:22:31\,UT \citep{gcn15915}. The burst duration in the 15--350\,keV BAT energy band is 
$T_{90} = 15.6\pm1.9$\,s, with a fluence of $F_{\gamma} = (1.2 \pm 0.1) 
\times10^{-6}$\,erg\,\pcmsq\ \citep[15--150\,keV;][]{gcn15927}. A bright optical afterglow was 
detected by the MASTER robotic network \citep{gcn15914}, subsequently confirmed by other 
ground-based observatories \citep{gcn15916,gcn15917,gcn15918,gcn15921}. Spectroscopic observations 
8.2\,hr after the burst at the 10.4\,m Gran Telescope Canarias (GTC) provided a redshift of $z = 
5.283$ \citep{gcn15936}.

The burst also triggered the \Fermi\ Gamma-ray Burst Monitor (GBM) at 13:22:31.48\,UT 
\citep{gcn15923}. The burst duration in the 50--300\,keV GBM band is $T_{90} = 32\pm6$\,s 
with a fluence of $(2.0\pm0.2)\times10^{-6}$\,erg\,\pcmsq (10--1000\,keV). 
A Band-function fit to the time-averaged $\gamma$-ray spectrum\footnote{From the \Fermi\ GRB 
catalog for trigger 140304557}
yields a break energy, $E_{\rm peak}=123\pm27$\,keV, low energy index, $\alpha=-0.80\pm0.22$, and 
high-energy index, $\beta = -2.35\pm0.43$. Using the source redshift of $z=5.283$, the inferred 
isotropic equivalent $\gamma$-ray energy in the 1-$10^4$ keV rest frame energy band is 
$\Egammaiso=(1.2\pm0.2)\times10^{53}$\,erg.

\subsection{X-ray: \Swift/XRT}
\label{text:data_analysis:XRT}
The \Swift\ X-ray Telescope \citep[XRT,][]{bhn+05} began observing the field at 75.2 seconds after 
the BAT trigger, leading to the detection of an X-ray afterglow. The source was
localized to RA = 2h\,2m\,34.26s, Dec = +33d\,28\arcmin\,25.7\arcsec\ (J2000), with an
uncertainty radius of 1.5\arcsec (90\% 
containment)\footnote{\url{http://www.swift.ac.uk/xrt_positions/00590206/}}. XRT continued 
observing the afterglow for 5.3\,d in photon counting mode, with the last detection at 3.0\,d.

We extracted XRT PC-mode spectra using the on-line tool on the \Swift\ website 
\citep{ebp+07,ebp+09}\footnote{\url{http://www.swift.ac.uk/xrt_spectra/00590206/}} 
in the intervals 125\,s to 1557\,s 
(spectrum 1) and 5192\,s to $5.78\times10^{5}$\,s (spectrum 2)\footnote{All analysis reported in 
this section excludes the flare between 20\,ks and 23\,ks}. 
We downloaded the event and response 
files generated by the on-line tool in these time bins, and fit them using the 
HEASOFT (v6.16) and corresponding calibration files. 
We used Xspec to fit all available 
PC-mode data, assuming a photoelectrically absorbed power law model (\texttt{tbabs $\times$ ztbabs 
$\times$ pow}), fixing the galactic absorption column at $N_{\rm H, Gal} = 
7.68\times10^{20}\,\pcmsq$ 
\citep{wsb+13}, and tying the value of the intrinsic absorption in the host galaxy, $N_{\rm H, 
int}$, to be the same between the two spectra since we do not expect any evolution in the intrinsic 
absorption with time. We find marginal evidence for spectral evolution
between the two spectra across the orbital gap; the results are summarized in Table 
\ref{tab:xrtspect}. 
In the following analysis, we take the 0.3 -- 10\,keV count rate light curve from the 
\Swift\ website and compute the 1\,keV flux density using our spectral models,
with $\Gamma_{\rm X} = 1.87$ before 1557\,s and $\Gamma_{\rm X} = 2.27$ thereafter.
We combine the 
uncertainty in flux calibration based on our spectral analysis (7\% in spectrum 1 and 16\% in 
spectrum 2) in quadrature with the statistical uncertainty from the on-line light curve. 

For the WT-mode, we convert the count rate light curve to a flux-calibrated light curve using 
$\Gamma=2.5$ and an unabsorbed 
count-to-flux conversion factor of $3.9\times10^{-11}{\rm erg}\,\pcmsq {\rm 
ct}^{-1}$ as reported on the \Swift\ website. The WT-mode X-ray light curve declines rapidly as 
$t^{-4.0\pm0.8}$ to $1.3\times10^{-3}$\,d. Similar early, rapidly declining X-ray light curves
are frequently observed in XRT light curves, and have been speculated to arise from 
the high-latitude component of the prompt emission \citep{kp00,tgc+05,nkg+06,wggo10}.
Alternatively, this steep decay could also arise from the end of a preceding flare, 
the beginning of which was missed during spacecraft slew.
The PC mode data beginning at 146\,s are also dominated by flaring activity until 
$\approx2\times10^{-2}$\,d. We therefore do not consider the X-ray data before $2\times10^{-2}$\,d 
in our afterglow modeling.

\begin{deluxetable}{lcc}
 \tabletypesize{\footnotesize}
 \tablecolumns{3}
 \tablecaption{XRT Spectral Analysis for GRB 140304A\label{tab:xrtspect}}
 \tablehead{   
   \colhead{Parameter} &
   \colhead{Spectrum 1} &
   \colhead{Spectrum 2} \\   
   }
 \startdata 
 $T_{\rm start}$ (s) & 125  & 5192\\
 $T_{\rm end}$ (s)   & 1557 & $5.78\times10^{5}$\\
 $N_{\rm H, gal}$ ($10^{20}$ \pcmsq)
                     & \multicolumn{2}{c}{$5.98$} \\
 $N_{\rm H, int}$ ($10^{22}$ \pcmsq) &
				      \multicolumn{2}{c}{$3.8^{+1.5}_{-1.4}$} \\
 Photon index, $\Gamma$ & $1.87\pm0.07$ & $2.27\pm0.15$\\
 Flux (obs\tablenotemark{\dag}) & 
                   $1.61^{+0.11}_{-0.10}\times10^{-10}$ & $1.62^{+0.30}_{-0.23}\times10^{-13}$\\
 Flux (unabs\tablenotemark{\ddag})  & 
                   $2.0\times10^{-10}$ & $2.3\times10^{-13}$\\
 Counts to flux (obs\tablenotemark{\dag}) & 
                   $2.6\times10^{-10}$ & $3.6\times10^{-11}$ \\
 Counts to flux (unabs\tablenotemark{\ddag}) & 
                   $3.3\times10^{-10}$ & $5.1\times10^{-11}$ \\
 C statistic (dof) & \multicolumn{2}{c}{425 (469)}
 \enddata
 \tablecomments{$^{\dag}$0.3--10\,keV, observed (erg\,\pcmsq\,s$^{-1}$);
 $^{\ddag}$0.3--10\,keV unabsorbed (erg\,\pcmsq\,ct$^{-1}$).}
\end{deluxetable}

\subsection{Optical}
\label{text:data_analysis:optical}
The \Swift\ UV/Optical Telescope \citep[UVOT,][]{rkm+05} observed \me\ beginning 138\,s after 
the burst \citep{gcn15920}. We analyzed the UVOT data using HEASOFT (v. 6.16) and 
corresponding calibration files and list our derived upper limits in Table \ref{tab:data:UVOT}. We 
compiled all observations reported in GCN circulars and present the compilation in Table 
\ref{tab:data:GCN}.

\begin{deluxetable}{ccc}
\tabletypesize{\scriptsize}
\tablecaption{\Swift\ UVOT Observations of GRB~140304A \label{tab:data:UVOT}}
\tablehead{
\colhead{$\Delta t$} & \colhead{Filter} & \colhead{$3\sigma$ Flux Upper 
Limit\tablenotemark{a}} \\ 
\colhead{(d)} & \colhead{} & \colhead{(mJy)} }

\startdata
$7.08\times10^{-2}$ &	\it{white} &	$2.38\times10^{1}$ \\
$7.60\times10^{-2}$ &	\it{b}     &	$3.70\times10^{1}$ \\
$2.67\times10^{-1}$ &	\it{b}     &	$7.24\times10^{1}$ \\
\ldots & \ldots & \ldots
\enddata
\tablecomments{This is a sample of the full table available on-line.}
\end{deluxetable}

\begin{deluxetable*}{ccccccccc}
\tabletypesize{\scriptsize}
\tablecaption{Optical Observations of GRB~140304A \label{tab:data:GCN}}
\tablehead{
\colhead{$\Delta t$} & \colhead{Observatory} & \colhead{Instrument} &
\colhead{Filter} & \colhead{Frequency} & \colhead{Flux density} & 
\colhead{Uncertainty\tablenotemark{\dag}} & \colhead{Detection?} & \colhead{Reference}\\ 
\colhead{(d)} & & & & \colhead{(Hz)} & \colhead{(mJy)} & \colhead{(mJy)} & \colhead{1=Yes} }

\startdata
$9.49\times10^{-4}$ & ICATE & MASTER & \textit{CR} & $4.56\times10^{14}$ & $7.70\times10^{-1}$ & 
    $1.56\times10^{-1}$ & 1 & \cite{gcn15932} \\
$1.99\times10^{-3}$ & ICATE & MASTER & \textit{CR} & $4.56\times10^{14}$ & $4.64\times10^{-1}$ & 
    $9.38\times10^{-2}$ & 1 & \cite{gcn15932} \\
$3.40\times10^{-3}$ & ICATE & MASTER & \textit{CR} & $4.56\times10^{14}$ & $2.32\times10^{-1}$ & 
    $4.70\times10^{-2}$ & 1 & \cite{gcn15932} \\
\ldots & \ldots & \ldots & \ldots & \ldots & \ldots & \ldots & \ldots & \ldots
\enddata
\tablecomments{\dag An uncertainty of 0.2~AB mag is assumed where not provided. The data have 
not been corrected for Galactic extinction. This is a sample of the full table available on-line.}
\end{deluxetable*}

\subsection{Millimeter: CARMA}
\label{text:data_analysis:millimeter}
We observed GRB\,140304A with the Combined Array for Research in Millimeter Astronomy (CARMA) 
beginning on 2014 March 04.02 UT (0.54 d after the burst) in continuum wideband mode with 8 
GHz bandwidth (16 windows, 487.5 MHz each) at a mean frequency of 85.5 GHz. Following an initial 
detection \citep{gcn15931}, we obtained two additional epochs. All observations utilized J0237+288 
as phase calibrator.  The first two epochs additionally utilized 3C84 as bandpass calibrator and 
Uranus as flux calibrator. For the third epoch, the array shut down due to high winds, truncating 
observations at a total track length of 1.9~h and preventing observations of the flux calibrator. 

We derived a linelength calibration to account for thermal changes in the delays through the optical
fibers connecting the CARMA antennas to the correlator using MIRIAD \citep{stw95}, 
and performed the rest of the data 
reduction using the Common Astronomy Software Applications (CASA; \citealt{mws+07}). For the third 
epoch, we used the flux density per spectral window and mean spectral index ($-0.81\pm0.02$) of 
3C84 
derived from the first two epochs for flux calibration. Our derived flux density values for the 
gain 
calibrator in the third epoch are consistent at all spectral windows with the values obtained from 
the first two epochs, where Uranus was available as a flux calibrator. We summarize our 
mm-band observations in Table \ref{tab:data:radio}.

\subsection{Centimeter: VLA}
\label{text:data_analysis:radio}
We observed the afterglow using the Karl G. Jansky Very Large Array (VLA) starting 0.45\,d after 
the 
burst. We detected and tracked the flux density of the afterglow from 1.2\,GHz to 
33.5\,GHz over seven epochs until 89\,d after the burst, when it faded beyond detection at all 
frequencies. We used 
3C48 as the flux and bandpass calibrator and J0205+3212 as gain calibrator. We carried out data 
reduction using CASA, and list the results of our VLA 
observations in Table \ref{tab:data:radio}.

\begin{deluxetable}{clcccc}
\tabletypesize{\scriptsize}
\tablecaption{GRB\,140304A: Log of radio observations \label{tab:data:radio}}
\tablehead{
\colhead{$\Delta t$} & \colhead{Facility} & \colhead{Frequency} & \colhead{Flux density} & 
\colhead{Uncertainty} & \colhead{Det.?}\\
\colhead{(d)} & & \colhead{(GHz)} & \colhead{(mJy)} & \colhead{(mJy)}
}
\startdata
0.45  &  VLA          &  4.9   &  0.036    &  0.012    &  1\\
0.45  &  VLA          &  7.0   &  0.073    &  0.011    &  1\\
0.54  &  CARMA        &  85.5  &  0.656    &  0.235    &  1\\
\ldots & \ldots & \ldots & \ldots & \ldots & \ldots
\enddata
\tablecomments{The last column indicates a detection (1) or non-detection (0). This is a sample of 
the full table available on-line.}
\end{deluxetable}

\section{Basic Considerations}
\subsection{Optical and X-rays}
\label{text:basic_considerations}
The X-ray light curve exhibits a steep decline followed by flaring behavior until the first 
\Swift\ orbital gap beginning at $1.7\times10^{-2}$~d. Such flaring behavior in the early X-ray 
light curve is often attributed to prolonged central engine activity \citep{brf+05}, and we 
therefore do not consider it further in the context of the afterglow. The subsequent X-ray light 
curve exhibits a flare or re-brightening event at $\approx 0.26$~d, where the light curve 
rises steeply by a factor of $\approx4$ between 0.13 and 0.26~d, 
corresponding to a rise rate\footnote{We employ the convention, 
$F_{\nu} \propto t^{\alpha}\nu^{\beta}$ throughout.}, $\alpha=2.0\pm0.3$. 
Such late-time flares and 
re-brightenings are less common in GRB X-ray light curves \citep{cso+08,bmc+11} and have variously 
been ascribed to instabilities in the accreting system \citep{paz06,knj08a,ros07}, magnetic 
field-driven turbulence \citep{pz06}, magnetic reconnection \citep{gia06} or by energy 
injection due to low-Lorentz factor ejecta \citep{mgg+10,hdm12,xpk+12,lbm+15}.

\begin{figure}
 \includegraphics[width=\columnwidth]{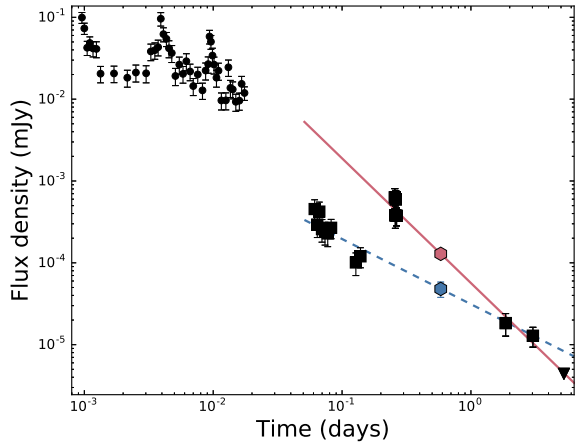}
 \caption{\Swift\ XRT light curve of GRB~140304A at 1\,keV (black points). Data before 0.02~d are 
dominated by flaring activity, while after the first orbital gap (starting at 0.02~d) 
the light curve exhibits a large flare / 
re-brightening event at 0.26~d. The lines are fits to data from 0.2 to 4.0~d (red) and from 0.04 
to 4.0~d excluding the flare (blue). The latter would require an additional break at $\approx4$~d 
to account for the upper limit at 5.3~d. The colored points at 0.58~d are the inferred interpolated 
flux density at 0.58~d from the two power law fits, derived for the purpose of comparing with 
multi-wavelength RATIR observations at this time (Section \ref{text:basic_considerations}
and Figure \ref{fig:betanirx}).}
\label{fig:XRT-lc}
\end{figure}

In the scenario where the XRT data at 0.26~d are dominated by a flare, a fit to the X-ray light 
curve between 0.04 and 4.0~d ignoring the flare
yields a power law decay rate of $\alpha_{\rm X} = -0.80\pm0.12$. 
Interpolating the X-ray light curve using this value to 0.58~d 
(the time of the first RATIR optical/near-IR observation;
\citealt{gcn15928,gcn15937})
yields a flux density of $F_{\nu,\rm X}(0.58~{\rm d}) = (4.8\pm1.0)\times10^{-5}$~mJy 
(Figure \ref{fig:XRT-lc}). On the other hand, assuming that the XRT 
data at 0.26~d are part of a re-brightening event (and dominated by afterglow emission), 
a fit to the light curve at $\gtrsim0.24$\,d yields a decay rate of $\alpha_{\rm X} = -1.5\pm0.1$ 
with an interpolated flux density of $F_{\nu,\rm X}(0.58~{\rm d}) = (12.9\pm1.4)\times10^{-5}$~mJy.

\begin{figure}
 \includegraphics[width=\columnwidth]{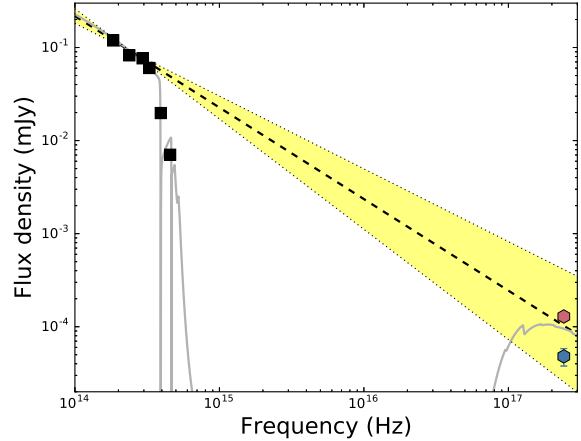}
 \caption{NIR to X-ray spectral energy distribution of the afterglow of GRB~140304A at 
0.58~d. The NIR and optical data (black points) are from RATIR \citep{gcn15928}, 
while the X-ray data 
(red and blue points) have been interpolated using two different fits to the \Swift\ XRT 
light curve (Figure \ref{fig:XRT-lc}). The dashed line and the shaded regions indicate the best fit 
power law to the RATIR \textit{zYJH} data and its $1\sigma$ error bound. The X-ray to NIR SED is 
consistent with a single power law, though the X-ray spectrum suggests the cooling frequency 
may be located between the NIR and X-ray bands at 0.58~d. The grey solid line is the best-fit
model (Section \ref{text:wind_fs}).}
\label{fig:betanirx}
\end{figure}

We plot these interpolated X-ray flux density measurements together with the RATIR optical and 
near-IR (NIR) observations at 0.58~d in Figure \ref{fig:betanirx}. A power law fit to the four 
longest wavelength RATIR observations (\textit{zYJH}) after correction for Galactic extinction 
yields a spectral index of $\beta_{\rm NIR-opt}=-0.98\pm0.20$. The flux density in the $r'$ and $i'$ 
bands is considerably lower than the extrapolation of this power law, consistent with intergalactic 
medium (IGM) absorption given the redshift of $z=5.283$. In comparison, the spectral index 
between the RATIR $H$-band measurement and the interpolated X-ray flux density is $\beta_{\rm NIR-X} 
= -0.96\pm0.02$ or $\beta_{\rm NIR-X} = -1.09\pm0.03$.
The X-ray spectral index, $\beta_{\rm X} = -1.29\pm0.25$ is marginally steeper,
suggesting a break frequency may lie between the optical and X-rays. 

RATIR claimed a second detection of the optical afterglow in $z'$-band at 1.5\,d at low 
significance ($4.9\sigma$). We note that the measured flux density is greater than the upper 
limit in the adjacent $Y$-band in the same epoch. Assuming this second $z'$-band detection is real, 
the decay rate between 0.58\,d and 1.59\,d in this band is $\alpha_{\rm z} = -1.3\pm0.2$. On the 
other hand, the constraints on the decay rate from the other bands with a detection at 0.58\,d are 
$\alpha_{\rm i} \lesssim -1.8$, $\alpha_{\rm Y} \lesssim -1.6$, 
$\alpha_{\rm J} \lesssim -1.4$, and $\alpha_{\rm H} \lesssim -1.5$.
Further analysis requires a simultaneous understanding of the radio light curves, 
and we return to this point in Section \ref{text:singlecompradio}.

\begin{figure} 
 \includegraphics[width=\columnwidth]{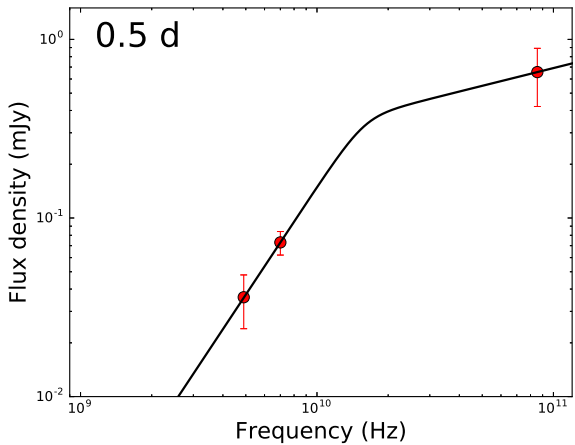}
 \caption{Radio SED of GRB 140304A at 0.5\,d with a single broken power law fit. The low- 
and high-frequency spectral indices are fixed at 2 and $1/3$, respectively (Section 
\ref{text:radio_bplfits_0}).}
\label{fig:radio_bplfits_0}
\end{figure}

\begin{figure*} 
 \begin{tabular}{cc}
  \includegraphics[width=\columnwidth]{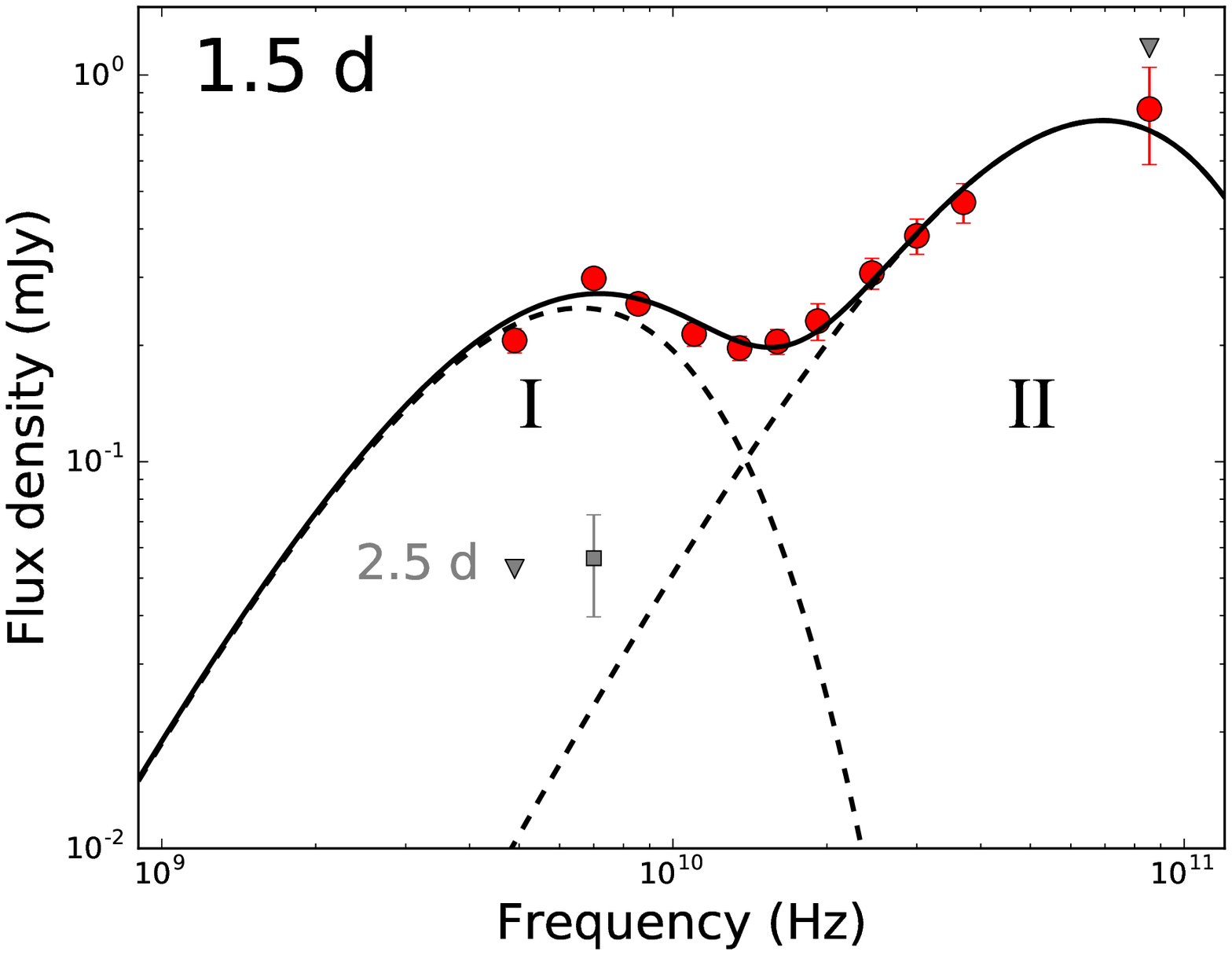} &
  \includegraphics[width=\columnwidth]{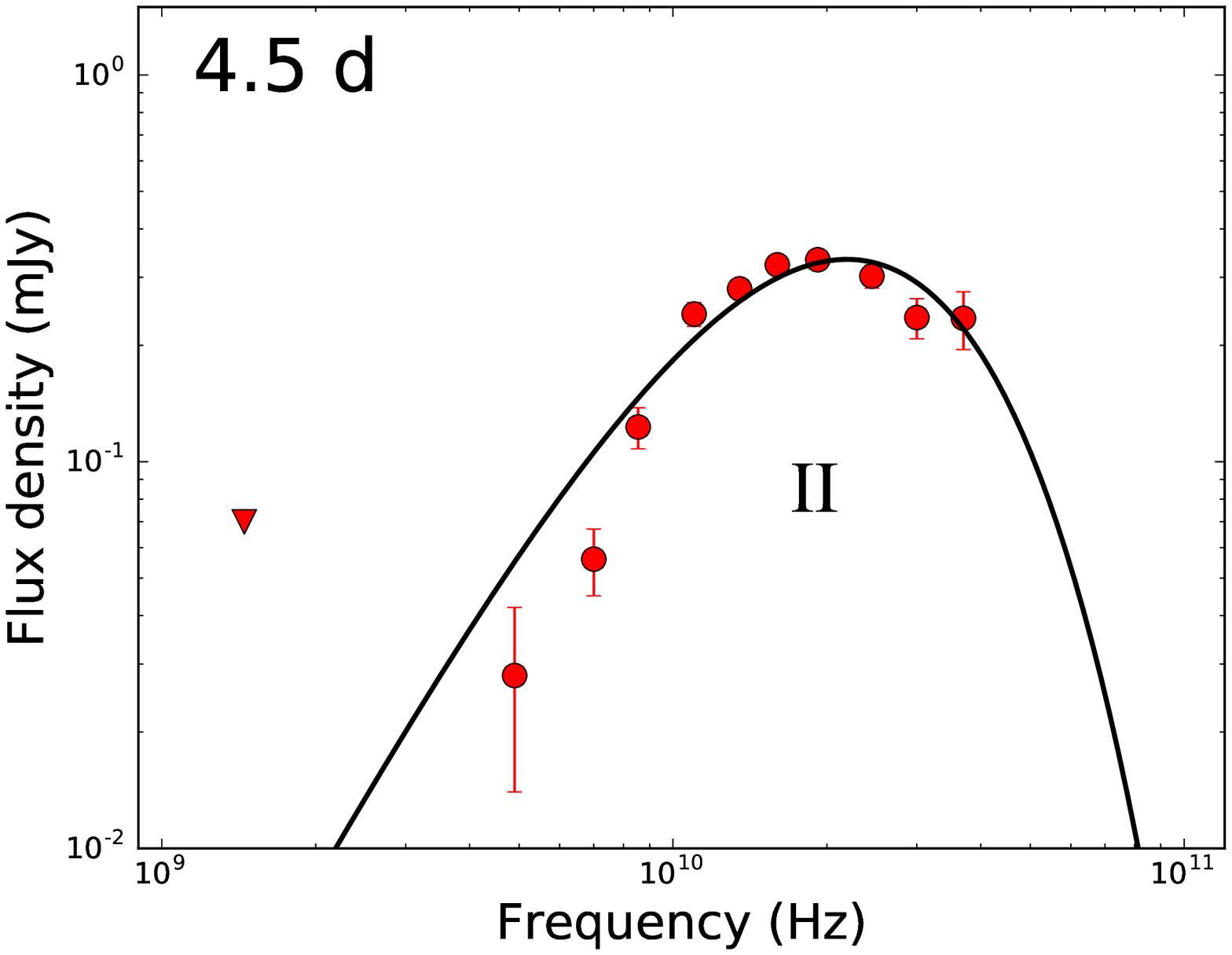} \\
  \includegraphics[width=\columnwidth]{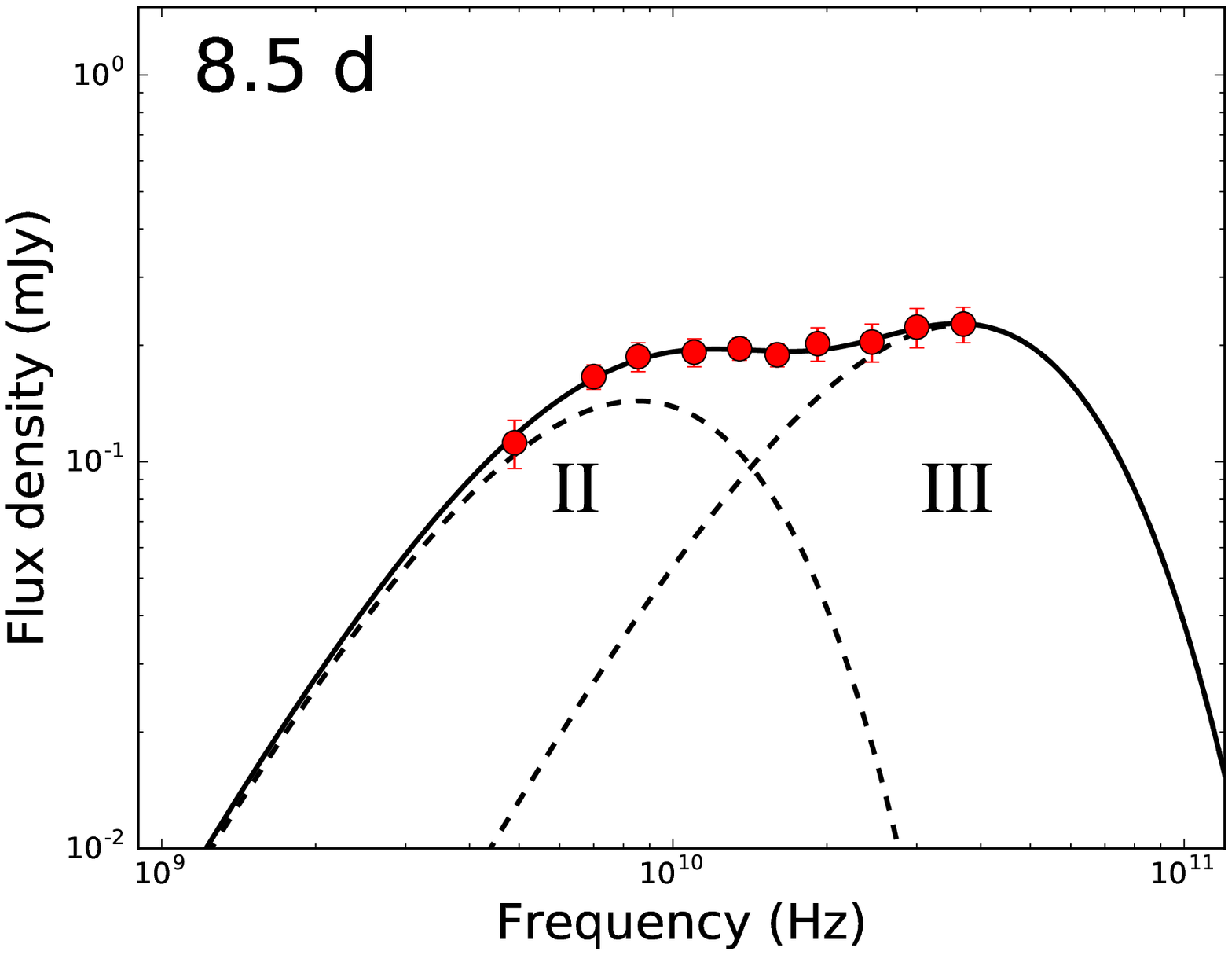} &
  \includegraphics[width=\columnwidth]{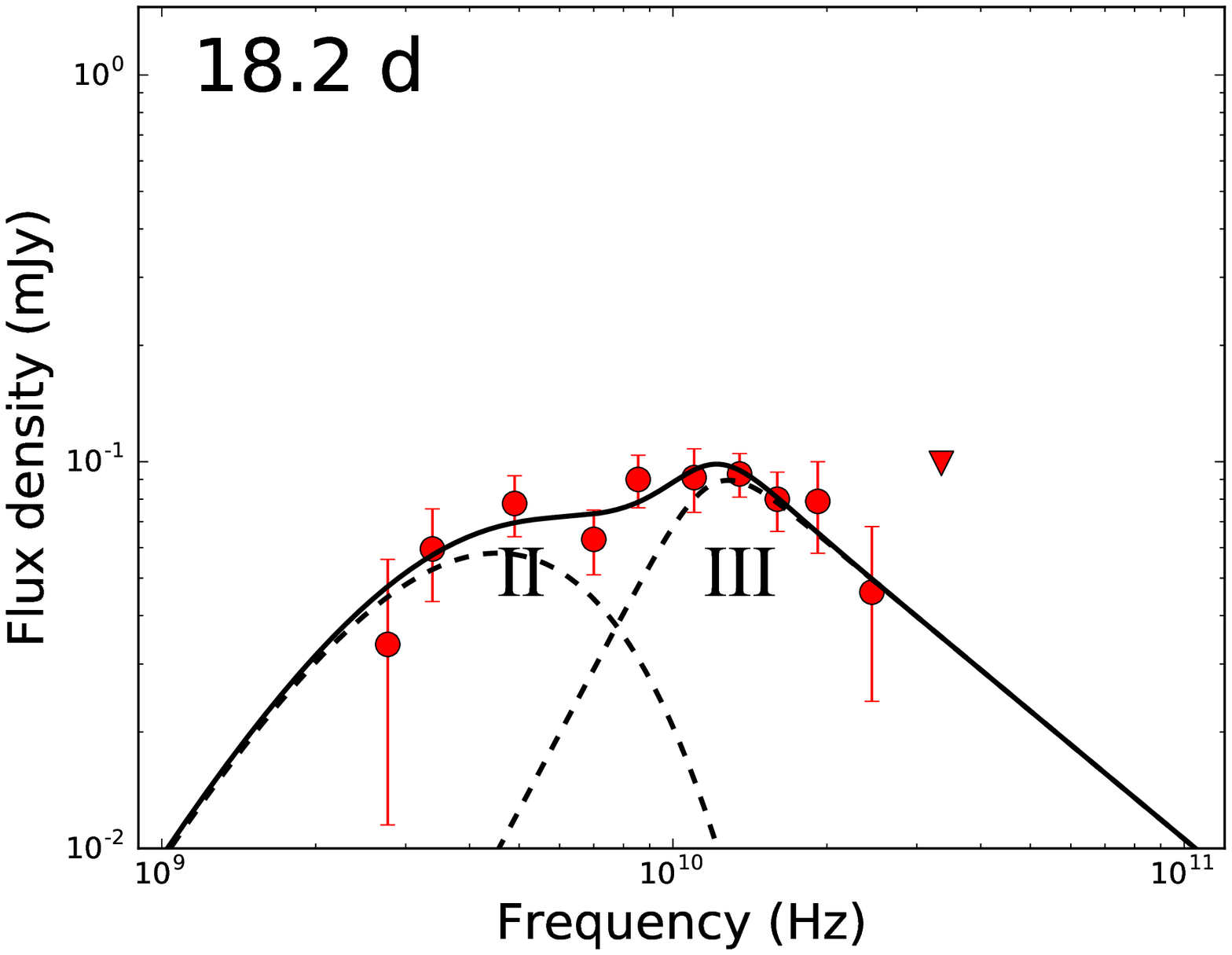} \\
  \includegraphics[width=\columnwidth]{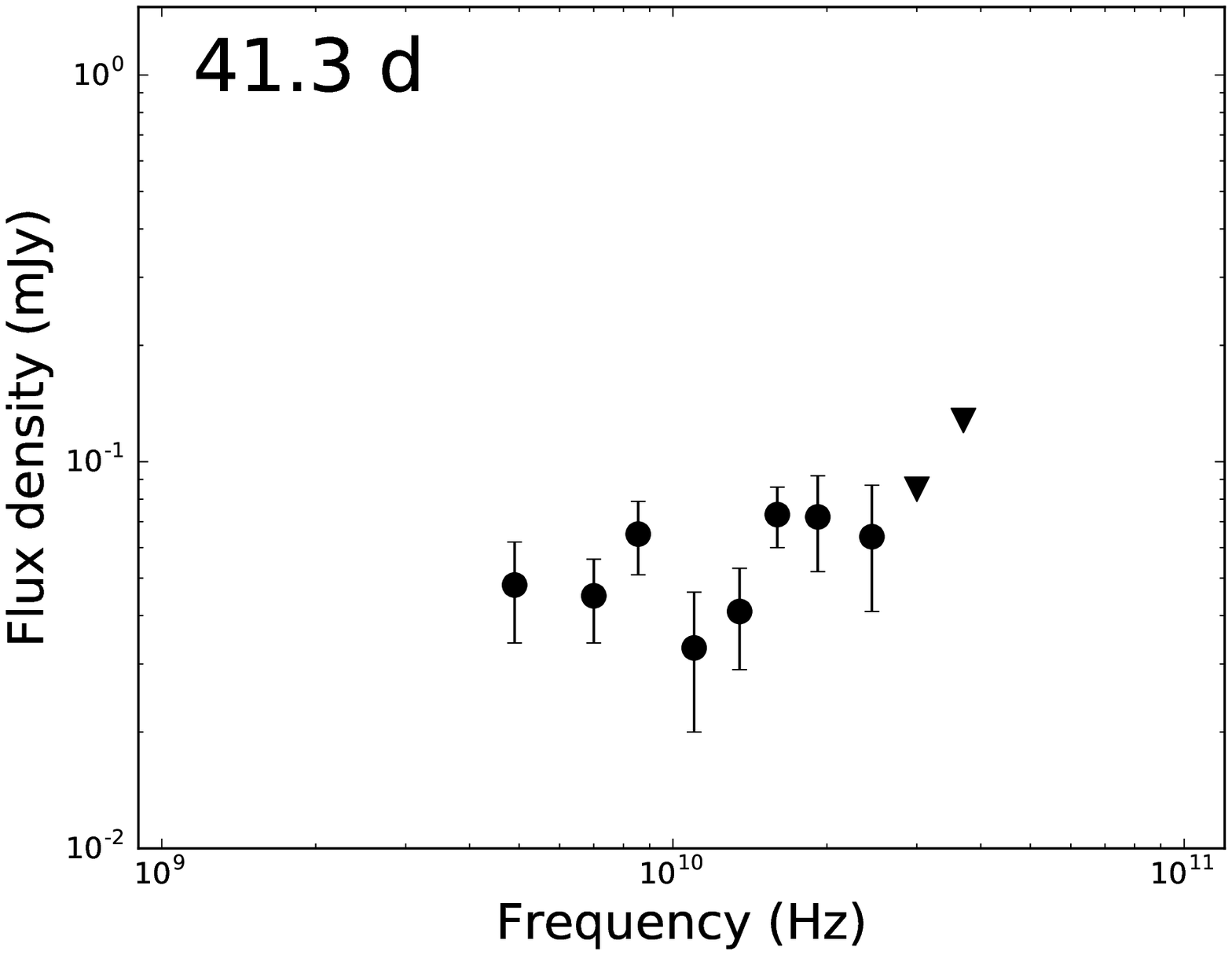} &
  \includegraphics[width=\columnwidth]{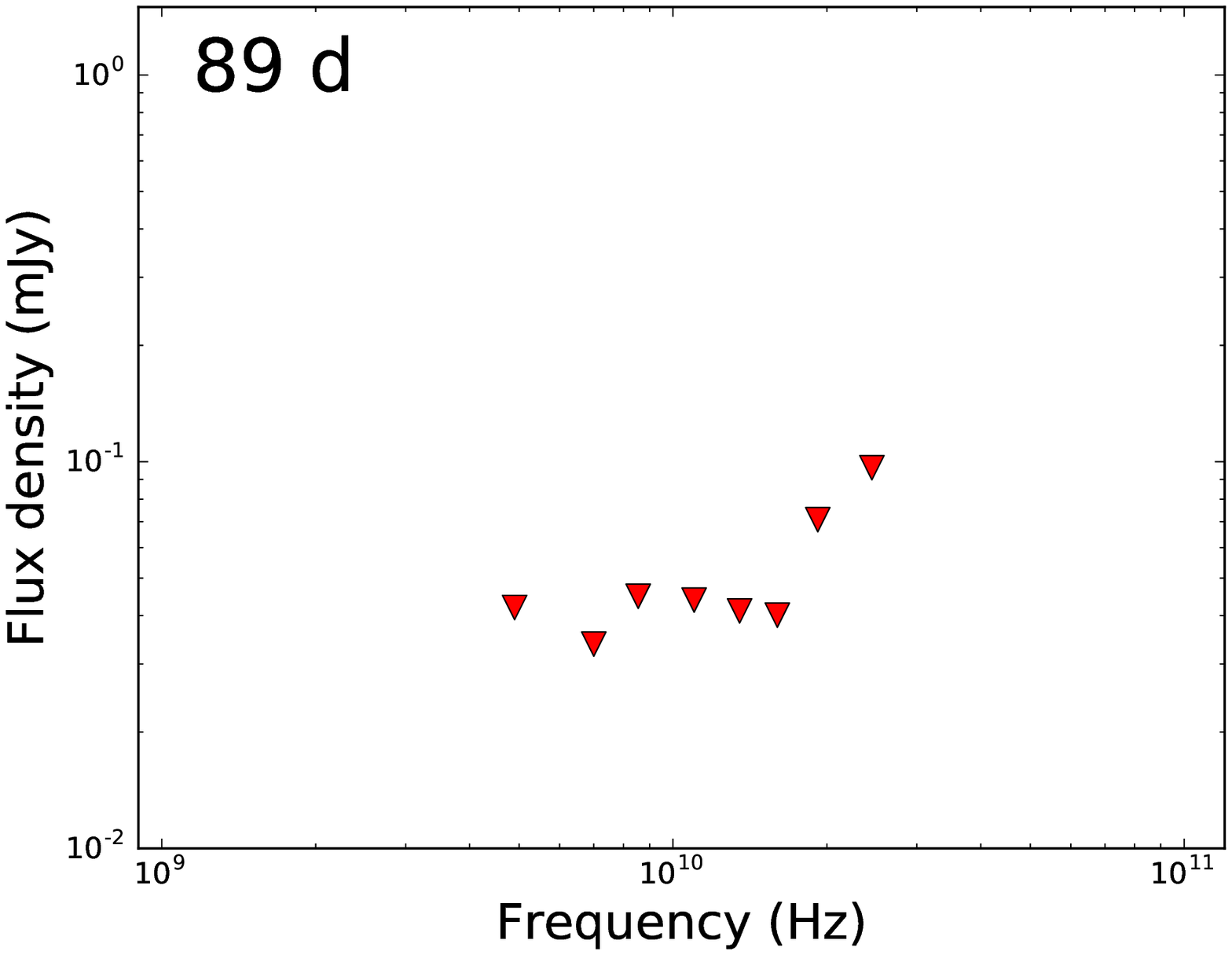} 
 \end{tabular}
 \caption{Radio SEDs of GRB 140304A (red points) at four epochs.
Top left -- SED at 1.5\,d with sum of two 
power law plus exponential cut-off models (Section \ref{text:radio_bplfits_1}). Also shown are the 
available radio data at 2.5\,d (grey square and upper limits; Section \ref{text:radio_bplfits_2}). 
Top right -- SED at 4.5\,d with a broken power law fit (Section \ref{text:radio_bplfits_4}).
Center left -- SED at 8.5\,d with sum of two power law plus exponential cut-off models (Section 
\ref{text:radio_bplfits_8}).
Center right -- SED at 18.2\,d with sum of a power law plus exponential cut-off and broken power 
law model. The high-frequency spectral index of the higher frequency component is not well 
constrained; the model shown here is plotted with $\beta_2=-1.1$ and $y=3$ (Section 
\ref{text:radio_bplfits_18}).
The low-frequency spectral index is fixed at 2.5 for both components in all four epochs from 
1.5\,d to 18.2\,d. Bottom panel -- SEDs at a mean time of 41.3\,d (left) 
and 89\,d (right; Section \ref{text:radio_bplfits_4080}).
}
\label{fig:radio_bplfits}
\end{figure*}

\begin{deluxetable*}{lccccc}
 \tabletypesize{\footnotesize}
 \tablecolumns{14}
 \tablecaption{Radio spectral fits for GRB 140304A\label{tab:radio_bplfits}}
 \tablehead{   
    \colhead{$\Delta T$ (days)} &
    \colhead{1.5} &
    \colhead{4.5} &
    \colhead{8.5} &
    \colhead{18.2} &
    \colhead{18.2}
   }
 \startdata 
   Ncomp      & 2   & 1   & 2   & 1${}^\S$    & 2${}^\S$    \\
   Type\tablenotemark{\dag} & exp & exp & exp & bpl & exp \\
   $\nu_{\rm{peak}}$ (GHz) & $6.6\pm0.4$ & $21.9\pm1.6$ & $8.5\pm0.3$ & $15.6\pm3.2$ & $4.5\pm0.8$\\
   $F_{\rm{peak}}$ (mJy) & $0.25\pm0.01$ & $0.33\pm0.02$ & $0.143\pm0.005$ & $0.086\pm0.007$ & 
   $0.058\pm0.07$\\
   $\beta_1$ & 2.5\tablenotemark{\ddag} & 2.5\tablenotemark{\ddag} 
             & 2.5\tablenotemark{\ddag} & $0.45\pm0.18$ & 2.5\tablenotemark{\ddag} \\
   $\beta_2$ & \ldots & \ldots & \ldots & $-1.7\pm1.0$ & \ldots \\
   $y$ & \ldots & \ldots & \ldots & 3.0\tablenotemark{\ddag} & \ldots \\ 
   Type\tablenotemark{\dag} & exp & \ldots & exp & \ldots & bpl \\
   $\nu_{\rm{peak}}$ (GHz) & $69\pm13$ & \ldots & $36.9\pm2.1$ & \ldots & $11.9\pm1.1$ \\
   $F_{\rm{peak}}$ (mJy) & $0.76\pm0.20$ & \ldots & $0.226\pm0.006$ & \ldots & $0.09\pm0.01$\\
   $\beta_1$ & 2.5\tablenotemark{\ddag} & \ldots 
             & 2.5\tablenotemark{\ddag} & \ldots & 2.5\tablenotemark{\ddag} \\
   $\beta_2$ & \ldots & \ldots & \ldots & \ldots & -1.1\tablenotemark{\ddag} \\ 
   $y$        & \ldots & \ldots & \ldots & \ldots & 3.0\tablenotemark{\ddag} 
 \enddata
 \tablecomments{
 ${}^\S$The SED at 18.2\,d can be fit with one or two components. 
 ${}^\dag$Type = `bpl' indicates a broken power law model (equation 
\ref{eqn:model1}), and type = `exp' indicates an exponential cut-off model (equation 
\ref{eqn:model2}). ${}^{\ddag}$Held fixed.}
\end{deluxetable*}

\subsection{Radio}
\label{text:radio_bplfits}
We now discuss the radio SED at each epoch from 0.45\,d to 89\,d. 
The radio emission is expected to arise from a combination of FS radiation 
from the interaction of the relativistic GRB ejecta with the circumburst environment, and RS
radiation from within the ejecta itself. Whereas the FS continually accelerates 
electrons, resulting in radio spectra comprising smoothly joined broken power law 
components, radiation from the RS arises from a cooling population of electrons and 
declines exponentially above the so-called cut-off frequency.
Motivated by these
physical possibilities, we fit the data with a combination of the following models in each 
instance wherever data quality allows:

Model 1 -- Broken power law:
\begin{equation}
\label{eqn:model1}
F_{\nu} = F_{\rm b}
\left(\frac{(\nu/\nu_{\rm b})^{-y\beta_1} + (\nu/\nu_{\rm b})^{-y\beta_2}}{2}\right)^{-1/y}.
\end{equation}
This model has the property that $F_{\nu}(\nu_{\rm b}) = F_{\rm b}$; however, the exact location 
of the peak (where ${\partial F_{\nu}}/{\partial{\nu}} = 0$) depends on $\beta_1$ and $\beta_2$.

Model 2 -- Power law with exponential cut-off:
\begin{equation}
\label{eqn:model2}
F_{\nu} = F_{\rm b}
\left(\frac{\nu}{\nu_{\rm b}}\right)^{\beta_1}e^{\beta_1\left(1-\nu/\nu_{\rm b}\right)}.
\end{equation}
This model has the properties that $F_{\nu}(\nu_{\rm b}) = F_{\rm b}$ and $\frac{\partial 
F_{\nu}}{\partial{\nu}}\rvert_{\nu_{\rm b}} = 0$.

\subsubsection{SED at 0.45 d}
\label{text:radio_bplfits_0}
Rapid response observations at the VLA and CARMA yielded C-band ($5$ and $7$\,GHz) and 
85.5\,GHz detections of the afterglow at 0.45\,d (1.7\,h in the rest frame).
The 5\,GHz to 7\,GHz spectral index is steep, $\beta = 2\pm1$, while the 7\,GHz to 
85.5\,GHz spectrum is shallower, $\beta = 0.9\pm0.2$. We fit the data at 0.5\,d with a broken power
law model (model 1; equation \ref{eqn:model1}), with the spectral indices fixed at $\beta = 2$ at 
low frequencies. We note that the data do not allow us to constrain both the high frequency 
spectral index and the location of the spectral break simultaneously. We set the 
high-frequency spectral index to $\beta = 1/3$, corresponding to the optically thin low-frequency
tail of synchrotron emission from shock-accelerated electrons, and fix the smoothness, $y = 5$. 
In the best 
fit model (Figure \ref{fig:radio_bplfits_0}), the spectral break occurs at $16$\,GHz and a flux 
density of $0.24$\,mJy with uncertainty $\approx10\%$, dominated by the uncertain value of $y$.

\subsubsection{SED at 1.5 d}
\label{text:radio_bplfits_1}
We sampled the afterglow radio spectrum at 11 approximately evenly logarithmically-spaced 
frequencies spanning 5\,GHz to 90\,GHz with the VLA and CARMA at 1.5\,d, yielding the most detailed
spectral radio coverage of any GRB radio afterglow at the time of acquisition.
The SED at 1.5\,d exhibits two emission peaks, a feature unexpected in GRB radio afterglows. 
The spectral index between 5\,GHz and 7\,GHz remains steep, with $\beta = 1.0\pm0.2$.

We fit the spectrum with a sum of two exponential cut-off models (each model 2; equation 
\ref{eqn:model2}). To account for the steep spectral index at the lower frequency end of 
both observed peaks in the spectrum, we fix the model spectral index at $\beta = 2.5$ for both 
components. The peak frequencies in our best fit model are located at $(6.6\pm0.4)$\,GHz and 
$(69\pm13)$\,GHz, with flux densities of $(0.25\pm0.01)$\,mJy and $(0.76\pm0.20)$\,mJy, 
respectively (Figure \ref{fig:radio_bplfits}). We summarize these results in Table 
\ref{tab:radio_bplfits}.

\subsubsection{SED at 2.5 d}
\label{text:radio_bplfits_2}
Only C-band data were obtained at 2.5\,d due to a scheduling constraint. 
The afterglow was only marginally-detected in the 
upper side band (7\,GHz; Figure \ref{fig:radio_bplfits}), implying that the emission component 
creating the spectrum peak at $6.4$\,GHz had faded away by 2.5\,d with a decline rate, 
$\alpha \approx -3$. 

\subsubsection{SED at 4.5 d}
\label{text:radio_bplfits_4}

We obtained another VLA radio SED spanning 1.4\,GHz to 40\,GHz at 4.5\,d, the mm-band afterglow 
having faded beyond the detection limit of CARMA. No source was detected at L-band (1.4\,GHz), 
while the remainder of the radio SED exhibits a steep low-frequency spectrum and a clear 
single peak. The data can be fit with a single broken power law 
model with $\beta_1 = 2.5$ (fixed), $\beta_2 = -0.69\pm0.39$, $\nu_{\rm b} = (14\pm1)$\,GHz, and 
$F_{\rm b} = 0.30\pm0.02$\,mJy (Figure \ref{fig:radio_bplfits}).

\subsubsection{SED at 8.5 d}
\label{text:radio_bplfits_8}
Our next full radio SED at 8.5\,d 
spanning 4.9\,GHz to 40\,GHz 
yields a spectrum that rises steeply from 4.9\,GHz to 7\,GHz, with $\beta = 1.1\pm0.4$ 
and exhibits a plateau to 40\,GHz, with marginal evidence for two components. We fit the 
spectrum with a sum of two exponential cut-off models, fixing the slopes of the power law 
components at $\beta = 2.5$. The peak frequencies in our best fit model are located at 
$8.5\pm0.3$\,GHz and $37\pm2$\,GHz, with flux densities of $(0.143\pm0.005)$\,mJy and 
$(0.226\pm0.006)$\,mJy, respectively (Figure \ref{fig:radio_bplfits}).

\subsubsection{SED at 18.2 d}
\label{text:radio_bplfits_18}
Due to the faintness of the radio emission, the VLA SED at 18.2\,d can be fit with a variety of 
different models. We present a fit with a sum of a cut-off power law model and a broken power 
law model in Figure \ref{fig:radio_bplfits}, fixing the lower-frequency slopes of both components at 
$\beta = 2.5$. Since the high-frequency spectral index of the higher frequency component ($\beta_2$) 
and the smoothness of the break ($y$) are not well constrained, we fix $\beta_2=-1.1$ and $y=3$. The 
peak frequencies in this model are located at $4.4\pm0.9$\,GHz and $11.8\pm1.2$\,GHz, with flux 
densities of $(0.056\pm0.08)$\,mJy and $(0.086\pm0.009)$\,mJy, respectively. 

\subsubsection{SEDs at 41.3 d and 89 d}
\label{text:radio_bplfits_4080}
The final two epochs of VLA radio observations comprise radio spectra sampled at 8--10 
approximately evenly logarithmically-spaced frequencies spanning 5\,GHz to 35\,GHz.
The afterglow fades from the previous epoch at 18.2 d to 41.3 at all observed radio frequencies. 
The large error bars 
at 41.3\,d do not allow for an unambiguous model fit. The afterglow was not detected at any 
observing frequency in the final epoch at about 89\,d. We include the observed SEDs at these last 
two 
epochs in Figure \ref{fig:radio_bplfits}.

\subsection{Radio light curves}
\begin{figure} 
 \includegraphics[width=\columnwidth]{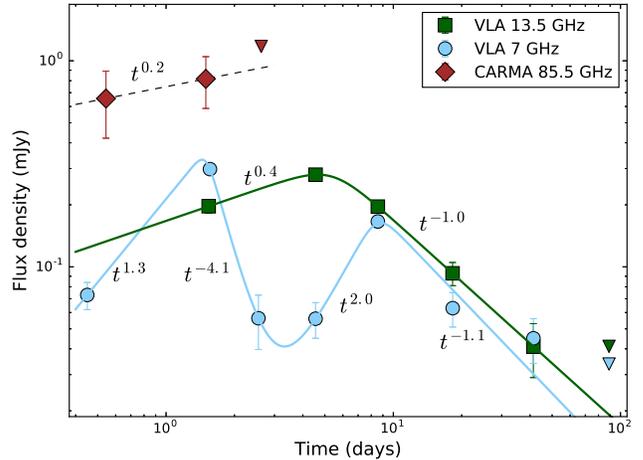}
 \caption{VLA 13.5 GHz (green) and 7 GHz (blue) light curves, together with best
 fit broken power law models and the resulting temporal slopes indicated. 
 The smoothness of the breaks are
 fixed to $y=5$, and the break times for the 7 GHz light curve are fixed to 1.5\,d and 8.5\,d, 
 respectively. The CARMA 85.5\,GHz light curve, which exhibits weak evidence
 for a rise, is shown for comparison (brown points).}
\label{fig:radiolc_bpl}
\end{figure}

The multiple components observed in the radio SEDs are also evident in the light curves. 
The 7\,GHz light curve exhibits a rapid brightening from 
0.45\,d followed by a fading to 4.5\,d. The precise rise and decline rates 
are not well constrained.
Simple power law fits yield a rise rate of $1.1\pm0.1$ and decline rate of 
$-3.4\pm0.6$, while a broken power law fit with the peak time fixed at 1.5\,d
yields a rise rate of $\approx 1.3$ from 0.45\,d to 1.5\,d followed by a decline 
at the rate $\approx -4.1$ for $y=5$. 
The light curve exhibits a re-brightening from 4.5\,d to 8.5\,d, with rise rate
$\approx2$ followed by a decline at the rate $\approx-1.1$ for a 
broken power law fit with peak time
fixed at 8.5\,d and $y=5$ (Figure~\ref{fig:radiolc_bpl}).

On the other hand, the 13.5\,GHz light curve can be fit with a single broken power law model, 
with a rise rate, $\alpha_{\rm Ku, 1}\approx0.4$, decline rate, $\alpha_{\rm Ku, 2}\approx-1.0$,
and break time, $t_{\rm b, Ku}\approx5.4$\,d, for fixed smoothness, $y=5$. The CARMA 85.5\,GHz
light curve exhibits modest evidence for a rise from 0.45\,d to 1.5\,d at the rate
$\approx0.2$; however, the low signal-to-noise of the detections preclude a more detailed 
analysis. 

\subsection{Unexpected behavior: multiple radio components}
\label{text:multicompradio}
Radio synchrotron radiation from relativistic shocks expanding adiabatically yield
spectra where the peaks move to lower frequencies and fade with time \citep{spn98}.
In this paradigm, the radio SEDs and light curves suggest that three 
distinct emission components contribute to the radio emission. We 
characterize and discuss each component in turn, followed by a critical discussion in section 
\ref{text:doublecompradio} on the physical nature of the multiple emission components.

\subsubsection{Component I}
Component I creates the low frequency peak in the spectrum at 1.5\,d and fades rapidly, 
disappearing by the time of the following observations at 2.5\,d (Figure \ref{fig:radio_bplfits}).
Since this component is not detected at
any other time, its evolution cannot be further constrained in a simple manner
by the radio observations; however, any model explaining this component
must account for the rapid rise observed at C-band from 0.45 to 1.5\,d ($\alpha\approx1.3$), and 
the fast fading ($\alpha\approx-3.4$ to $\alpha\approx-4.1$) thereafter. 

\begin{figure*}
\begin{tabular}{cc}
\centering
\includegraphics[width=\columnwidth]{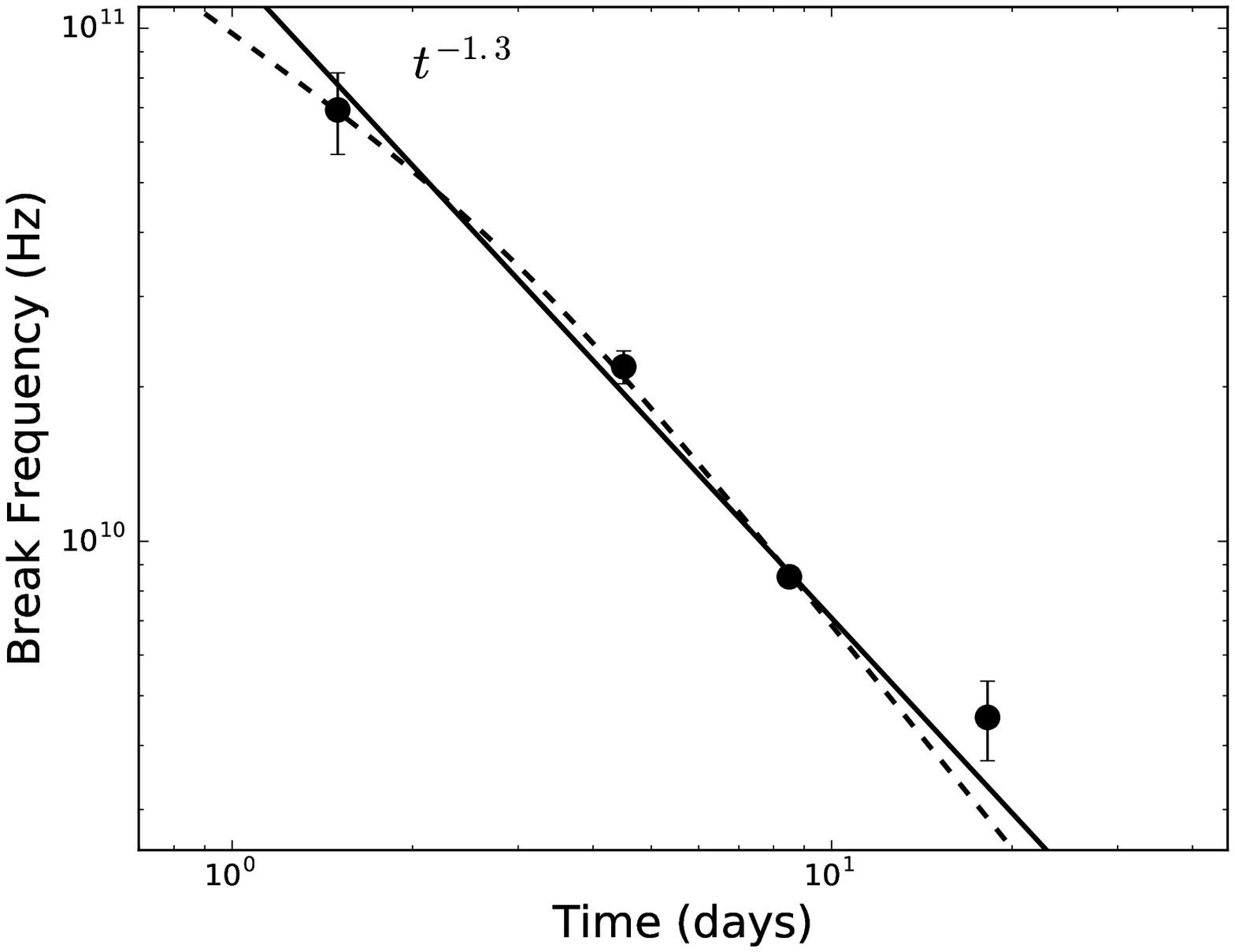} &
\includegraphics[width=\columnwidth]{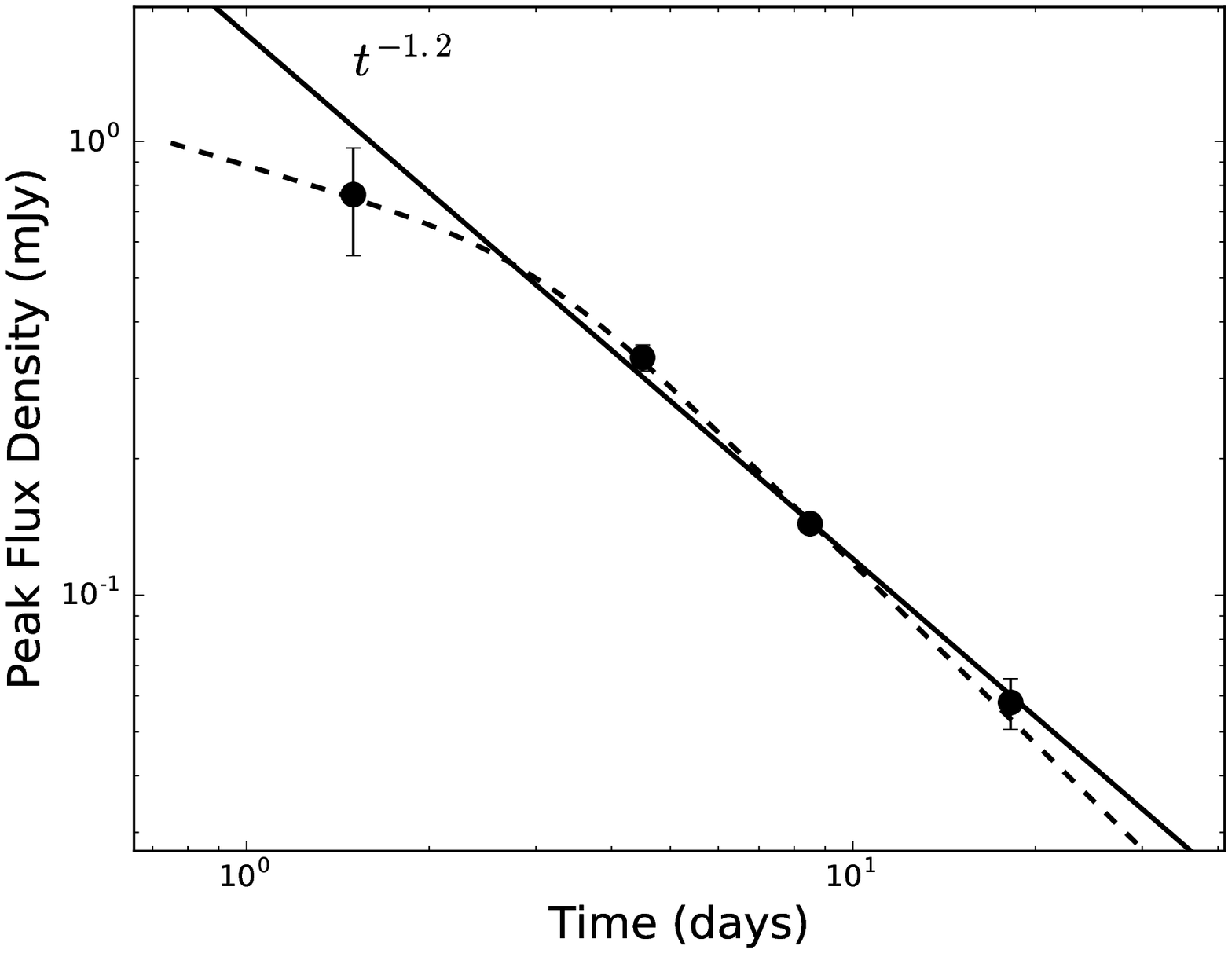} 
\end{tabular}
\caption{Evolution of the break frequency (left) and peak flux (right) of Component II in the 
radio SED of GRB 140304A for single power law (solid) and broken power law (dashed) models. For 
the broken power law fit to the evolution of the peak flux density, the break time and break 
sharpness have been fixed to $3$\,d and $y=5$, respectively. The two models provide equally good 
fits for the peak frequency, and we prefer the single power law evolution for simplicity. The 
broken 
power law model is a better fit for the evolution of the peak flux; however, the break time and 
pre-break evolution are degenerate. Annotations in the figures refer to the single power law fits.
See Section \ref{text:radio_bplfits_componentII} for details. 
\label{fig:component_II_evolution}}
\end{figure*}

\subsubsection{Component II}
\label{text:radio_bplfits_componentII}
This component is identified as the high-frequency peak in the spectrum at 1.5\,d. 
The peak of the radio emission 
for this component is $\nupk=69\pm13$\,GHz with flux density, $\fnupk=0.76\pm0.20$\,mJy at 
1.5\,d. Component II comprises the entirety of the radio emission at 4.5\,d, with a peak at 
$14.2\pm0.1$\,GHz and flux density $\fnupk=0.30\pm0.02$, corresponding to a decline rate of 
$-1.4\pm0.2$ in \nupk\ and $-0.8\pm0.2$ in \fnupk\ between 1.5\,d and 4.5\,d. This component 
additionally creates the low-frequency peak at 8.5\,d ($\nupk=8.5\pm0.3$\,GHz) and 18.2\,d 
($\nupk=4.4\pm0.9$\,GHz), and is marked ``II'' in Figure \ref{fig:radio_bplfits}.

Since we observe Component II at four separate epochs (1.5, 4.5, 8.5, and 18.2\,d), we can fit for 
the temporal evolution of its peak frequency and peak flux density. 
If this component shares the same start time as the afterglow 
(we relax this assumption later), the evolution of the 
spectral peak can be fit with a single power law, 
$\nupk=(1.3\pm0.2)\times10^{11}\td^{-1.25\pm0.08}$\,Hz. A 
broken power law model does not significantly improve the fit for $\nupk$. 
Fitting a single power law to \fnupk\ for all four epochs yields 
$\fnupk=(1.69\pm0.26)\td^{-1.15\pm0.07}$\,mJy ($\chi^2$/dof\,$= 4.5$).
A broken power law yields a better fit,
but the time of this break is degenerate with the pre-break decline rate. 
For example, for a break time of 3\,d and smoothness, $y=5$, the 
best fit requires $\fnupk\propto t^{-0.4\pm0.4}$ transitioning into 
$\fnupk\propto t^{-1.30\pm0.09}$ (Figure \ref{fig:component_II_evolution}).

We note that evolving \nupk\ and \fnupk\ as a single power law to earlier time
over-predicts the clear band flux density observed by MASTER at $9.5\times10^{-4}$\,d 
by a factor of 1400. Imposing a break in \fnupk\ leads to a
discrepancy of a factor of 14. We return to this point in Section \ref{text:injectionRS}. 
These models do not produce significant X-ray flux. 
We conclude that Component II does not connect 
spectrally and temporally to the X-ray and optical bands, and therefore forms a distinct emission 
component independent of the mechanism producing the X-ray and optical radiation.

\subsubsection{Component III}
\label{text:radio_bplfits_componentIII}
This component first appears in the radio SED at 8.5\,d with a peak around 37\,GHz and contributes 
the bulk of the observed flux above 10\,GHz at 18.2\,d. The observed SED 
at 41.3 d is also expected to include significant contribution from this component. 
Comparing the observed SEDs at 8.5\,d and 18.2\,d, we derive the temporal evolution of the 
spectral break frequency and peak flux density to be 
$\alpha_{\nu,\rm pk} = -1.5\pm0.1$ and $\alpha_{\rm F,pk} = -1.2\pm0.2$ 
between these two epochs, respectively.
In the next section,
we show that this component \textit{does} connect with the optical and X-ray observations
and therefore likely arises from the FS. The rapid decline of the peak flux density
at $\gtrsim 8.5$\,d suggests that a jet break occurs after $8.5$\,d.

\subsection{Summary}
To summarize, the radio data exhibit three distinct spectral components. Components I and II appear 
at 0.5~d -- 18.2~d and do not connect with the optical and X-ray SED. 
We consider physical models for their origin in Section \ref{text:doublecompradio}.
Component III appears at 
8.5~d at the highest cm-band frequencies ($\approx 30$\,GHz) and likely arises from the FS. 

\section{Single-component models}
\label{text:singlecompradio}
In the above discussion, we have argued for the presence of multiple spectral components in the VLA 
radio observations. For simplicity, we begin with a search for single-component radio models that 
explain the gross features of the radio SEDs and light curves in this section, and discuss 
multi-component radio models in Section \ref{text:doublecompradio}. We interpret the radio 
observations together with the X-ray and optical/NIR data in the framework of the standard 
synchrotron model, where the observed SED is characterized by three spectral break frequencies -- 
the self-absorption frequency, $\nua$, the characteristic synchrotron frequency, $\numax$, and the 
cooling frequency, $\nuc$ -- and an overall flux normalization, $f_{\rm peak}$ 
\citep{spn98,cl00,gs02}. 

Single-component radio models for GRB\,140304A can be divided into two categories based on the 
interpretation of the X-ray light curve, in particular, the rapid re-brightening at 0.26\,d. If 
this excess is ascribed to a flare, then the underlying X-ray light curve decline rate is 
$\alpha_{\rm X}\approx-0.8$. 
Over this same period, however, the optical light curve
is declining at least as steeply as $\alpha_{\rm opt} \sim \alpha_{\rm z'}\approx-1.3$, and perhaps 
as steeply as $\alpha_{\rm opt} \sim \alpha_{\rm i} \lesssim -1.8$. In the standard synchrotron 
model, 
the largest difference between $\alpha_{\rm X}$ and $\alpha_{\rm NIR}$ (when both are steeper than 
$\alpha=-2/3$) is $\Delta\alpha = -1/4$, which occurs on either side of the cooling frequency. It 
is thus impossible to arrange this scenario where a higher frequency light curve is decaying at a 
shallower rate than at lower frequencies, and both are declining faster than $\alpha = -2/3$, if 
the two light curves are dominated by radiation produced by the same shock. Thus in this situation, 
the X-ray and optical light curves after 0.26\,d must arise from different emission regions.

We therefore consider two possible scenarios: (i) the X-ray excess at 0.26\,d is a flare caused 
perhaps by extended central engine activity, while the near-IR 
radiation between 0.58\,d and 1.59\,d is produced by a different mechanism, or (ii) the X-ray and 
optical light curves are dominated by the FS and the X-ray excess at 0.26\,d is a 
re-brightening of the FS radiation. We investigate both models in detail, beginning with the first 
scenario.

\begin{figure}
 \includegraphics[width=\columnwidth]{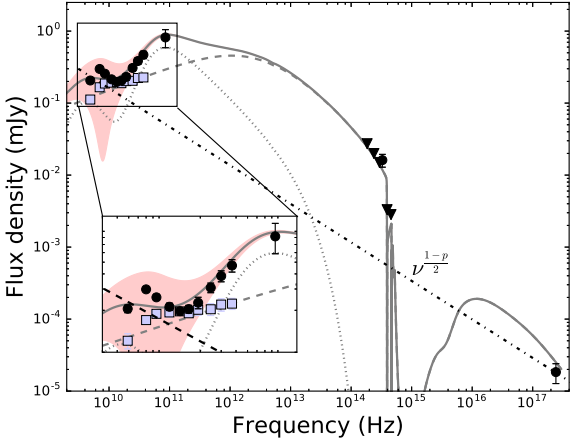}
 \caption{Radio to X-ray SED of GRB~140304A at 1.6\,d (black points) and 8.5\,d (blue points), with 
a zoom in to the radio section (inset), together with the best fit model (grey, solid), 
decomposed into FS (dashed) and double RS (dotted) contributions (Section 
\ref{text:doublecompradio}). 
The dash-dot line 
is a single power law extrapolation from 
the 
X-ray to the radio, demonstrating that $\nu_{\rm radio} \approx \numax < \nux < \nuc$ with $p=2.07$ 
cannot match the radio and X-ray observations (Section \ref{text:singlecompradio_flare}). In this 
plot, the interpolations required to plot the SED at a common time of $1.6$\,d are $\lesssim10\%$ 
and therefore ignored.}
\label{fig:ISM9_sketch}
\end{figure}

\subsection{X-ray excess at 0.26\,d due to flare}
\label{text:singlecompradio_flare}
If we ascribe the X-ray excess at 0.26\,d to a flare, the optical and X-ray emission must arise 
from distinct emission regions as discussed above. Under the assumption that the X-rays are 
dominated by the FS, the decline rate $\alpha_{\rm X} = -0.8\pm0.12$ 
implies $\alpha_{\rm X} = $
\begin{equation}
  \begin{cases}
      \frac{2-3p}{4}  \implies p = 1.73\pm0.16; & \numax,\nuc<\nux  \\
      \frac{3(1-p)}{4} \implies p = 2.07\pm0.16; & \numax<\nux<\nuc\ (\rm ISM)  \\
      \frac{1-3p}{4}  \implies p = 1.40\pm0.16; & \numax<\nux<\nuc\ (\rm wind) \\
  \end{cases}
\end{equation}
For the standard assumption of $2 < p < 3$, the only viable scenario is the second case above; 
however, this model has several shortcomings. First, $\nux < \nuc$ would predict $\beta_{\rm X} = 
(1-p)/2 \approx -0.54$, which is not consistent with the observed value, $\beta_{\rm X} \approx 
-1$. 
Second, if we anchor the theoretical SED to $F_{\rm X} \approx 2\times10^{-5}$\,mJy at 
$\approx1.5$\,d and extend this spectrum to lower frequencies, we underpredict the optical by a 
factor of $\approx30$, although this was already expected. Third, this model also underpredicts the 
CARMA detection by a factor of $\approx80$; in fact it is impossible to reconcile the 
$\nu^{(1-p)/2}$ spectrum with the radio observations without additional radio components (Figure 
\ref{fig:ISM9_sketch}).

We conclude that interpreting the X-ray excess as due to a flare would require additional 
components in both the optical and radio at $1.5$\,d. This is driven by the unexpectedly 
shallow decline in the X-rays, $\alpha_{\rm x} \approx -0.8$ after 0.1\,d, combined with a steep 
decline in the optical and a bright radio afterglow. Thus single-component radio models, which are 
the focus of this section, cannot explain the X-ray and radio observations under the assumption 
that 
the X-ray excess at 0.26\,d is due to a flare. 

\subsection{All X-ray emission from FS}
\label{text:singlecompradio_nonflare}
\label{text:wind_fs}
We now consider the X-ray excess to be due to a re-brightening event, possibly arising from 
an episode of energy injection into the FS \citep[e.g.][]{sm00}. In this scenario, we ignore 
the X-ray data before $\approx0.2$\,d for the moment, and expect the X-ray and optical observations 
after this time to match a single FS model. We note that $\beta_{\rm X} = -1.29\pm0.25$
is marginally different from
$\beta_{\rm NIR-opt} = -0.98\pm0.20$ and 
$\beta_{\rm NIR-X} = -0.96\pm0.02$ at 0.58\,d; which results in three possible scenarios:
(i) $\nuc > \nux$ at 0.58\,d, implying $p = 1-2\beta \approx 3$, and predicting 
a common decline rate of $\alpha_{\rm X} \approx \alpha_{\rm NIR} \approx -1.5$ (ISM) or 
$\alpha\approx-2$ (wind). Since $\alpha_{\rm X} = 
-1.5\pm0.1$, the wind model is ruled out, but the $p\approx 3$ ISM model is viable.
(ii) $\nuc < \nunir$ at 0.58\,d. This implies $p = -2\beta \approx 2$, and requires a common 
decline rate of $\alpha_{\rm X}\approx\alpha_{\rm NIR}\approx -1$. The steeper observed decline 
rate could be explained by a jet break between 0.3 and 2\,d.
(iii) $\nunir < \nuc < \nux$ at 0.58\,d. The X-ray decline rate of $\alpha_{\rm X} =-1.5\pm0.1$ 
then implies $p\approx2.67\pm0.13$, predicting $\beta_{\rm X}=-1.3\pm0.06$, $\beta_{\rm 
NIR}=(1-p)/2=-0.84\pm0.07$, and $\alpha_{\rm NIR}=3(1-p)/4=-1.3\pm0.1$ (ISM) or $\alpha_{\rm NIR} = 
(1-3p)/4=-1.7\pm0.1$ (wind). The steeper observed spectral index in the near-IR may then be 
explained by a small amount of extinction in the host galaxy.

The observed X-ray and NIR spectral indices ($\beta_{\rm X}=-1.29\pm0.25$ 
and $\beta_{\rm NIR} =-0.98\pm0.20$) are not strongly constraining, and the model is consistent 
with the expected light curves in both ISM and wind scenarios. 
In Section \ref{text:radio_bplfits_componentIII}, we showed that Component III in the radio
behaves like an FS with a jet break between $\approx 8.5$ and $18.2$\,d.
Since coeval detections spanning the radio to X-ray bands only exist before $\approx3$\,d 
where the radio SEDs exhibit multiple components, a simple determination of the location
of \nuc\ is not straightforward. In the next section, we
construct a model explaining the X-ray, near-IR, and radio light curves and SEDs
under case (iii) above for the wind environment, and 
show that the other cases are disfavored. The remaining possibilities
are presented in appendix \ref{appendix:additional_fs} for completeness,
and their associated figures are available in the on-line version of this article.

For a wind environment, we expect $\alpha_{\rm NIR} = (1-3p)/4$. Thus $-1.8 \lesssim \alpha_{\rm 
NIR} \lesssim \alpha_{\rm z'} \approx -1.3$ implies $2.1\lesssim p \lesssim2.7$, which yields 
$-1.5\lesssim \alpha_{\rm X} \lesssim -1.1$, also consistent with observations. We find a good fit 
to the X-ray and optical light curves with $p\approx2.6$ and \nuc\ above the NIR band (Table 
\ref{tab:params}). 
This model fits the optical and light curves after 0.2\,d well, and also captures the 
evolution of the radio SED after $\approx18.2$\,d. 
Our best fit model has $\epse\approx2.5\times10^{-2}$, $\epsb\approx2.9\times10^{-2}$,
$\Astar\approx2.6\times10^{-2}$, and $\EKiso\approx4.9\times10^{54}$\,erg. The model requires
modest extinction, $\AV\approx0.1$\,mag, and
a jet break at $\tjet \approx10.6$\,d, corresponding to a jet opening angle of 1.1\,deg.

\subsection{Energy injection into the FS}
\label{text:injection}
In the preceding section, we have argued that the X-ray excess at $\approx0.26$\,d likely
arises from a re-brightening of the afterglow, rather than from late-time central engine activity. 
We now model the rapid rise in the X-ray light curve as energy injection into the 
FS using the methods described in \cite{lbm+15}. 

\begin{figure*} 
 \begin{tabular}{cc}
  \includegraphics[width=\columnwidth]{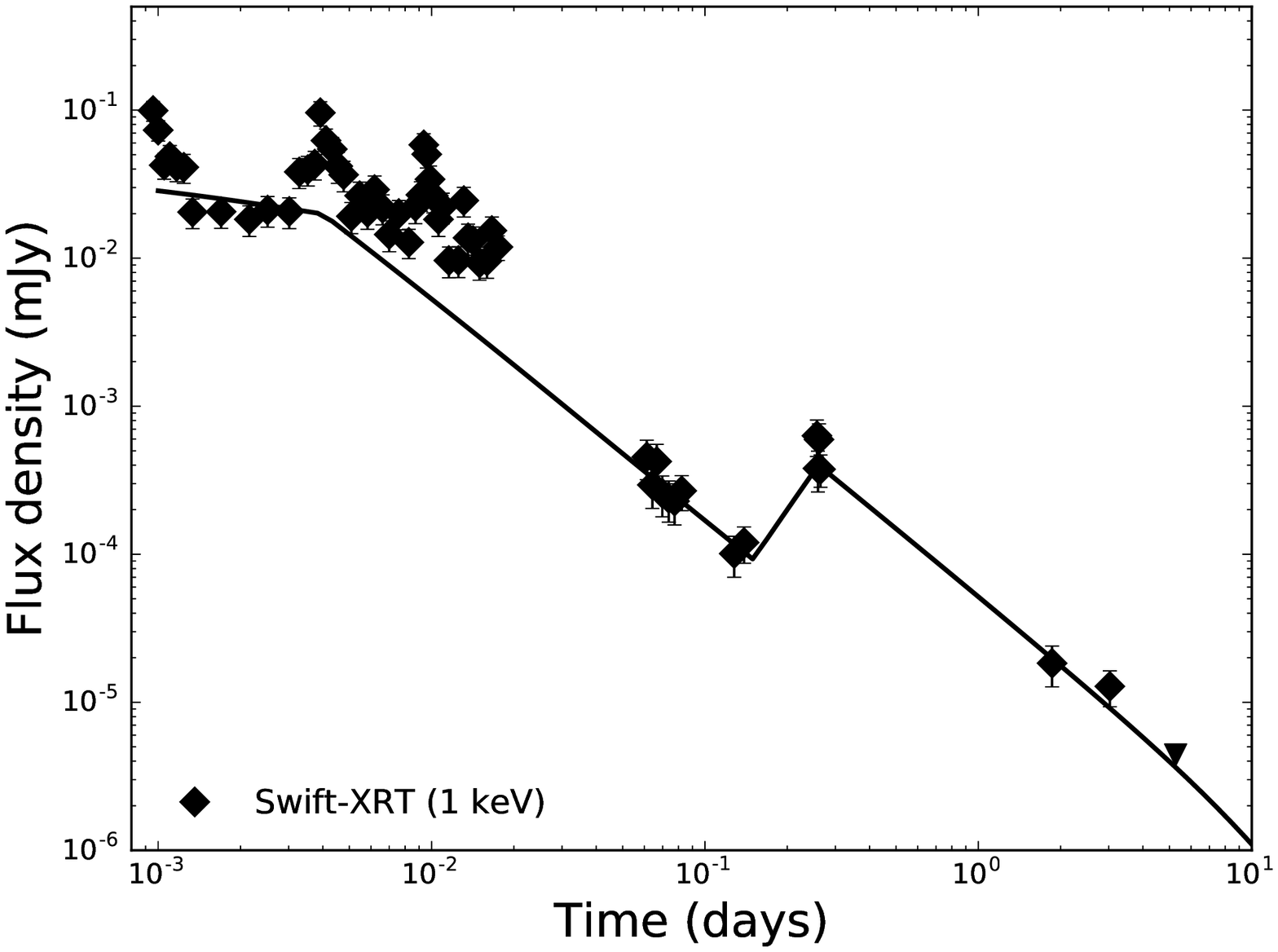} &
  \includegraphics[width=\columnwidth]{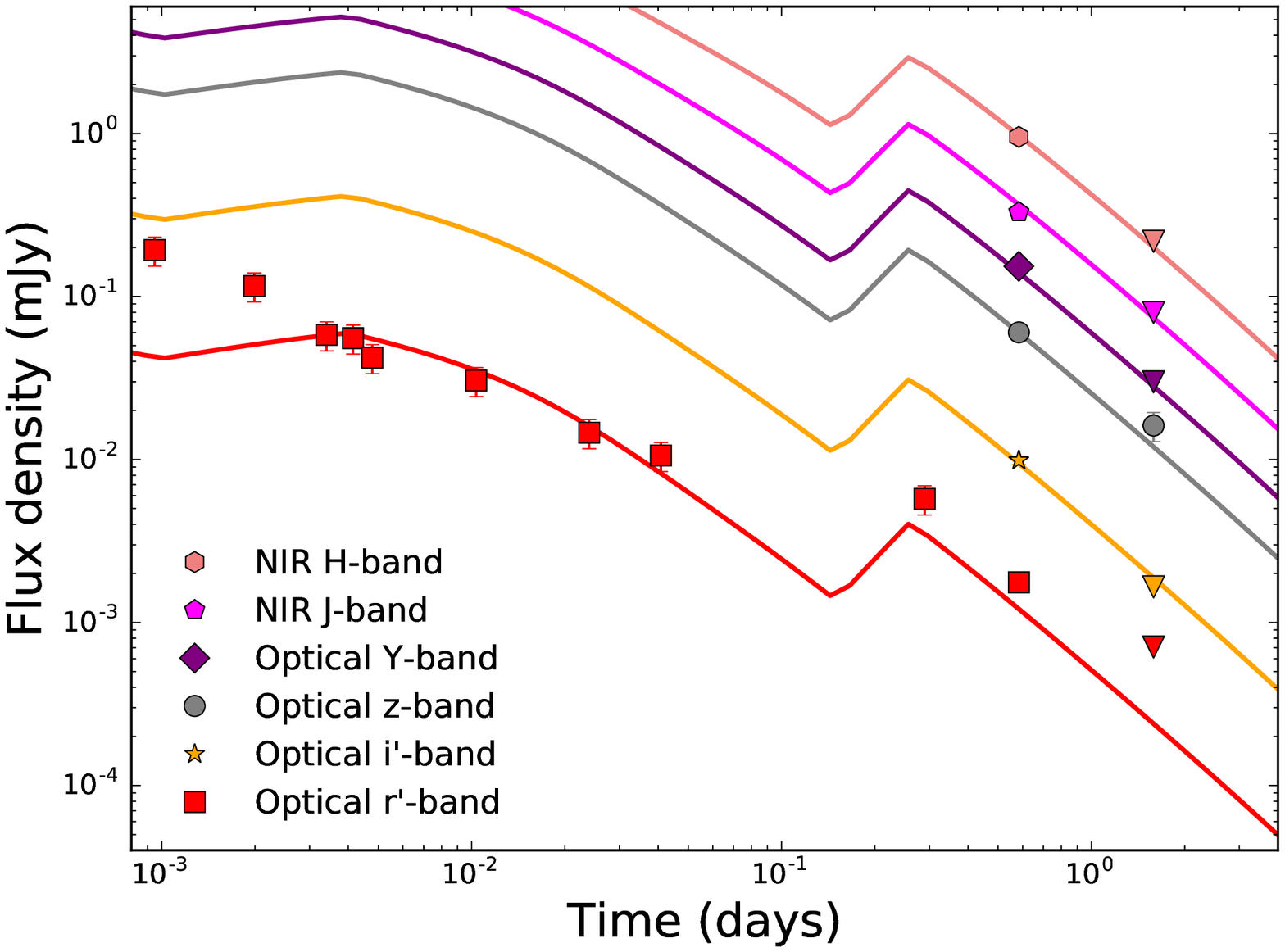} \\
  \includegraphics[width=\columnwidth]{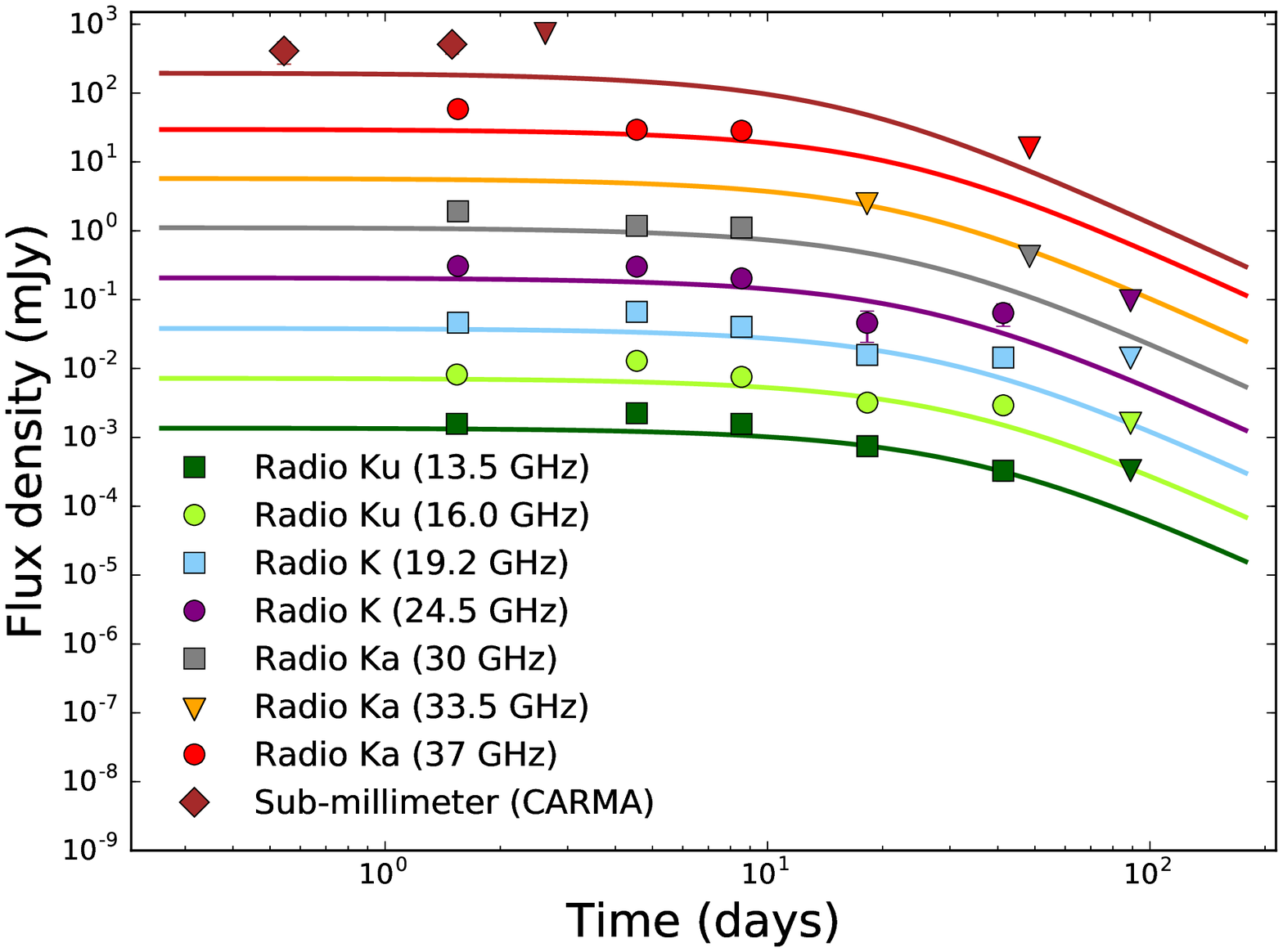} &
  \includegraphics[width=\columnwidth]{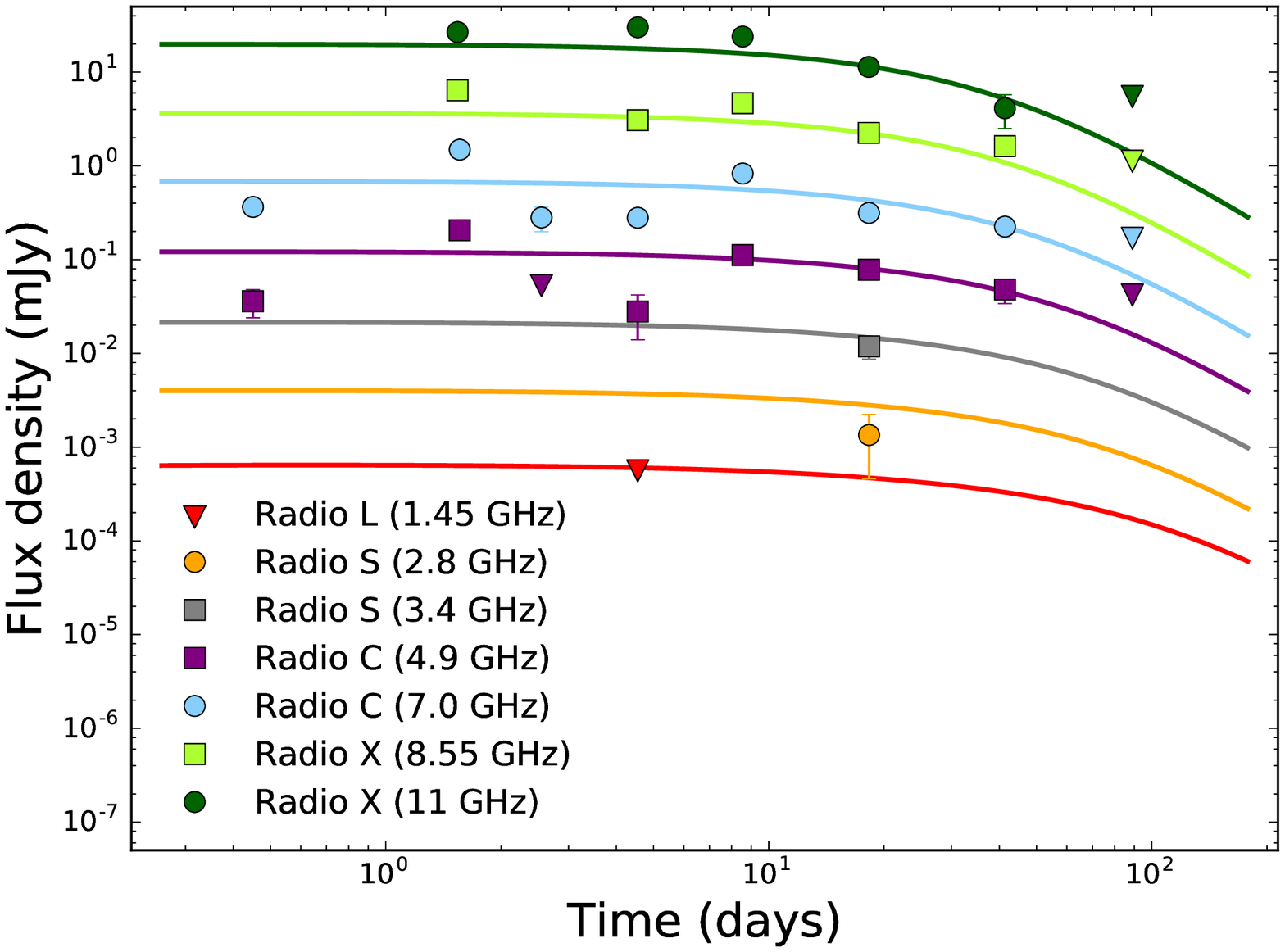} \\
 \end{tabular}
 \caption{X-ray (top left), optical/NIR (top right) and radio (bottom) light curves of the 
afterglow of GRB 140304A, together with a FS wind model, including energy injection
between 0.15\,d and 0.26\,d (Section \ref{text:injection}). 
The model matches the X-ray light curve after $5\times10^{-2}$\,d,
the optical observations, and the overall features of the radio light curves.
}
\label{fig:modellc_wind10}
\end{figure*}

\begin{figure*}
\begin{tabular}{ccc}
 \centering
 \includegraphics[width=0.31\textwidth]{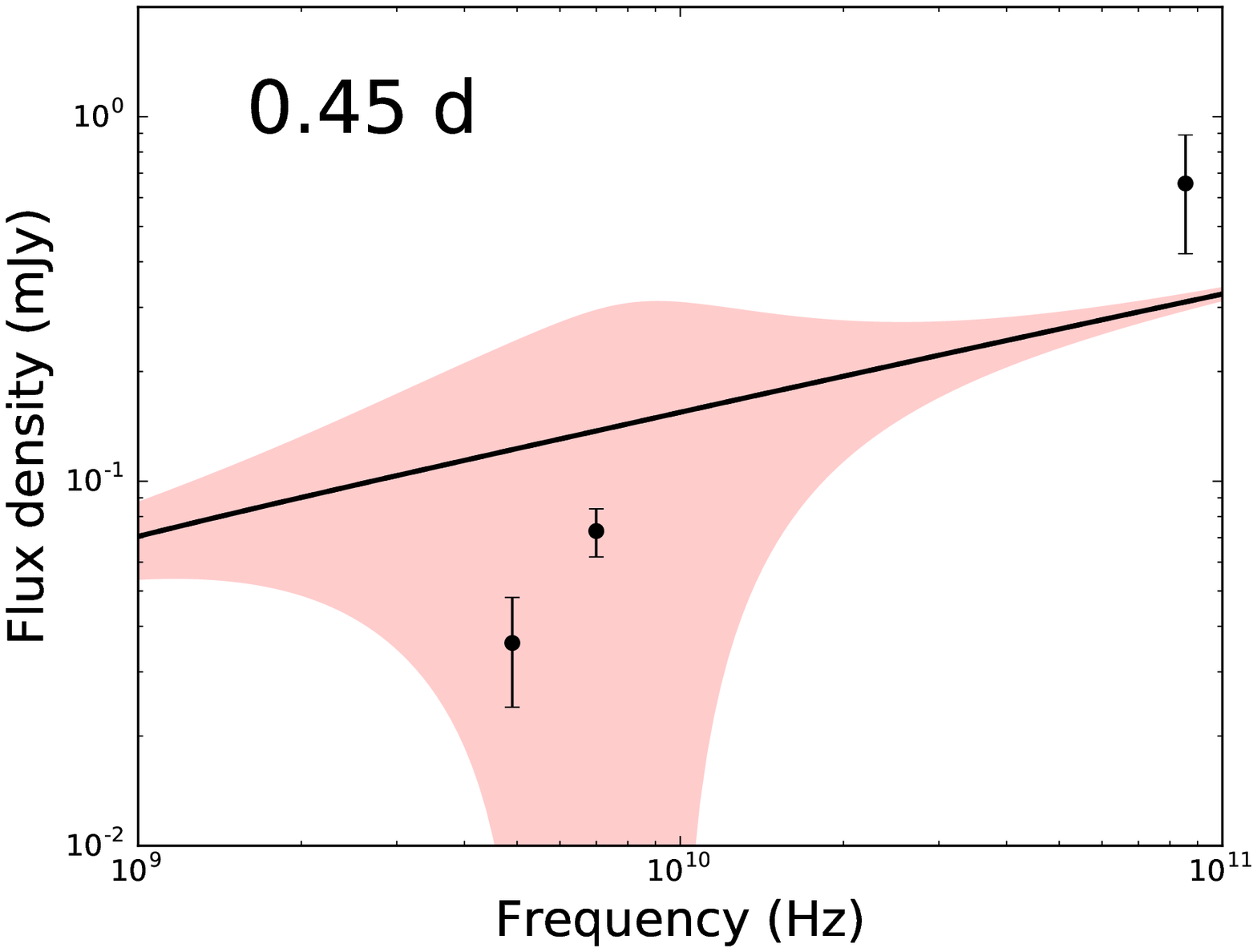} &
 \includegraphics[width=0.31\textwidth]{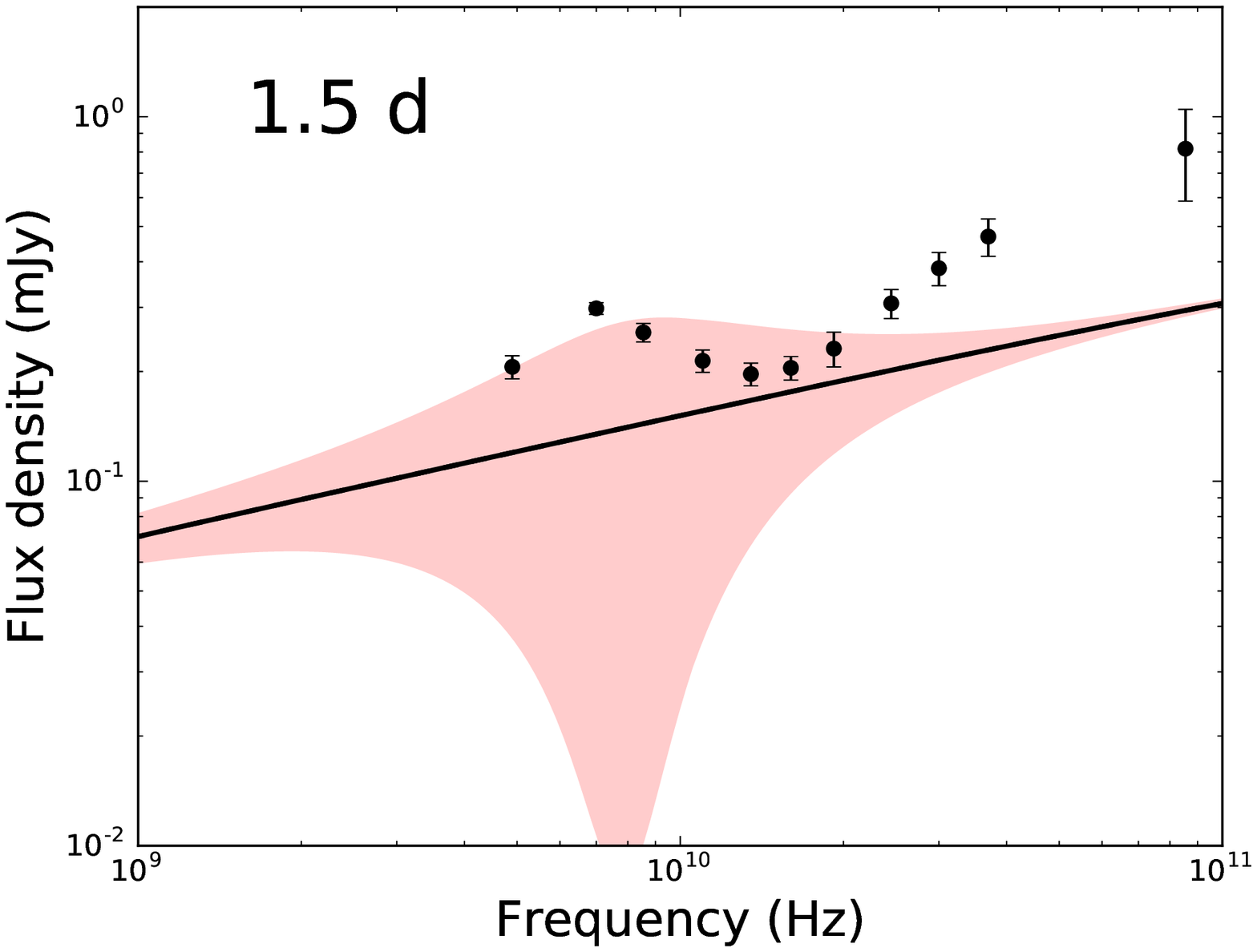} &
 \includegraphics[width=0.31\textwidth]{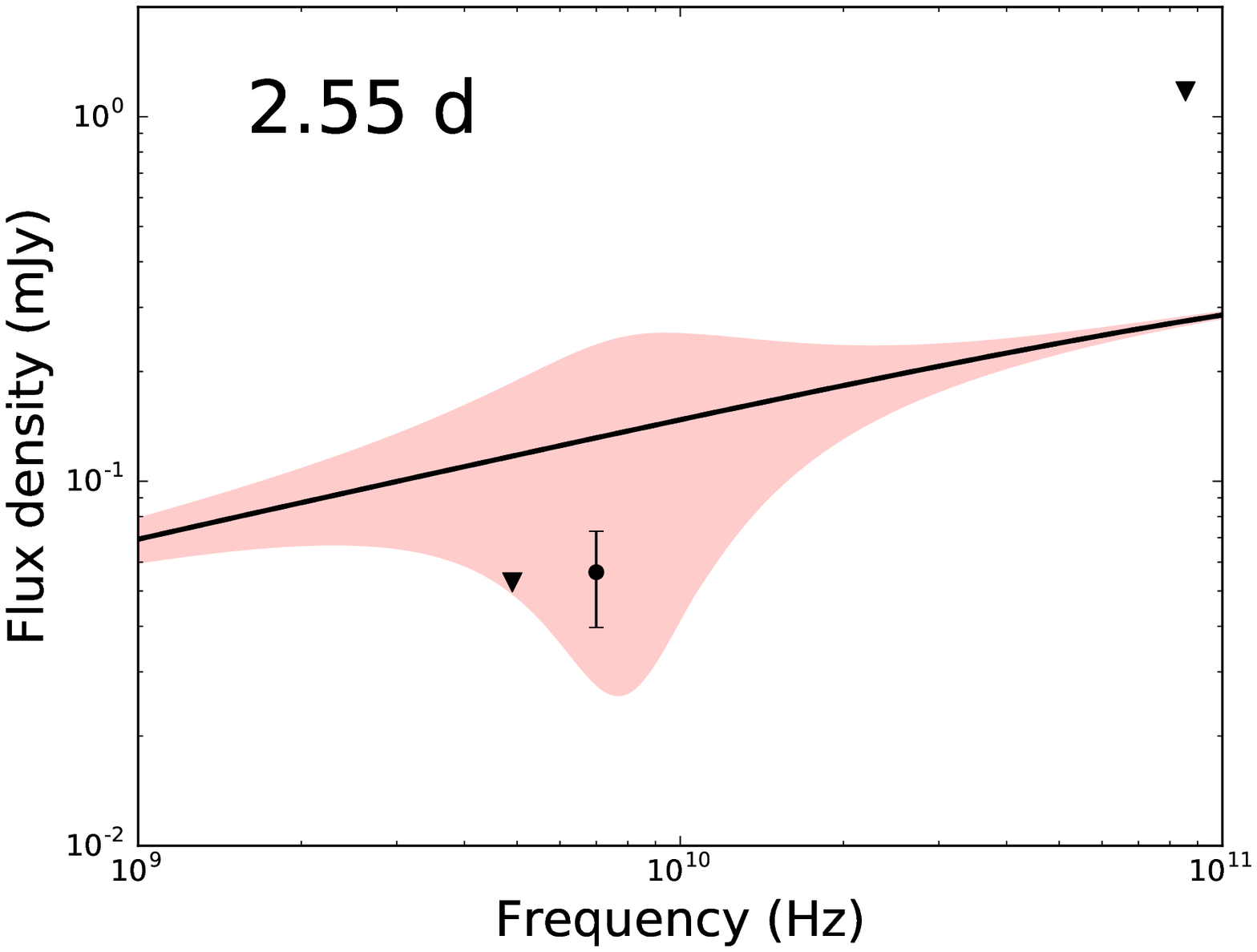} \\
 \includegraphics[width=0.31\textwidth]{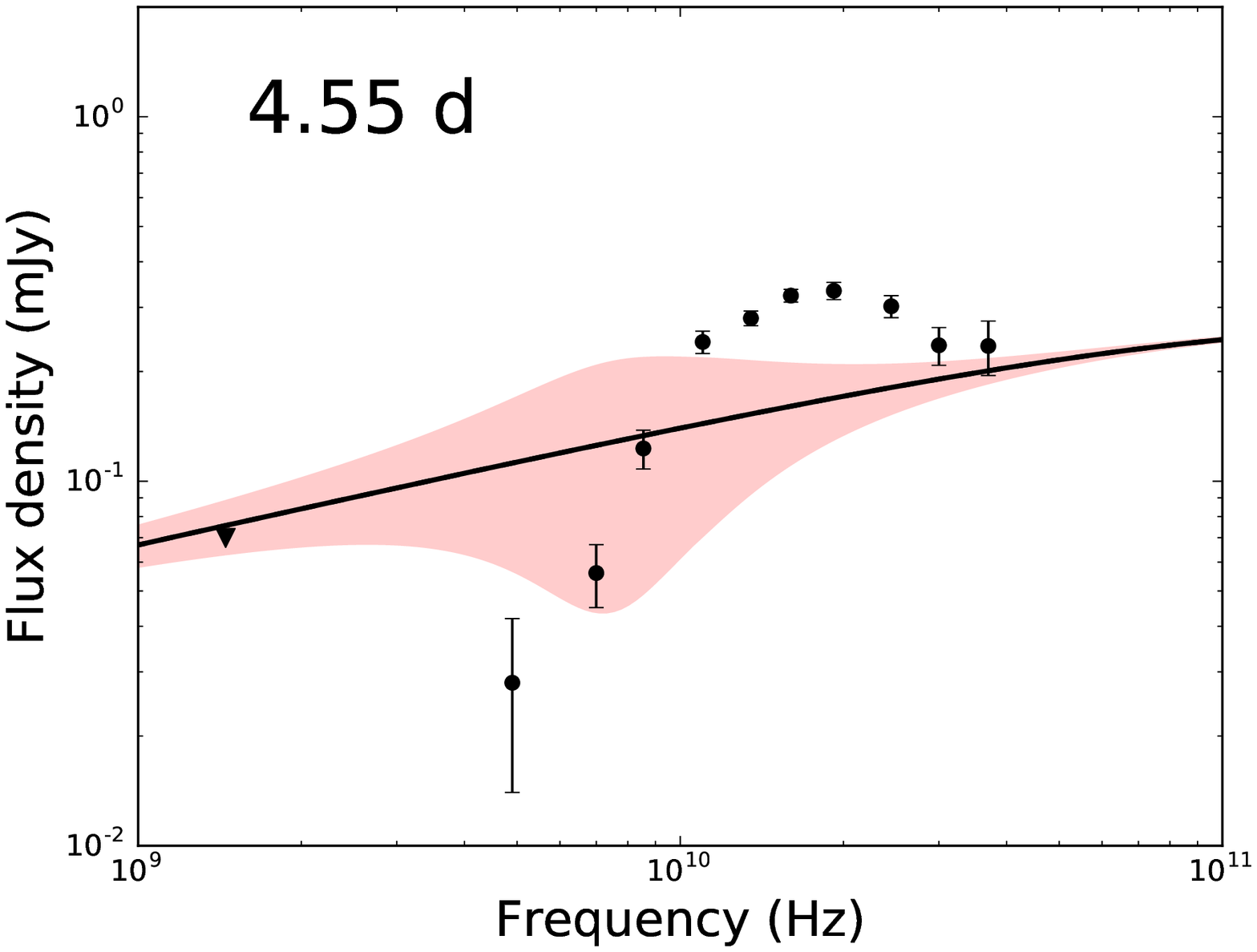} &
 \includegraphics[width=0.31\textwidth]{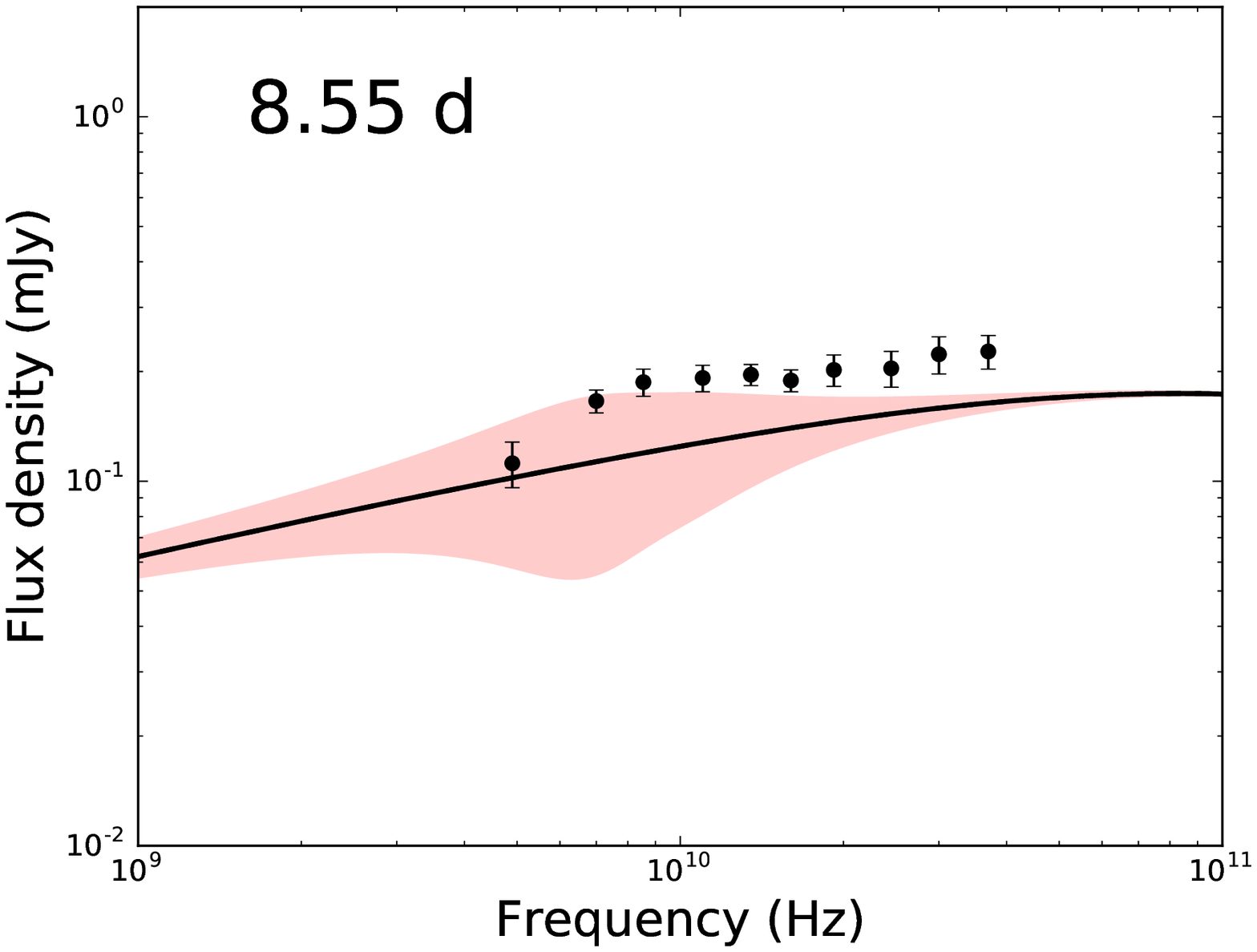} &
 \includegraphics[width=0.31\textwidth]{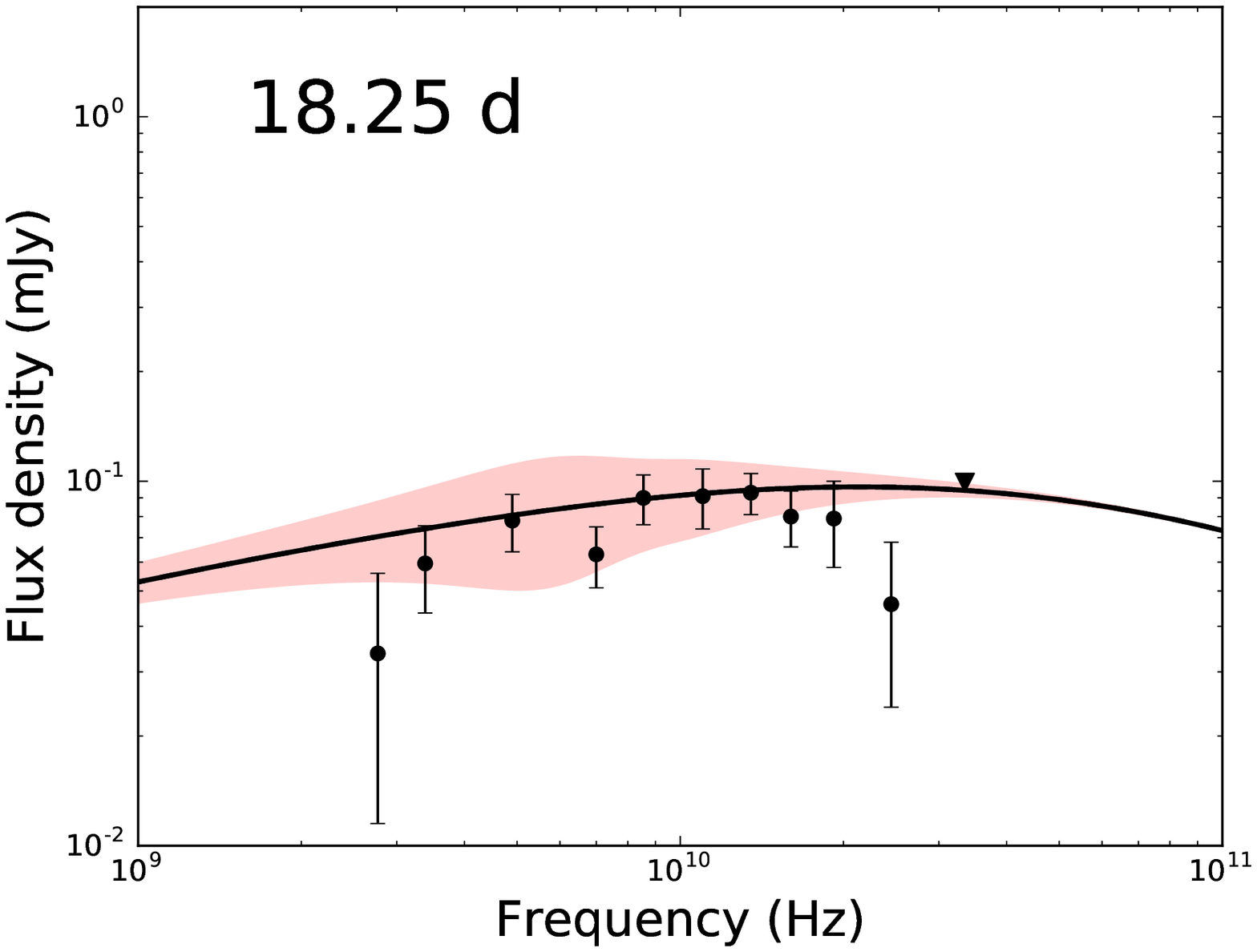} \\
 \includegraphics[width=0.31\textwidth]{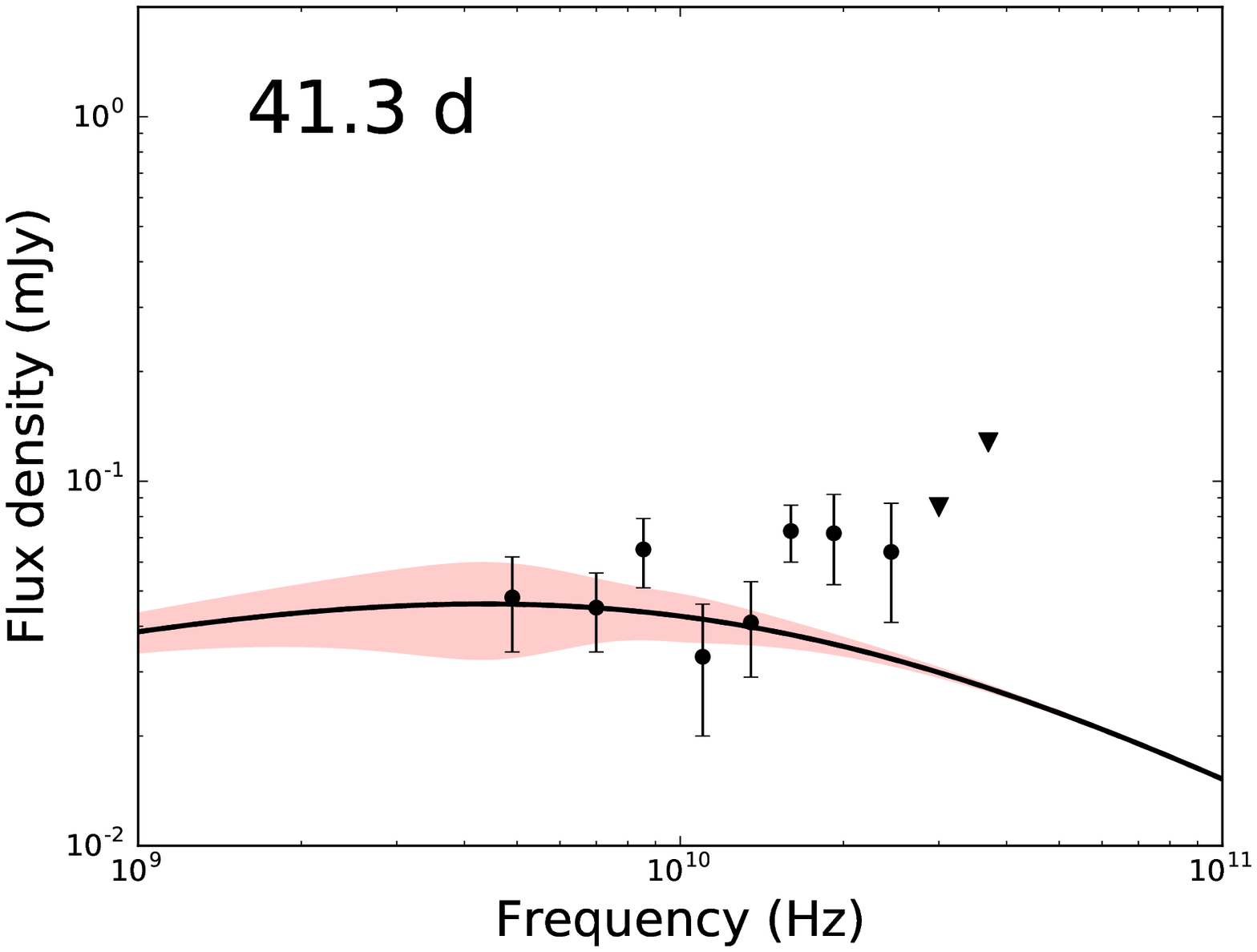} &
 \includegraphics[width=0.31\textwidth]{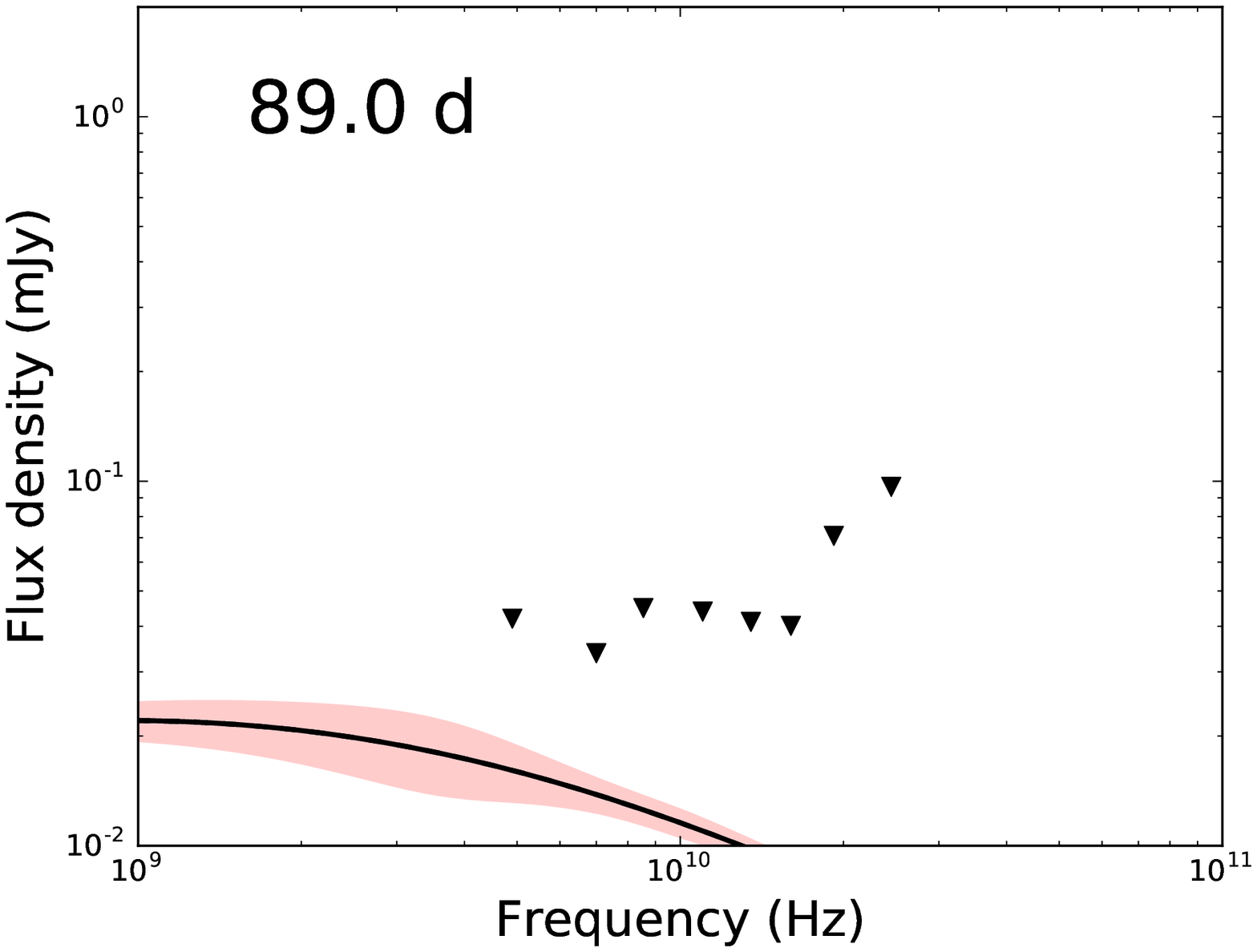} & 
\end{tabular}
\caption{Radio spectral energy distributions of the afterglow of \me\ at multiple epochs starting 
at 0.45~d, together with the same FS wind model in Figure 
\ref{fig:modellc_wind10}. The red shaded regions represent the expected 
variability due to scintillation. The model captures the evolution after 18.25\,d, but 
under-predicts the mm-band data and the observations above $\approx10$\,GHz before 8.55\,d.
}
\label{fig:modelsed_wind10}
\end{figure*}

Since $\nuc<\nux$ at the time of the re-brightening, the X-ray flux density,
$F_{\rm X}\propto E^{\frac{2+p}{4}} t^{\frac{2-3p}{4}}$. The light curve rises
during this period as $t^{2.0\pm0.3}$, which yields $E\sim t^{3.0\pm0.3}$.
Starting with the afterglow model, we find a good fit to the re-brightening event
for $E\sim t^{3.8}$ between 0.15\,d and 0.26\,d, close to the value expected 
from simple considerations. 
This corresponds to an increase in $\EKiso$ by a factor of $\approx 8$
from $\EKiso\approx6.1\times10^{52}$\,erg at 0.15\,d
to $\approx4.9\times10^{54}$\,erg at 0.26\,d.
The resultant $r^{\prime}$-band light curve\footnote{Given 
the high redshift of the GRB, $z=5.283$, 
the optical $r^{\prime}$ and $i^{\prime}$-band observations are significantly
affected by IGM absorption. For our subsequent analysis, we integrate model spectra
over the SDSS $r^{\prime}$ and $i^{\prime}$ bandpasses. A more detailed analysis
would require knowledge of the individual response functions of each telescope 
and the spectra of the calibration stars used; however, this is not available and 
clearly beyond the scope of the present discussion.} 
also agrees with the MASTER optical 
observations between $3\times10^{-3}$\,d and $4\times10^{-2}$\,d.
Our best-fit wind model is presented in Figures \ref{fig:modellc_wind10} and 
\ref{fig:modelsed_wind10}.
We investigate the ISM case for the three possible locations of 
$\nuc$ relative to $\nunir$ and $\nux$ (as described in Section  
\ref{text:singlecompradio_nonflare}) in Appendix \ref{appendix:additional_fs}, and 
present the associated light curves and radio SEDs in Figures 
\ref{fig:modellc_ISM14}, \ref{fig:modelsed_ISM14},
\ref{fig:modellc_ISM16}, \ref{fig:modelsed_ISM16},
\ref{fig:modellc_ISM15}, and \ref{fig:modelsed_ISM15},
including the effects of energy injection.
However, in each case the models significantly under-predict the optical light curve 
before $4\times10^{-2}$\,d, and are therefore disfavored.

\begin{deluxetable}{lc}
 \tabletypesize{\footnotesize}
 \tablecolumns{2}
 \tablecaption{Parameters for best-fit wind model}
 \tablehead{   
           \colhead{Parameter} &
           \colhead{Value}
   }
 \startdata    
   Ordering at 0.1\,d   & $\nunir<\nuc<\nux$  \\   
   $p$                  & 2.59                \\
   \epse                & $2.5\times10^{-2}$  \\
   \epsb                & $5.9\times10^{-2}$  \\
   \Astar               & $2.6\times10^{-2}$  \\
   $E_{\rm K, iso, 52}$ & $4.9\times10^{2}$   \\ 
   \tjet\ (d)           & 10.6                \\
   \thetajet\ (deg)     & 1.13                \\
   \AV\ (mag)           & 0.09                \\    
   \nua\ (Hz)           & $3.0\times10^{8\dag}$   \\ 
   \numax\ (Hz)         & $2.0\times10^{14}$  \\
   \nuc\ (Hz)           & $6.4\times10^{15}$  \\
   $F_{\nu, \rm max}$ (mJy) & 4.2             \\
   $E_{\gamma}$ (erg)   & $2.3\times10^{49}$  \\
   $E_{\rm K}$ (erg)    & $9.5\times10^{50}$  \\
   $E_{\rm tot}$ (erg)  & $9.7\times10^{50}$  \\
   $\eta_{\rm rad}$     & $\approx2\%$
 \enddata
 \tablecomments{All break frequencies are listed at 0.1\,d. 
 \dag\ These break frequencies are not directly constrained by the data.}
\label{tab:params}
\end{deluxetable}

To summarize, 
the continued high flux density of the X-ray detections after 
$\approx1$\,d suggests that the X-ray excess at $\approx0.26$\,d is not due to late-time central 
engine activity, but is caused by a re-brightening of the FS radiation. 
The multi-band data are consistent with a wind-like circumburst medium, requiring 
a single episode of energy injection at $\approx0.2$\,d. The resultant model fits 
the optical and X-ray evolution and 
matches the radio SED after 18\,d. However, this model under-predicts the CARMA light 
curve, as well as the 5--83\,GHz SED at 1.5\,d and 8.55\,d. 
To account for these deficits, we next investigate the effect of including emission from
additional components. We delineate the evidence for their presence, and consider their 
possible physical origins.

\section{Multi-component models}
\label{text:doublecompradio}
\subsection{Reverse Shock}
\label{text:RS1}
In Section \ref{text:radio_bplfits}, we discussed the apparent multi-component structure of the
radio SEDs between 1.5\,d and 18.2\,d. In the previous section, we have shown that while a single
FS model can reproduce the gross features of the radio light curves, such a model
cannot explain the multiple peaks in the SED at 1.5\,d, the $\approx14\,$GHz peak at 4.5\,d, and
the SED at 8.55\,d. Of these, the greatest discrepancy between data and model arises in the 
radio SEDs at 1.5\,d and 4.55\,d. We now consider whether each spectral peak at 1.5\,d 
can in turn be ascribed to radiation from an RS.

An RS propagating into GRB ejecta is expected to produce synchrotron radiation with its own
set of characteristic frequencies, \nuar, \numr, and \nucr, and peak flux, \fnumr. 
These quantities are related to those of the FS at the deceleration time, \tdec, 
when the RS just crosses the ejecta, and the relation between the two sets of break frequencies
and fluxes allows for a determination of the ejecta Lorentz factor and magnetization.
After the RS crosses the ejecta, the flux above \nucr\ declines rapidly\footnote{The 
angular time delay effect prevents 
abrupt disappearance of flux above \nucr; instead,
we expect $F_{\nu > \nucr}\propto t^{-\frac{p+4}{2}}\approx t^{-3.3}$ \citep{kz03a}.}
as no electron is newly
accelerated within the ejecta. Since no radiation is expected above $\nucr$, a 
conservative lower limit to the optical light curve can be computed by taking \nucr\ to be 
located near each observed radio spectral peak at 1.5\,d in turn.

For the high frequency component, this occurs in the mm band ($85.5$\,GHz) at 1.5\,d. 
For a wind-like circumburst environment, 
$\nucr\propto t^{-15/8}$ for a relativistic RS and $\nucr\propto t^{-(15g+24)/(14g+7)}$ 
for a Newtonian RS, where $1/2 \lesssim g \lesssim 3/2$ from theoretical arguments 
\citep{mr99,ks00}. 
Therefore, the slowest expected evolution of this break frequency is $\approx t^{-1.7}$, 
whereupon it would have crossed the optical $r^{\prime}$-band at $\approx0.01$\,d.
The peak flux density evolves as $\fnupk \propto t^{-(11g+12)/(14g+17)}\propto t^{-1}$. 
The flux density in the CARMA 85.5\,GHz band at 1.5\,d is $\approx0.8$\,mJy, which yields
a peak flux density $\gtrsim120$\,mJy at 0.01\,d in the optical, which is two orders
of magnitude brighter than the MASTER observations. Therefore, a regular RS
cannot explain the high-frequency radio peak at 1.5\,d.

The low frequency spectral peak at 1.5\,d occurs at $\approx 7$\,GHz.
Taking $\nucr\approx7$\,GHz and $\fnumax\approx0.25$\,mJy at this time,
we can show that relativistic RS models over-predict the optical flux density before
$\approx4\times10^{-3}$\,d by two orders of magnitude and are therefore ruled out. 
On the other hand, a Newtonian RS with $g\approx2.3$, $\nuar\approx4.2\times10^{9}$,
$\nucr\approx7.7\times10^{9}$, and $F_{\nu,\rm a}\approx0.28$\,mJy
results in spectra and light curves that represent the data well. In this model,
$\numr\ll\nuar$, and is therefore unconstrained. 
Requiring that these values 
be consistent with the FS at the deceleration time (\tdec), we derive 
$\tdec \approx 1.2\times10^{-3}$\,d ($\approx 100$\,s $\approx6 T_{90}$), the Lorentz factor at the 
deceleration time,
$\Gamma(\tdec) \approx 300$, and the RS magnetization, 
$R_{\rm B}\equiv\epsilon_{\rm B,RS}/\epsilon_{\rm B,FS}\approx0.6$.
Here, \tdec\ is constrained to be between the first two MASTER
observations in order to not over-predict the flux at either time. 
Whereas our derived value of $g$ is higher than the theoretically expected bounds
for a wind environment, we note that previous
studies have found even higher values from observations and modeling of 
GRB~130427A \citep{lbz+13,pcc+14}.
The critical Lorentz factor separating the thick
and thin shell regimes is given by
\begin{equation}
\Gamma_{\rm crit} = 88\left[
                    \frac{(1+z)\Astar}{E_{\rm K,iso,52}T_{90}}
                    \right]^{1/4},
\end{equation}
with $\Gamma<\Gamma_{\rm crit}$ corresponding to the thin shell and Newtonian RS regime
\citep{kmz04}. For the FS parameters in Table \ref{tab:params},
$\Gamma_{\rm crit} \approx 690$ and $\Gamma(\tdec)<\Gamma_{\rm crit}$ as required; 
however, we caution that $\Gamma(\tdec)$, $R_{\rm B}$, and $\tdec$ 
are all degenerate with respect to $\numr$ in this model.

\begin{figure*} 
 \begin{tabular}{cc}
  \includegraphics[width=\columnwidth]{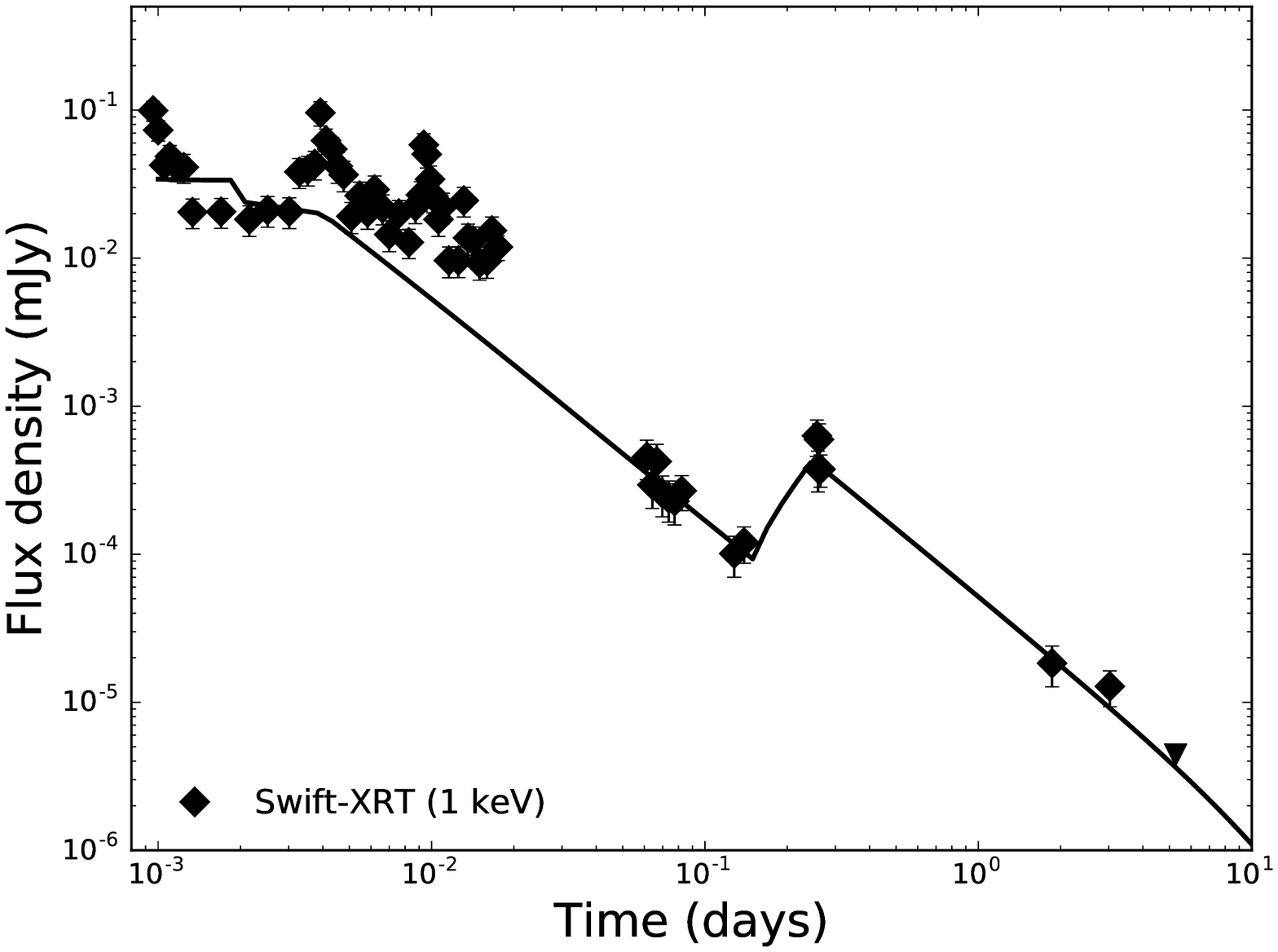} &
  \includegraphics[width=\columnwidth]{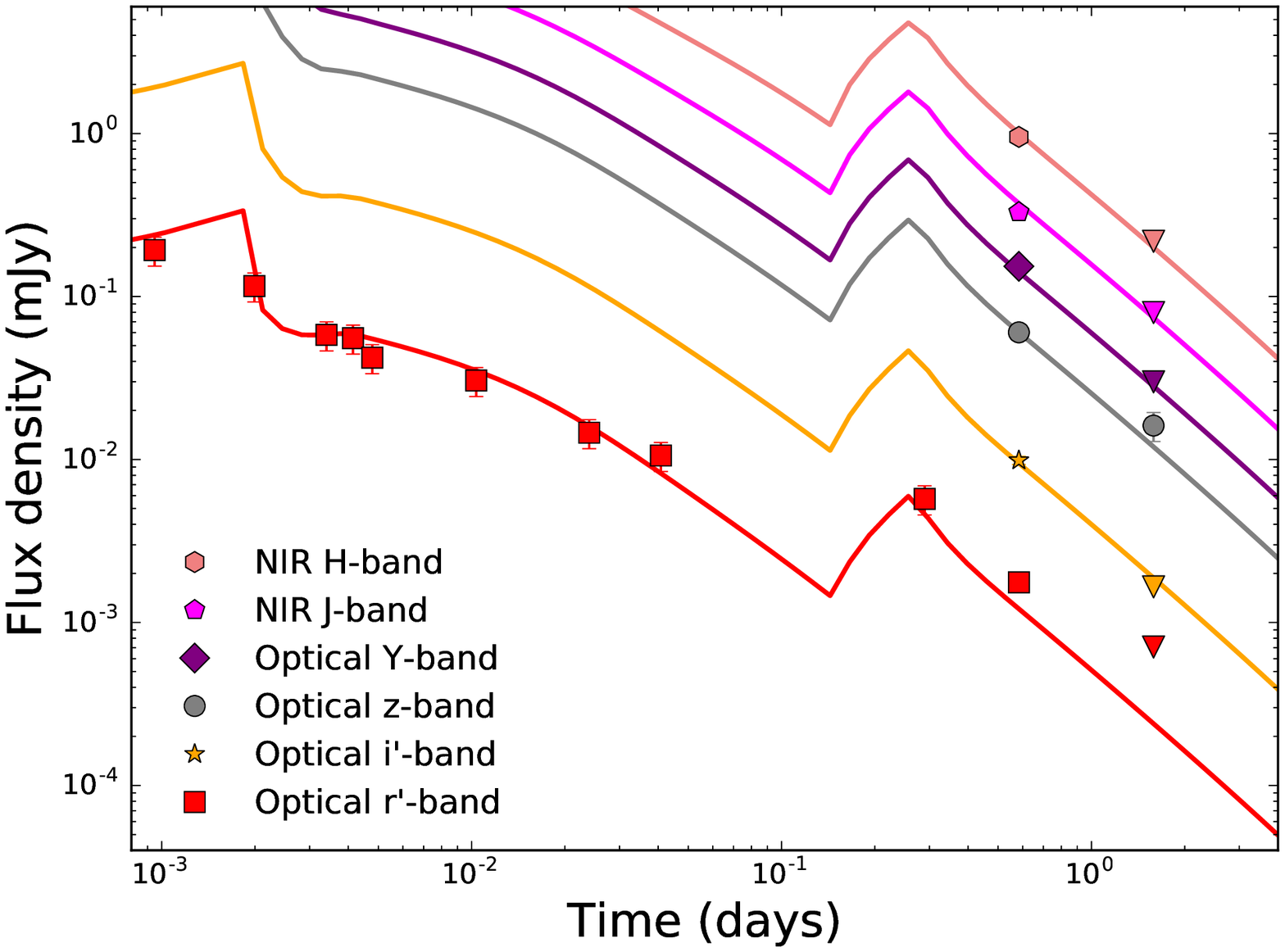} \\
  \includegraphics[width=\columnwidth]{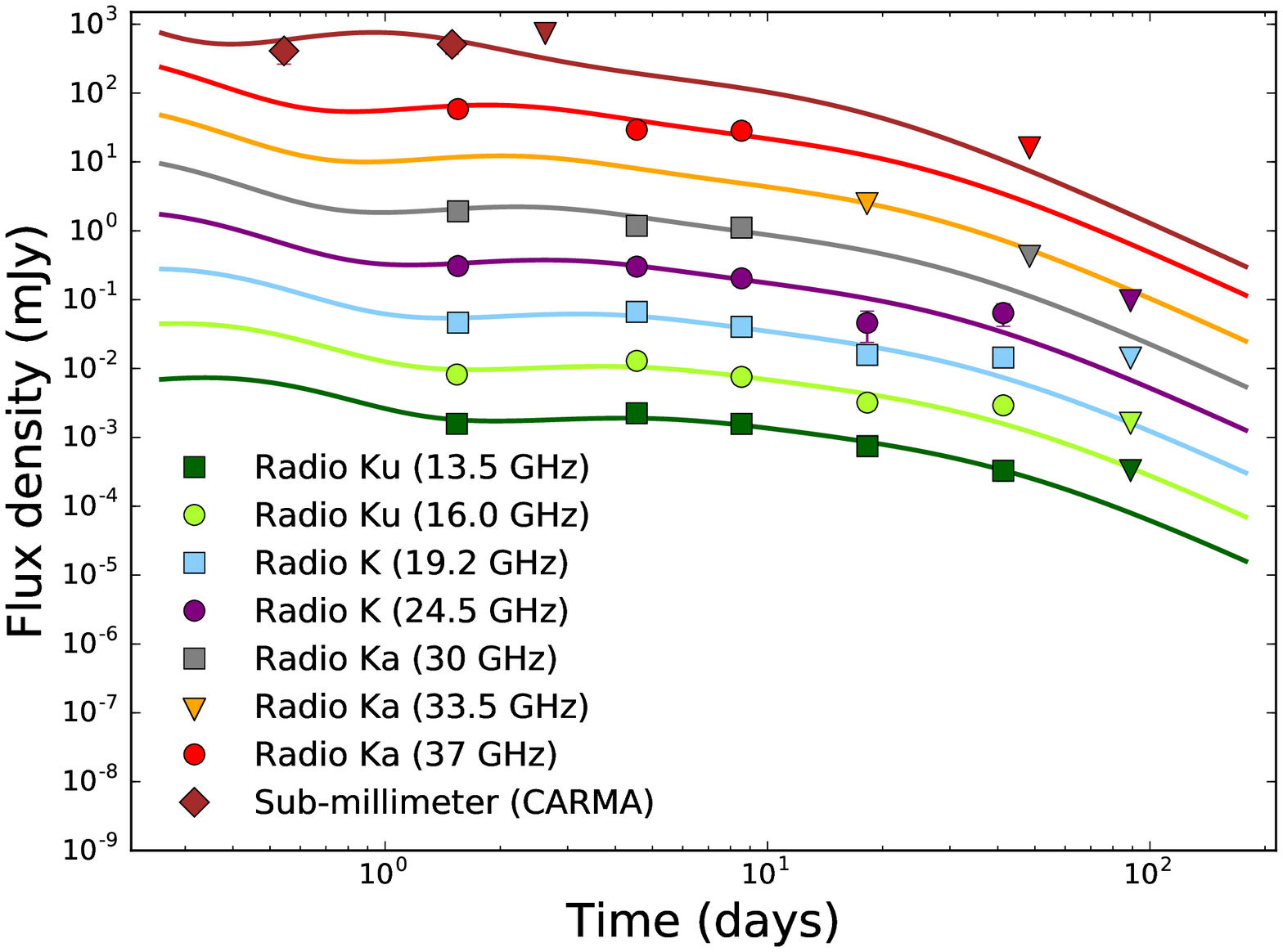} &
  \includegraphics[width=\columnwidth]{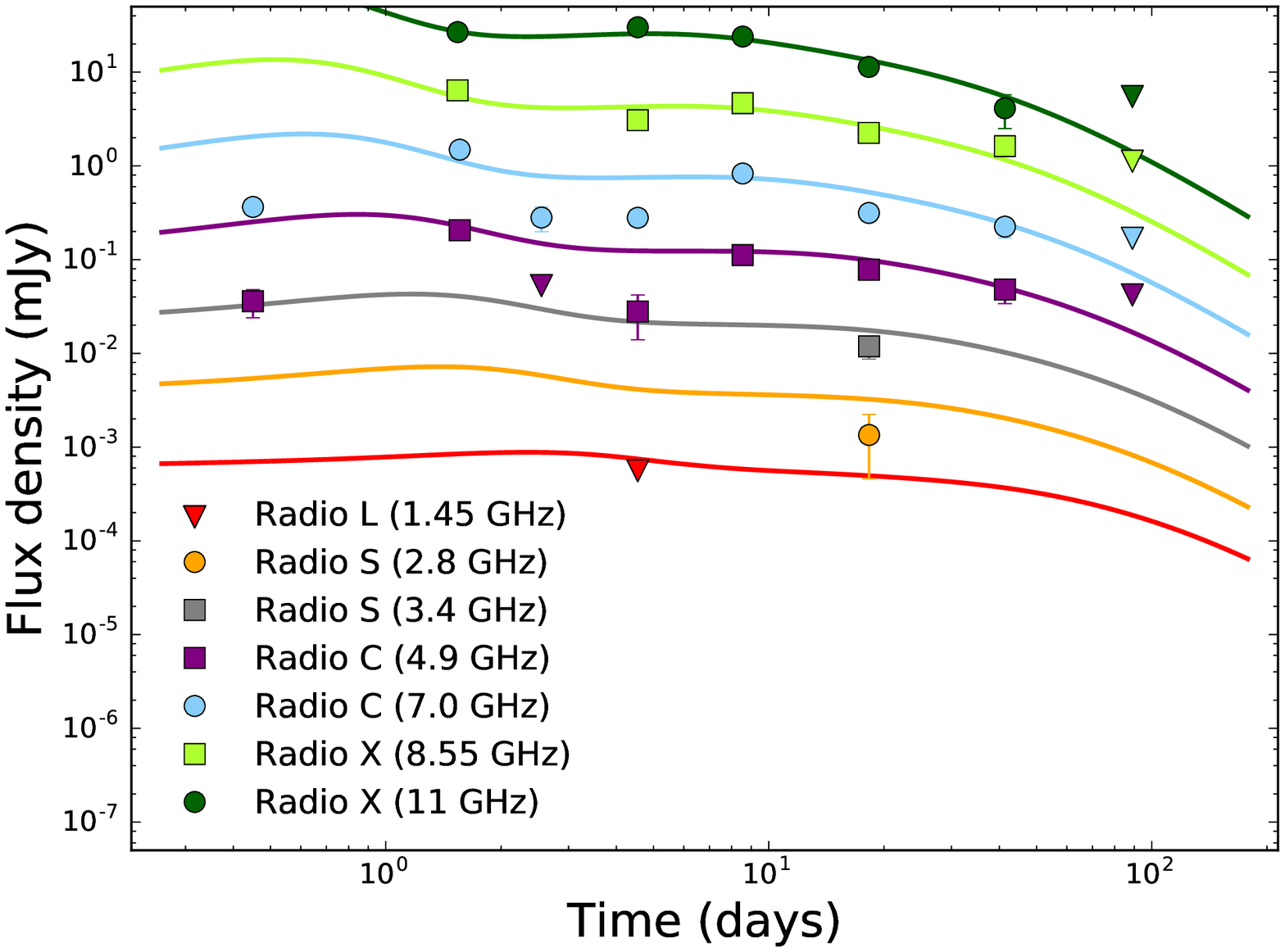} \\
 \end{tabular}
 \caption{Same as Figure \ref{fig:modellc_wind10}, now including energy injection
between 0.15\,d and 0.26\,d (Section \ref{text:injection}), 
a standard RS contributing to the optical/NIR light curve before $2\times10^{-3}$\,d
and to the radio at 1.5\,d (Section \ref{text:RS1}),
and a refreshed RS contributing to the mm-band light curve (Section \ref{text:injectionRS}).
}
\label{fig:modellc_wind13}
\end{figure*}

\begin{figure*}
\begin{tabular}{ccc}
 \centering
 \includegraphics[width=0.31\textwidth]{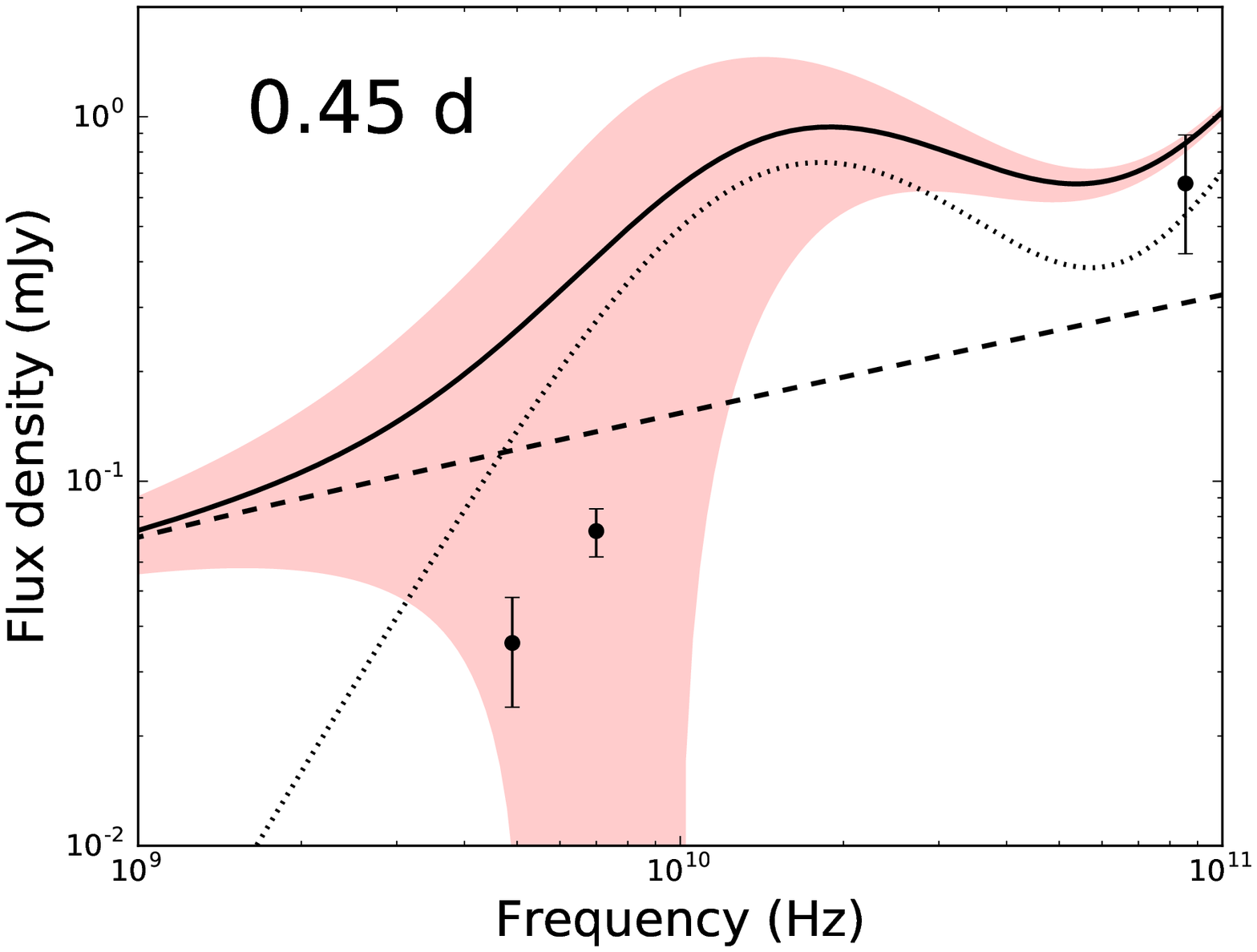} &
 \includegraphics[width=0.31\textwidth]{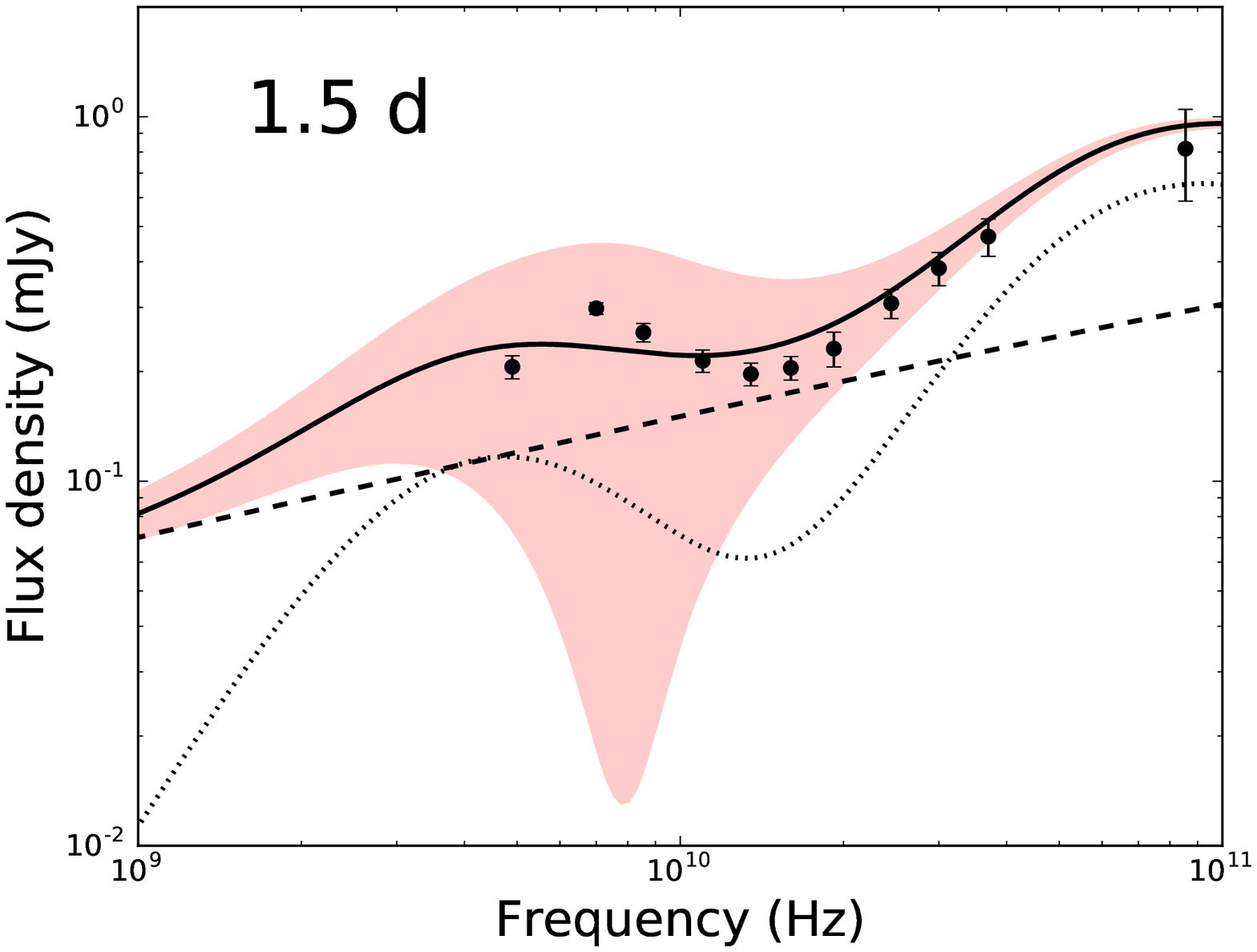} &
 \includegraphics[width=0.31\textwidth]{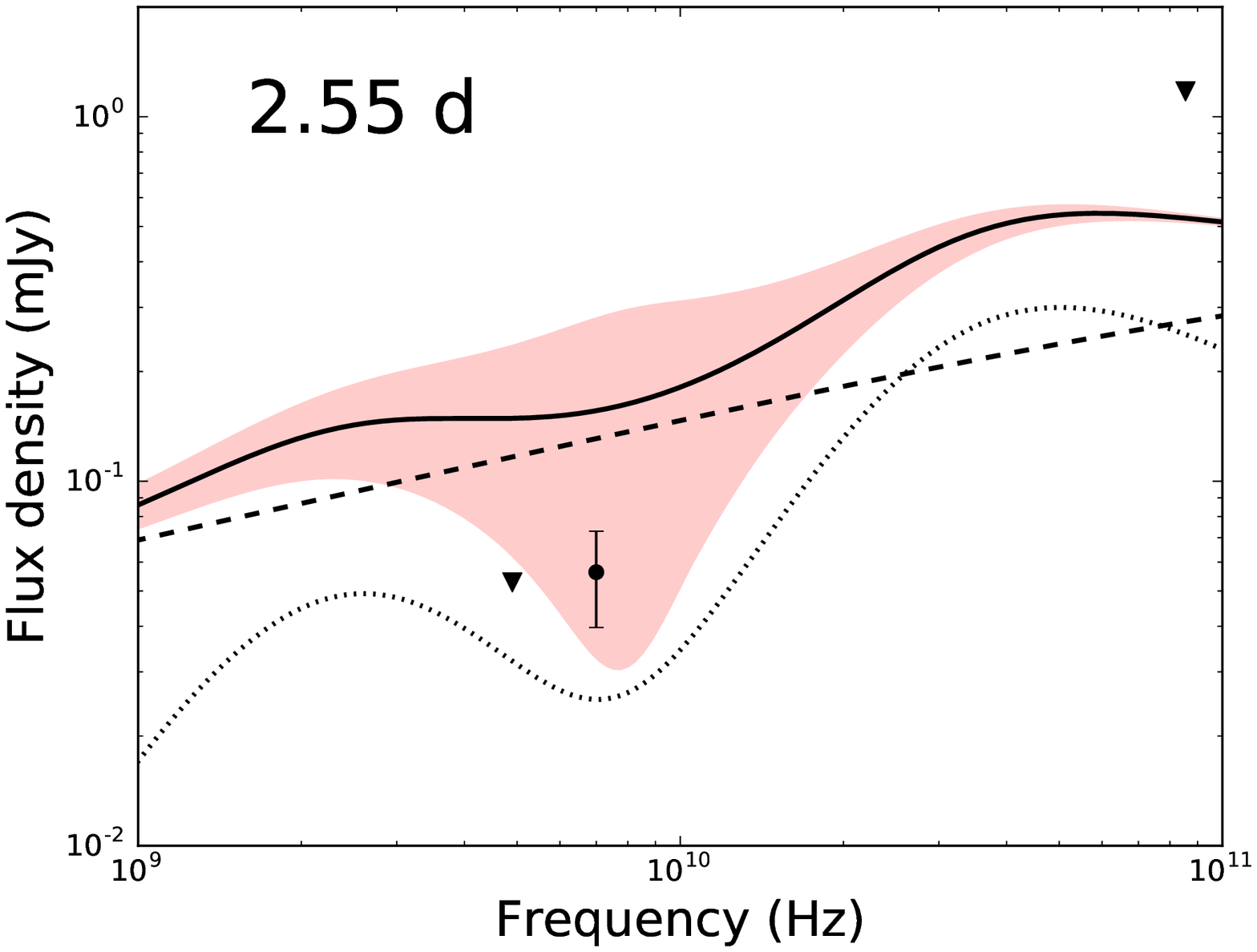} \\
 \includegraphics[width=0.31\textwidth]{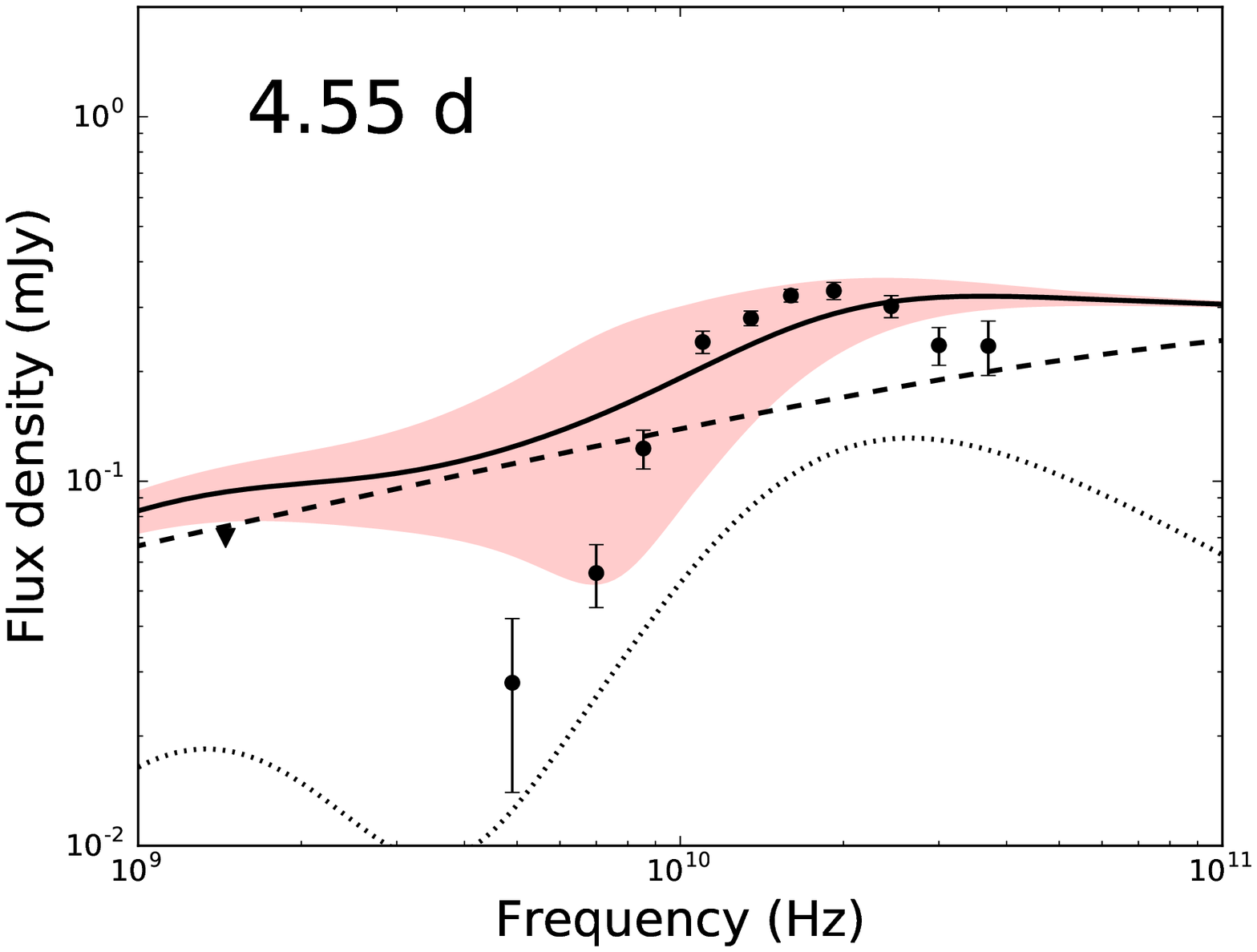} &
 \includegraphics[width=0.31\textwidth]{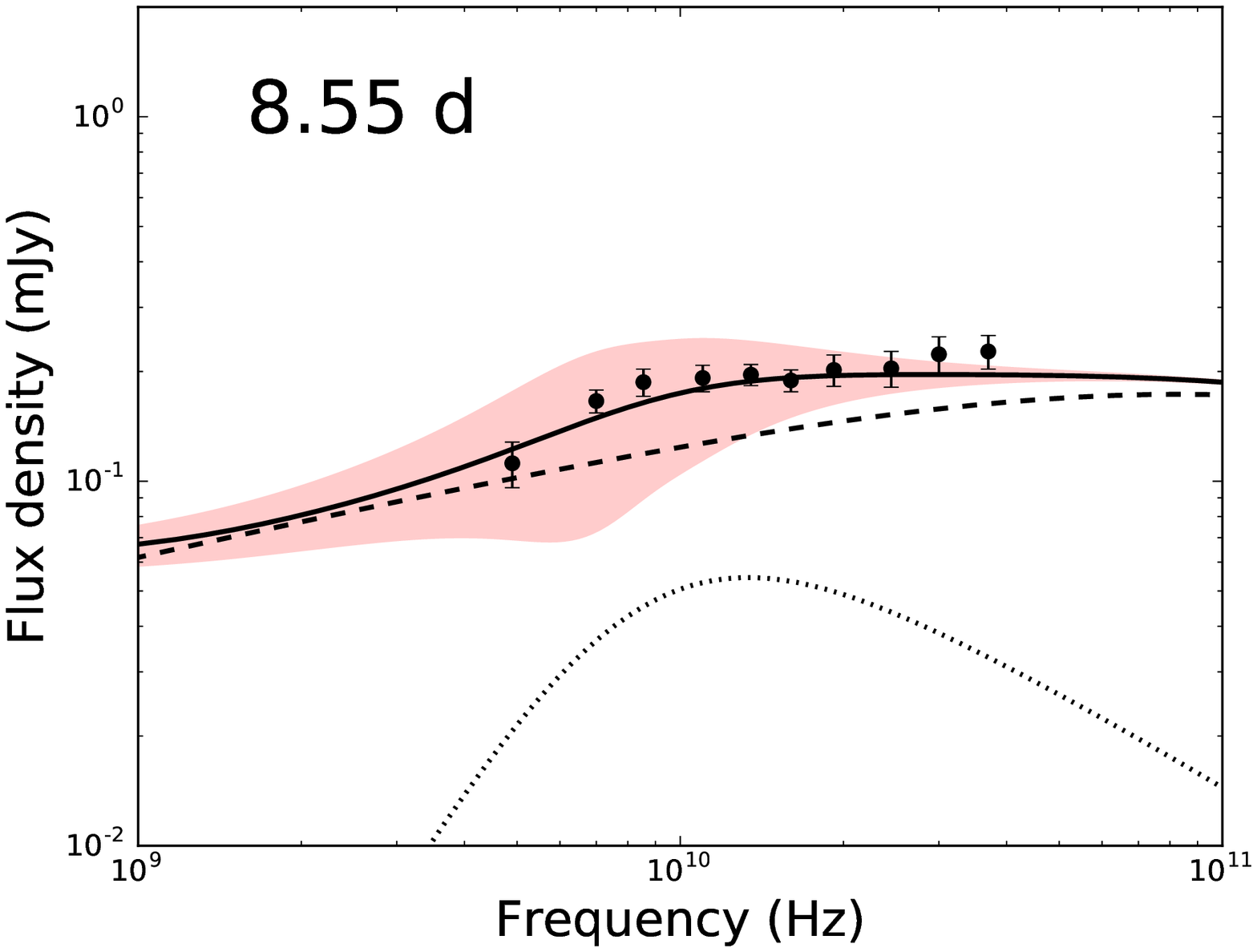} &
 \includegraphics[width=0.31\textwidth]{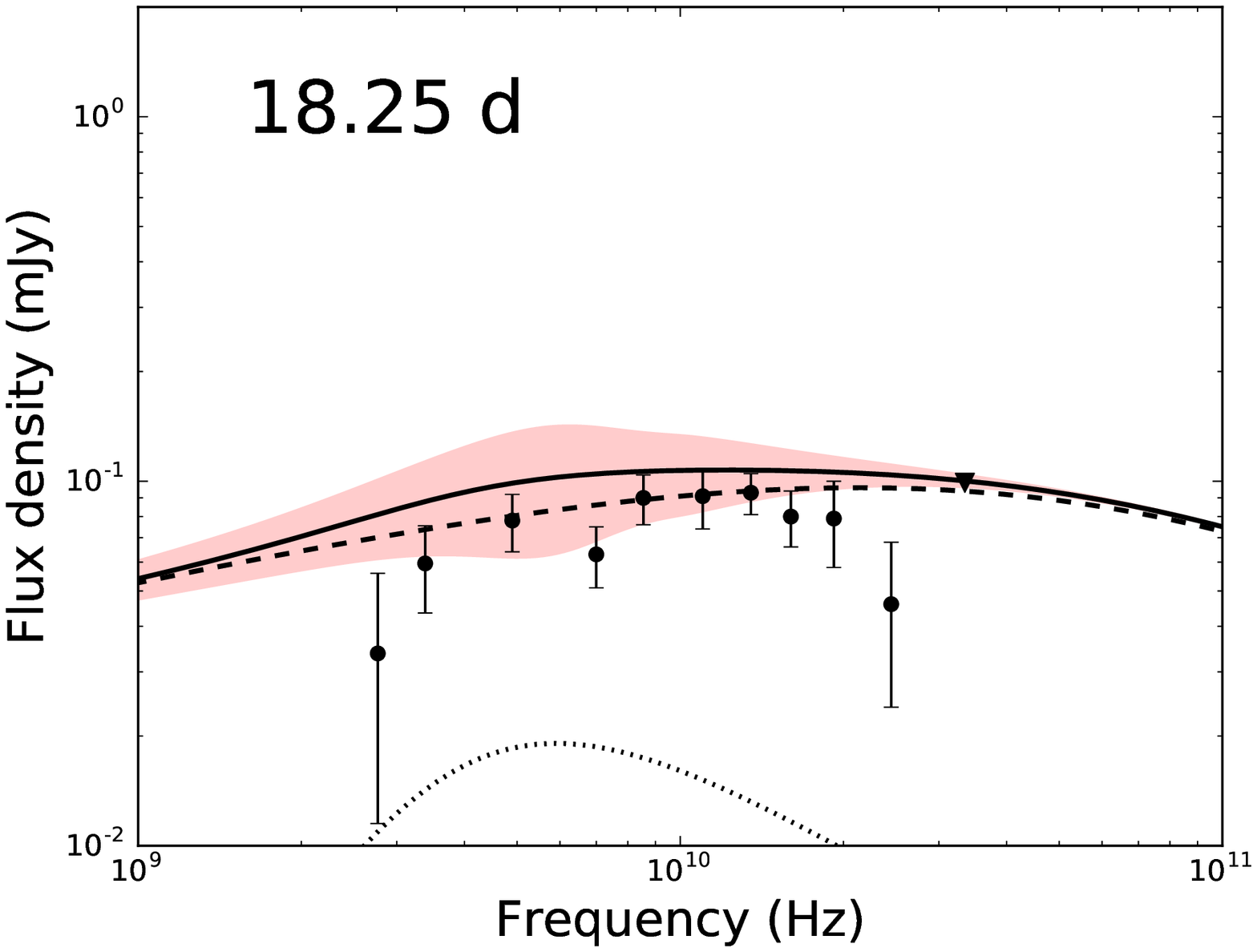} \\
 \includegraphics[width=0.31\textwidth]{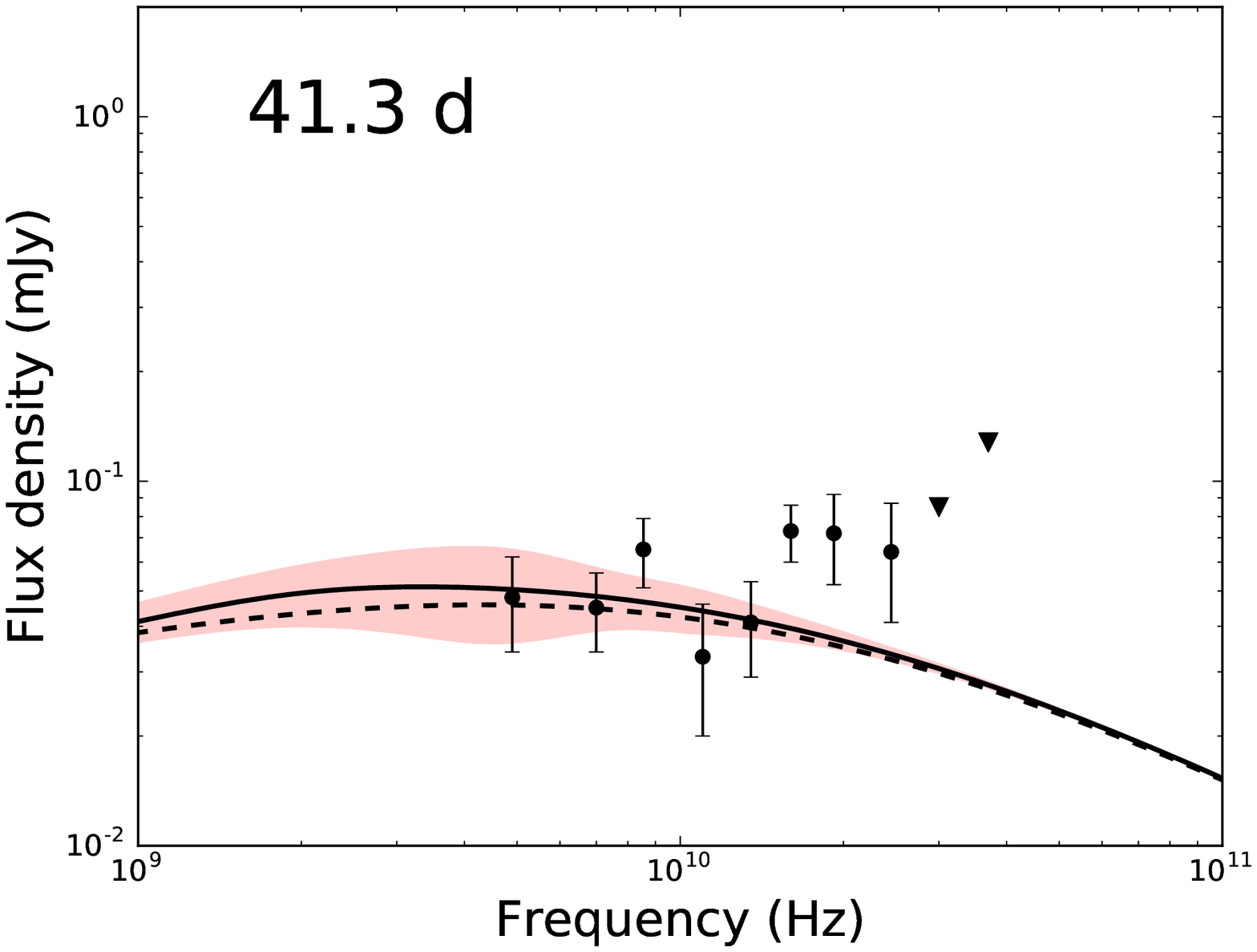} &
 \includegraphics[width=0.31\textwidth]{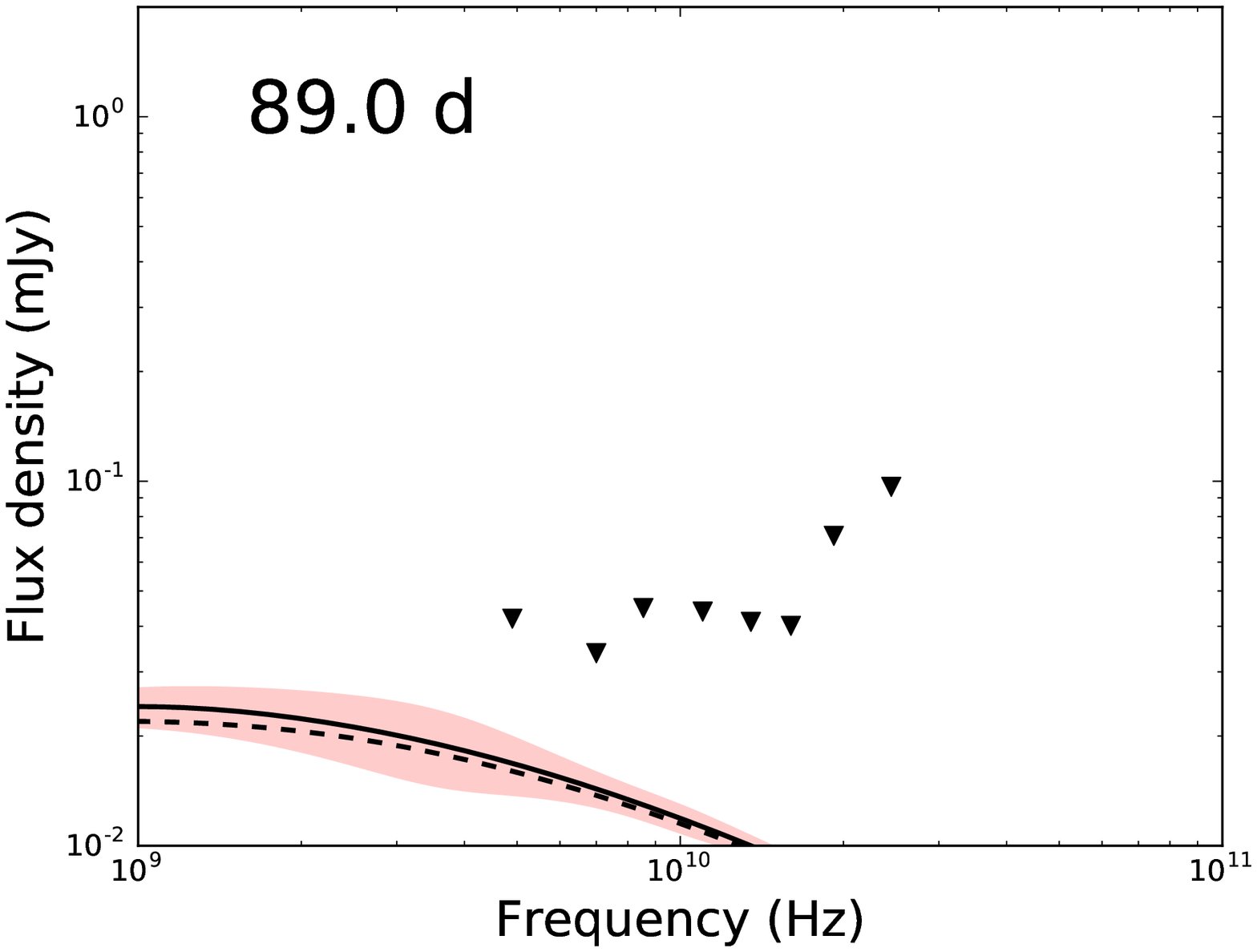} &  
\end{tabular}
\caption{Same as Figure \ref{fig:modelsed_wind10}, but for the model (solid) shown in Figure 
\ref{fig:modellc_wind13} decomposed into FS (dashed) and RS (dotted) contributions. The two RS 
components are best 
distinguished at 1.5\,d, where the lower frequency component arises from a RS in the original
shell producing the GRB, while the higher frequency component arises from
the collision of two shells at $0.15$\,d, which also produces the X-ray re-brightening.
The red shaded regions represent the expected variability due to scintillation. 
}
\label{fig:modelsed_wind13}
\end{figure*}

Before $\tdec$, the FS is expected to increase in energy as the ejecta
energy is transferred to the FS. A complete description of this process
requires knowledge of the ejecta Lorentz factor distribution and 
numerical simulations; however, for a single shell this 
process can be approximated by linear energy injection $E\propto t$ 
(see Appendix \ref{appendix:earlyFS}), which yields
a constant blast wave Lorentz factor akin to the coasting phase of jet evolution
\citep{dm15}. We plot the resulting light curves and radio SEDs
in Figures \ref{fig:modellc_wind13} and \ref{fig:modelsed_wind13}, respectively.

\subsection{Interstellar Scintillation}
We note that detailed analysis for the low-frequency radio spectral component 
at 1.5\,d is challenging due to the increased contribution of interstellar scintillation (ISS)
expected at frequencies below $\approx 10$\,GHz from the Milky Way interstellar medium (ISM).
It is possible that the entirety of this component is caused by an upward fluctuation due
to ISS. The correlation bandwidth for diffractive ISS is given by, 
\begin{equation}
\Delta\nu_{\rm D} \approx 2.8\times10^8(\nu/{5\,\rm GHz})^{4.4}\,{\rm GHz},
\end{equation}
toward this line of sight \citep{gn06}. 
This is of the same order as the observing bandwidth, $\approx 1$\,GHz. 
The diffractive scintillation time scale, 
\begin{equation}
t_{\rm diff}\approx80(\nu/5\,{\rm GHz})^{1.2}(v_{\perp}/30\,{\rm km\,s}^{-1})^{-1}\,{\rm min},
\end{equation}
where $v_{\perp}$ is the perpendicular velocity of the Earth relative to the line of sight.
Here we have taken the distance to the scattering screen of 
$d_{\rm scr} = 1.1$\,kpc for a transition frequency of $\nu_{\rm T}\approx11.6$\,GHz
and a scattering measure, ${\rm SM} = 3.5\times10^{-4}$\,kpc\,${\rm m}^{-20/3}$ from the
Galactic electron density model, NE2001 \citep{cl02}. Our C-band and X-band observations
at 1.5\,d span $\approx 20$\,min each. We attempted to test for short-time scale variability
by imaging each scan individually\footnote{The 
scan length is 510\,s at C band and 309\,s at X band.}.
The results do not reveal significant variability, suggesting that either 
the observed spectral feature at 1.5\,d is intrinsic to the source,
or the variability time scale is significantly longer than probed by our observations. 
Our subsequent analysis incorporates the expected contribution of both diffractive and refractive 
scintillation as described in \cite{lbt+14}.

\subsection{Reverse Shock from a shell collision}
\label{text:injectionRS}
We note that our multi-wavelength analysis in Section \ref{text:wind_fs} indicates a period
of significant energy injection between 0.15\,d and 0.26\,d. During this interval, the 
observed rate of energy increase ($E\propto t^{3.8}$) is greater than can be achieved from 
the gentle interaction of ejecta shells with a simple power law distribution of 
Lorentz factor. In particular, for injection due to ejecta mass distribution of 
$M(>\Gamma)\propto\Gamma^{-s+1}$, the energy of the FS increases as 
$E\propto t^m$, where $m=(s-1)/(s+3) < 1$. However, if the increase in energy is 
due to a violent interaction of two colliding shells, a greater rate of energy increase
is feasible \citep{lyu17,ld17}. Such an interaction would generate a RS propagating 
into the second shell and contribute to the observed synchrotron radiation.

Since the energy injection at 0.15\,d is rapid, we expect an RS to form at this time,
which propagates through the second shell while the injection process continues. After
the injection ends at 0.26\,d, radiation from the shock is expected to fade
as the shocked ejecta expand and cool due to synchrotron and adiabatic losses.
In Figures \ref{fig:modellc_wind13} and \ref{fig:modelsed_wind13}, 
we present a model with an additional Newtonian RS that 
is launched at a collision time of 
$\tcol \approx 0.15$\,d, and propagates through
the ejecta until the end of injection at 0.26\,d. This model requires $\nuar\approx60$\,GHz, 
$\nucr\approx2\times10^{13}$\,Hz, $g\approx 2$, and $F_{\nu, \rm a}\approx1.3$\,mJy, with
$\numr \ll \nuar$. For this value of $g$, 
$\nuc\sim t_{\rm x}^{-1.5}$ and $\fnupk\sim t_{\rm x}^{-1}$, 
where $t_{\rm x} = t-\tcol$,
which agree well with the basic properties of Component II derived in
Section \ref{text:radio_bplfits_componentII}.

The Lorentz factor of the second shell is then given by,
\begin{equation}
\Gamma_2 = 2^{1/2}\Gamma_1\left(1-\frac{\Delta t_{\rm L}}{\tcol}\right)^{-1/2},
\end{equation}
where $\Gamma_1$ is the Lorentz factor of the FS at the time of collision 
and $\Delta t_{\rm L}$ is the interval in the observer frame between the ejection
of the two shells (Appendix \ref{appendix:shellcollision}). 
From the energy injection model, $\tcol\approx 0.15$\,d. 
From the \citet[][BM]{bm76} solution, $\Gamma_1\approx 110$ for 
the FS at this time for the parameters derived in Section \ref{text:wind_fs}. 
The two quantities $\Delta t_{\rm L}$ and $\Gamma_2$ are degenerate -- 
a shell emitted at a later
time may catch up at the same collision time, $\tcol$ if $\Gamma_2$ is higher.
We can break this degeneracy by invoking additional information from the X-ray light curve. 
If we suppose the X-ray
flaring activity up to $\approx 10^{-2}$\,d is related to the ejection 
of this second shell, then we determine $\Gamma_2\approx 160$. 
While this computation relies on several assumptions, we note the resulting Lorentz factor is 
lower than $\Gamma(\tdec)\approx300$ for the ejecta derived 
in Section \ref{text:RS1}.

\section{Summary and Discussion}
\label{text:discussion}
We present detailed multi-frequency, multi-epoch radio observations of 
GRB~140304A at $z=5.283$ spanning 1\,GHz to 86\,GHz and 0.45\,d to 89\,d.
The radio and mm SEDs comprise at least three distinct spectral components. 
We investigate physical models responsible for each emission component
through detailed multi-wavelength analysis in the standard synchrotron 
emission paradigm.

The first component may arise either from extreme scintillation, 
or from a Newtonian RS propagating through the first ejecta shell. 
In the latter case, we derive a Lorentz factor of 
$\Gamma(\tdec)\approx300$, a deceleration time,
$\tdec \approx 1.2\times10^{-3}$\,d, and weak ejecta magnetization,
$R_{\rm B}\approx0.6$. However, these parameters are degenerate with respect
to the unknown value of the characteristic frequency, $\numr$, which is 
located below the radio band at all times.

The second component is consistent with emission from a refreshed RS
produced by the violent collision of two shells with different Lorentz factors
emitted at different times. The collision injects energy into the FS,
which manifests as a re-brightening in the X-ray light curve.
The initial Lorentz factor of the second shell is degenerate with its
launch time; if the flaring activity observed in the X-ray light curve
is associated with the creation of this shell, we can break this 
degeneracy and obtain $\Gamma\approx 160$.

The third component is consistent with synchrotron radiation from a forward shock 
propagating into a wind medium with 
$\Astar\approx2.6\times10^{-2}$, 
corresponding to a progenitor mass loss rate of 
$\dot{M}\approx3\times10^{-7} M_{\odot}\,{\rm yr}^{-1}$ for a wind velocity of
1000\,km\,s$^{-1}$. The total energy of the ejecta inferred from modeling the 
FS using the X-ray, optical/NIR and radio observations after 0.26\,d is 
$\EKiso\approx5\times10^{54}$\,erg. 
The inferred prompt efficiency derived by comparing \Egamma\ with
the final ejecta kinetic energy is low, $\eta\approx2\%$. 
However, the true prompt efficiency is expected to be related to the energy
of the first shell, $\EKiso\approx6\times10^{53}$\,erg.
Assuming this shell is responsible for the prompt $\gamma$-rays,
we obtain a prompt efficiency of $\eta \approx 17\%$, 
commensurate with the internal shock model \citep{kps97}. 
The radio observations suggest a jet break
at $\approx11$\,d, yielding a narrow jet opening angle of 
$\thetajet\approx1.1$\,deg. The small opening angle is in accordance with 
our previous work showing evidence of a narrower median opening angle
for GRBs at $z\gtrsim6$ compared to events at lower redshift \citep{lbt+14}.
The resulting beaming-corrected energies are
$\EK\approx10^{51}$\,erg and $\Egamma\approx 2\times10^{49}$\,erg.

The proposed model matches the X-ray, optical, NIR, and radio observations 
over five orders of magnitude in time from $10^{-3}$\,d to 89\,d.
For completeness, we have considered alternate scenarios that do not invoke 
multiple radio components. However, no scenario can simultaneously explain 
the X-ray re-brightening, together with the observed optical and radio evolution
as well as the three-component radio SEDs.

\section{Conclusions}
\label{text:conclusions}
The bright radio afterglow of GRB~140304A has yielded unexpected riches:
the presence of multiple radio spectral components and a late-time 
X-ray re-brightening. Together with optical observations, the radio components 
are suggestive of multiple shocks partly arising from a period of energy injection 
initiated by the collision of two relativistic shells. 
Whereas the details of the radio SEDs cannot be perfectly matched
even with multiple emission components, the residual offsets are 
consistent with an origin in interstellar diffractive scintillation.
These observations highlight the importance of disentangling the effects of scintillation 
from intrinsic physical processes, a process that is challenging given the current
uncertainty in scintillation theory. 
Further multi-frequency observations of GRB afterglows in the cm and mm-bands, 
coupled with advances in scintillation theory will be key to a detailed understanding
of these new physical effects.

\acknowledgements
We thank Paul Duffell for helpful discussions. 
TL is a Jansky Fellow of the National Radio Astronomy Observatory. 
The Berger Time-Domain Group at Harvard is supported in part by the NSF under grant 
AST-1411763 and by NASA under grant NNX15AE50G.
BAZ acknowledges support from NSF AST-1302954. 
VLA observations for this study were obtained via project 14A-344. 
The National Radio Astronomy Observatory is a facility of the National Science Foundation operated 
under cooperative agreement by Associated Universities, Inc.
This research has made use of data supplied by the UK Swift Science Data Centre at the
University of Leicester, and of data obtained through the High Energy Astrophysics Science Archive 
Research Center On-line Service, provided by the NASA/Goddard Space Flight Center.

\appendix

\section{Additional Forward Shock-only Models}
\label{appendix:additional_fs}
In the main text, we show that the radio to X-ray emission 
of the afterglow is consistent
with FS and RS radiation in a wind-like circumburst environment. 
We present ISM models here for completeness. 
There are three possible scenarios depending on the location of \nuc\ relative to 
\nunir\ and \nux\ as discussed in Section \ref{text:wind_fs},
and we describe each in turn. 
The resulting best-fit parameters are summarized in Table \ref{tab:params_ISM}. 
The figures and tables associated with these models are available
in the on-line version of the journal.
We note that these models significantly underpredict the $r^{\prime}$-band light curve before 
$\approx0.1$\,d and are therefore disfavored\footnote{Attempting to explain the deficit as RS 
emission either over-predicts the data at other wavelengths or yields unrealistic parameters in 
conjunction with the FS model, such as $\Gamma(\tdec)\approx 10$.}.

\subsection{High cooling frequency, $\nuc > \nux$}
\label{text:highnuc}
\begin{figure*} 
 \begin{tabular}{cc}
  \includegraphics[width=0.47\textwidth]{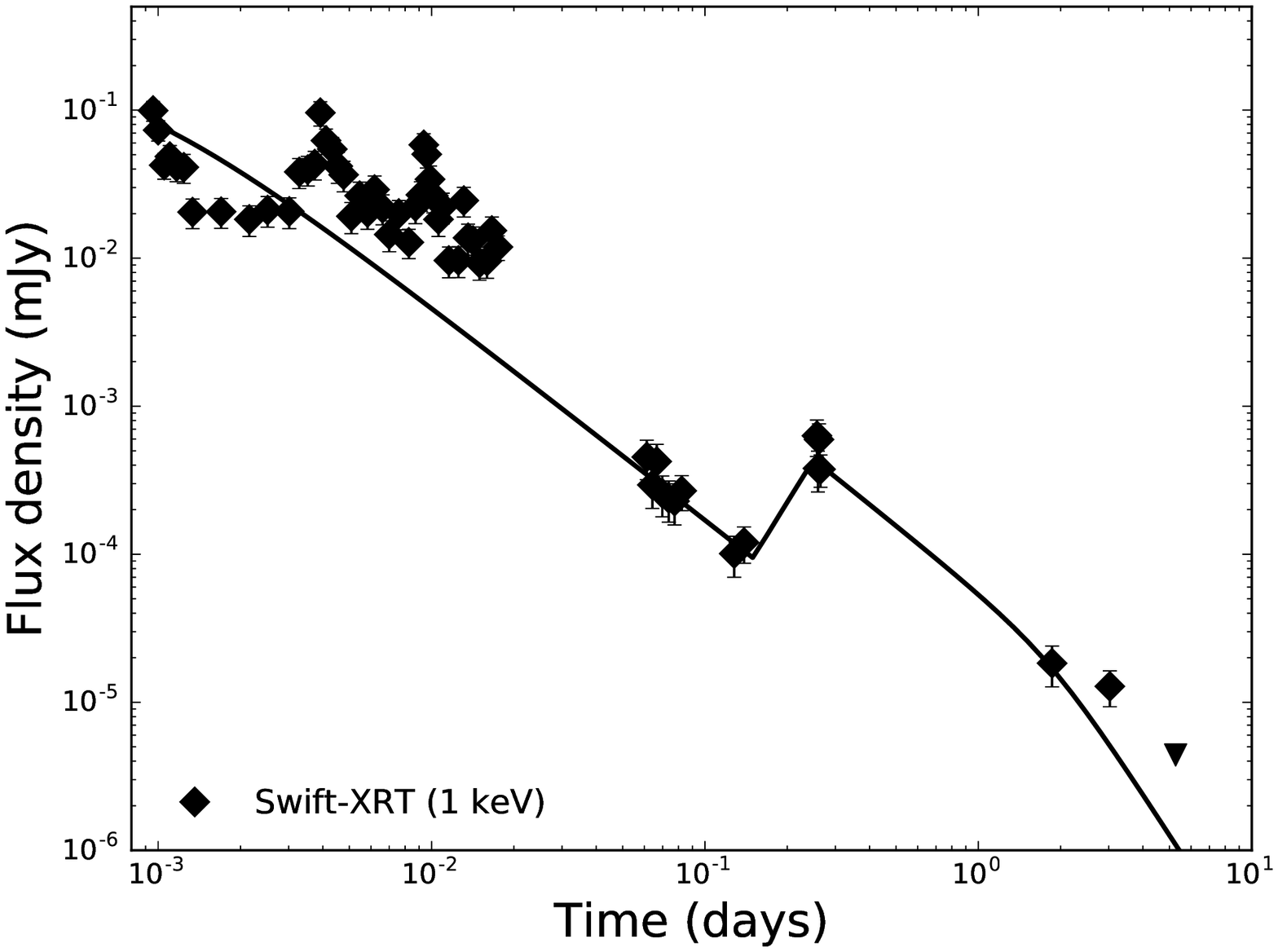} &
  \includegraphics[width=0.47\textwidth]{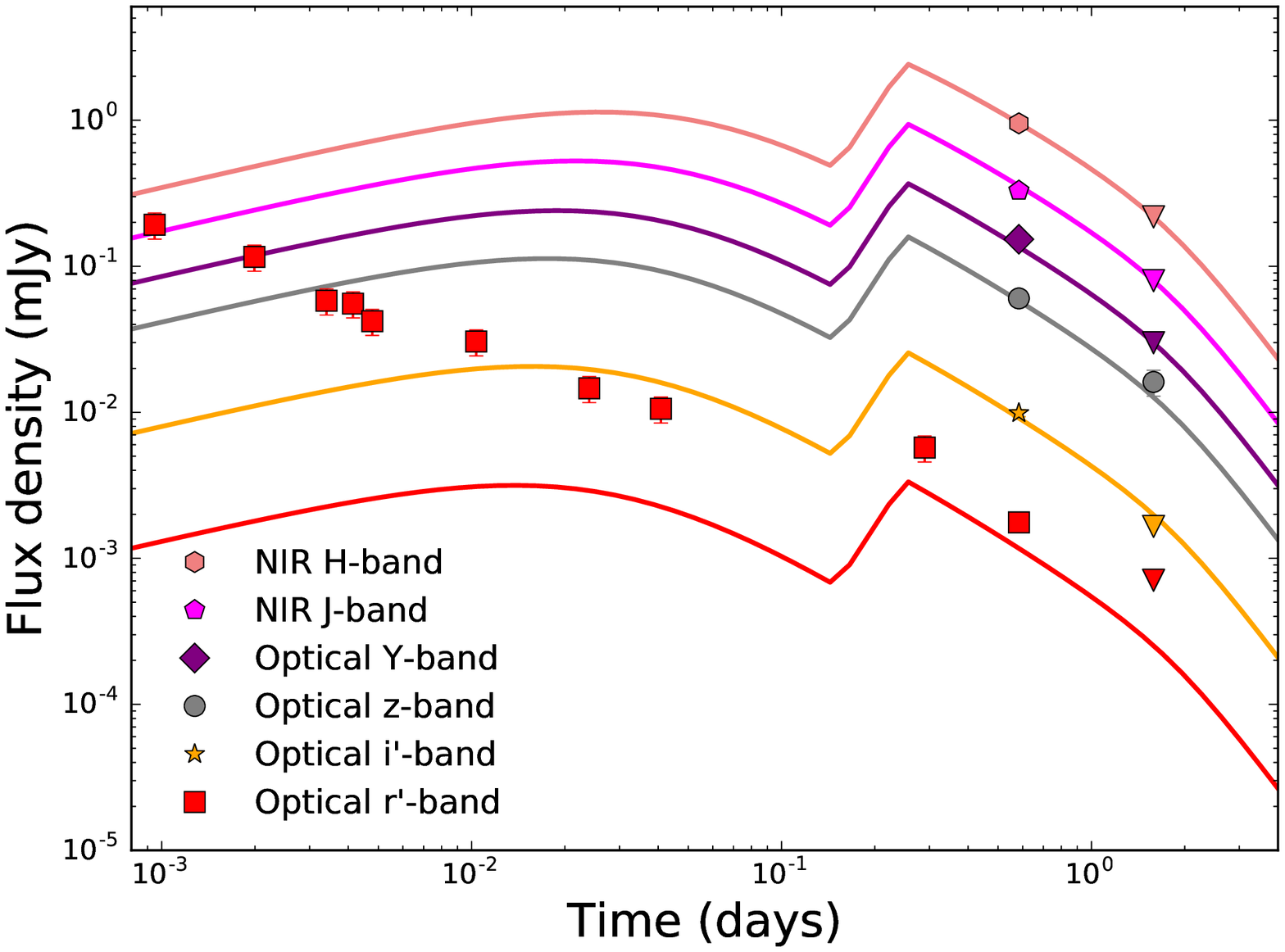} \\
  \includegraphics[width=0.47\textwidth]{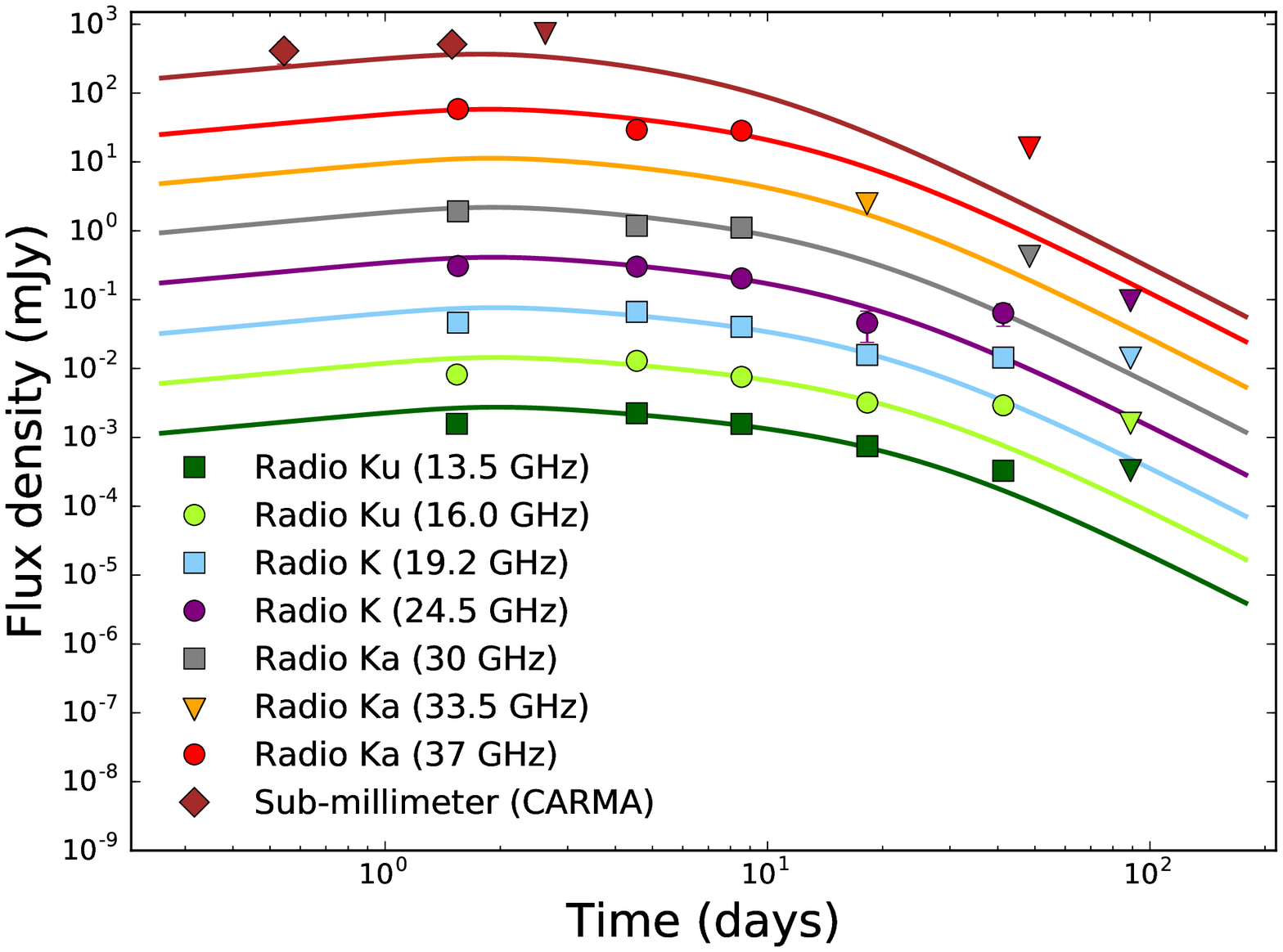} &
  \includegraphics[width=0.47\textwidth]{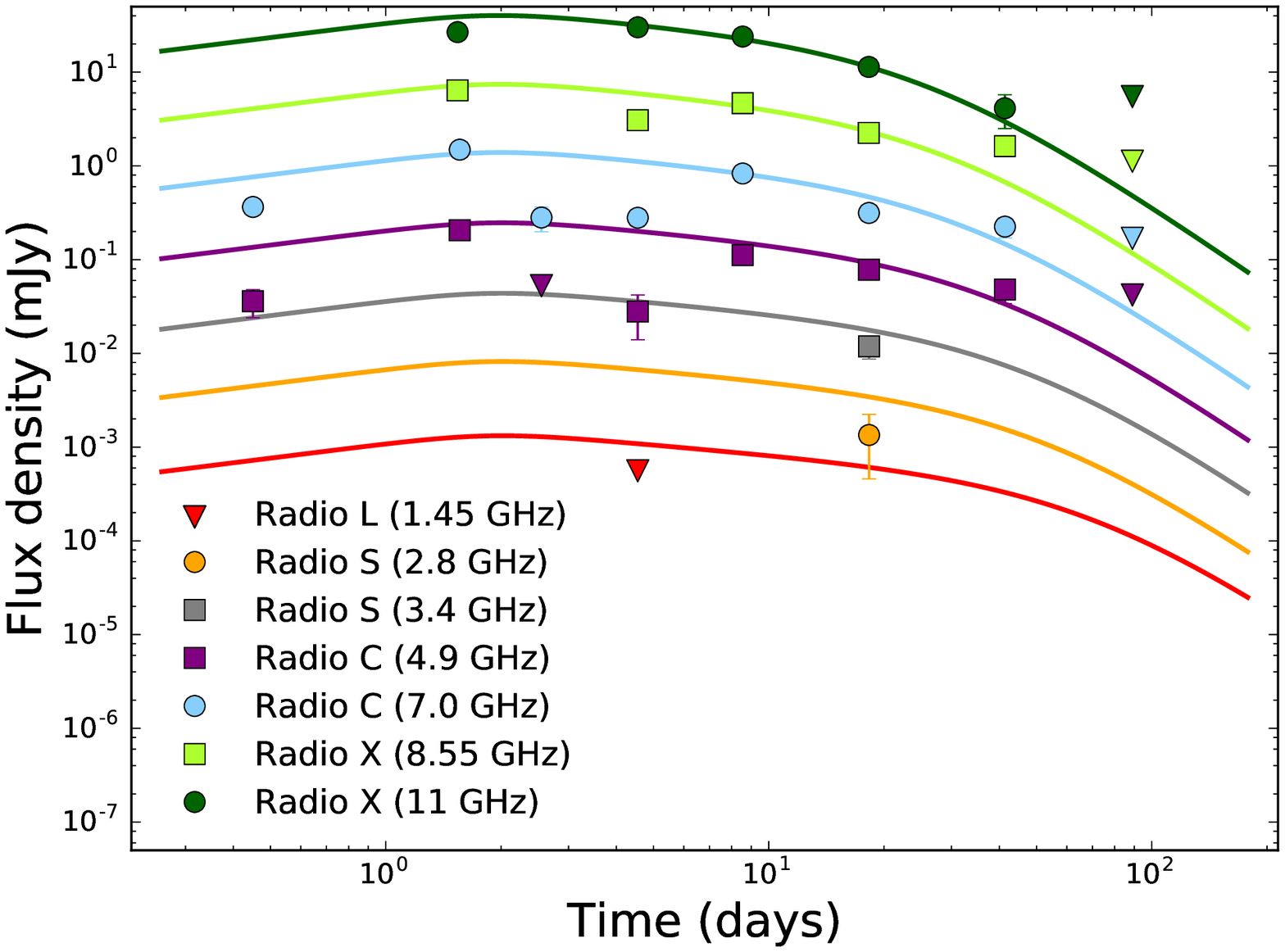} \\
 \end{tabular}
 \caption{X-ray (top left), optical/NIR (top right) and radio (bottom) light curves of the 
afterglow of GRB 140304A, together with an FS ISM model with $\nunir,\nux<\nuc$ including
energy injection between 0.15\,d and 0.26\,d (Section \ref{text:injection}). 
The model significantly under-predicts the optical light curve before $4\times10^{-2}$\,d,
and is therefore disfavored.
}
\label{fig:modellc_ISM14}
\end{figure*}

\begin{figure*}
\begin{tabular}{ccc}
 \centering
 \includegraphics[width=0.31\textwidth]{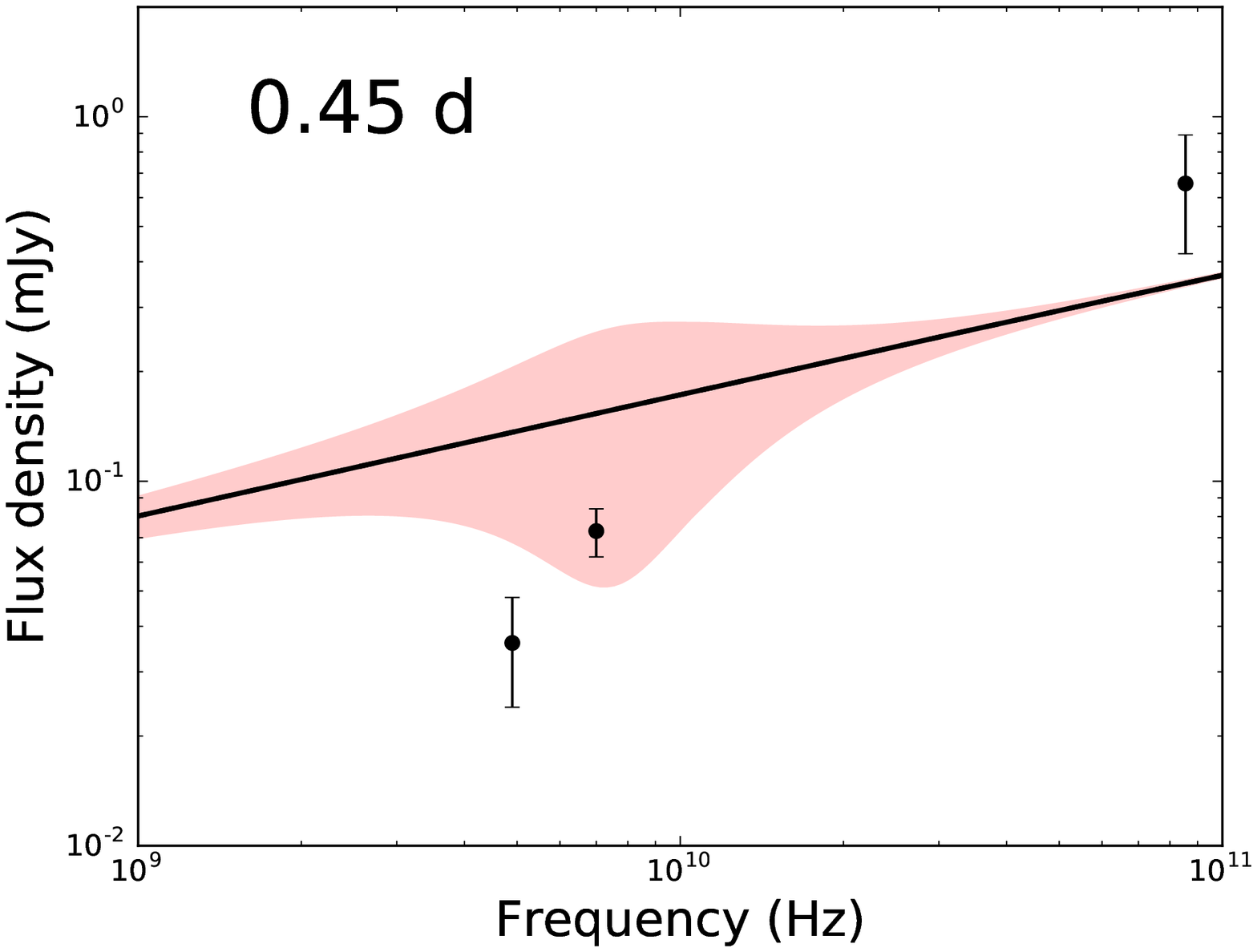} &
 \includegraphics[width=0.31\textwidth]{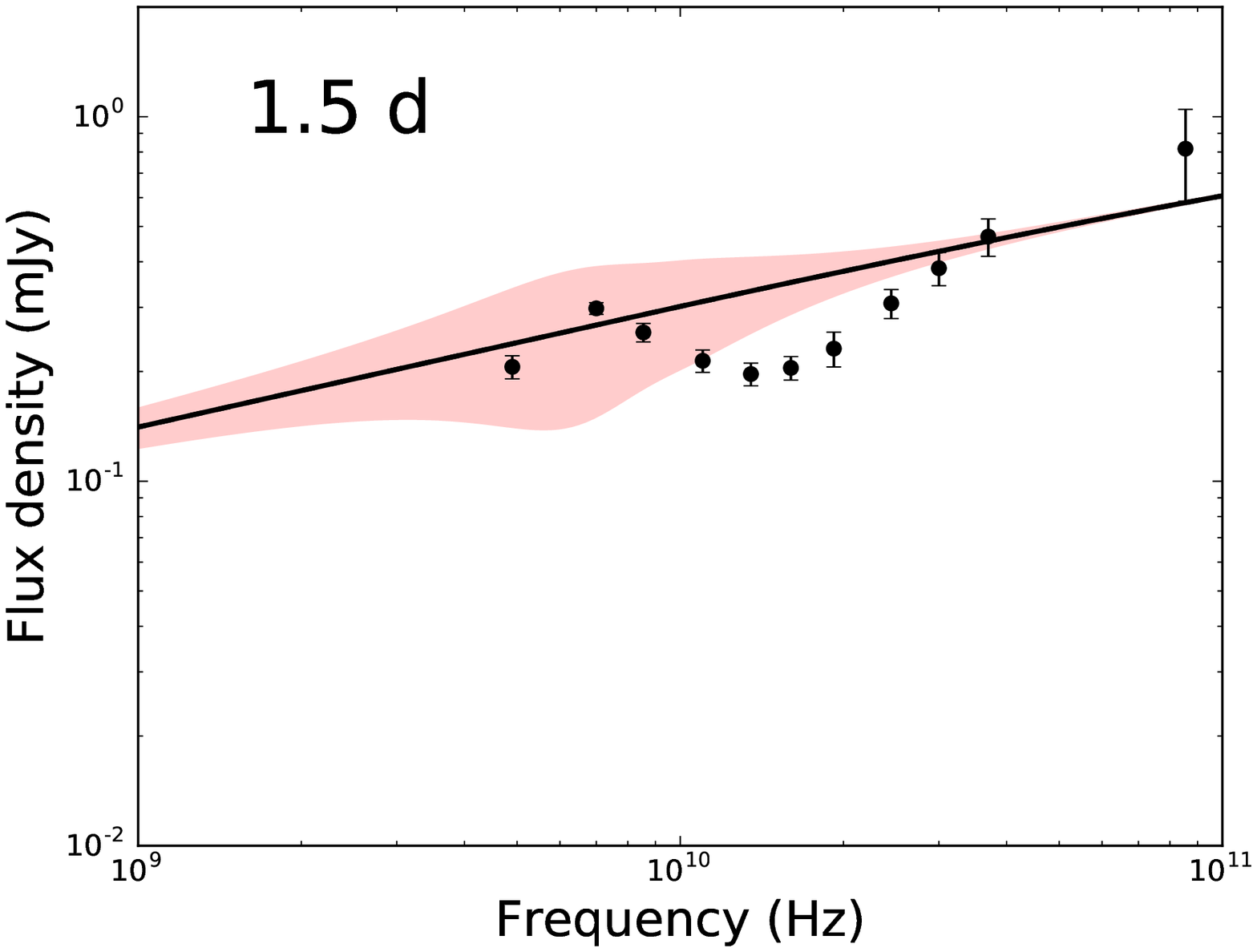} &
 \includegraphics[width=0.31\textwidth]{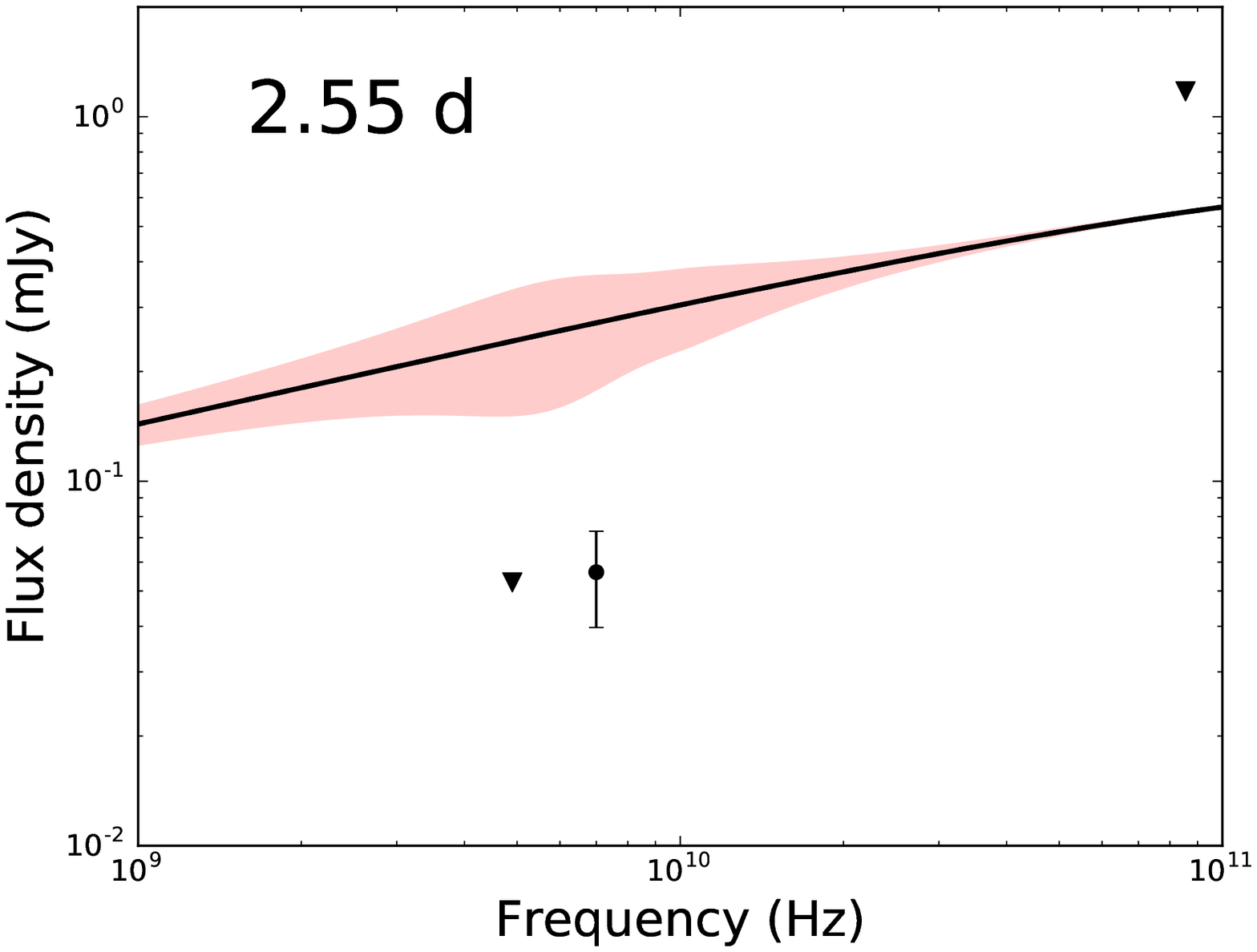} \\
 \includegraphics[width=0.31\textwidth]{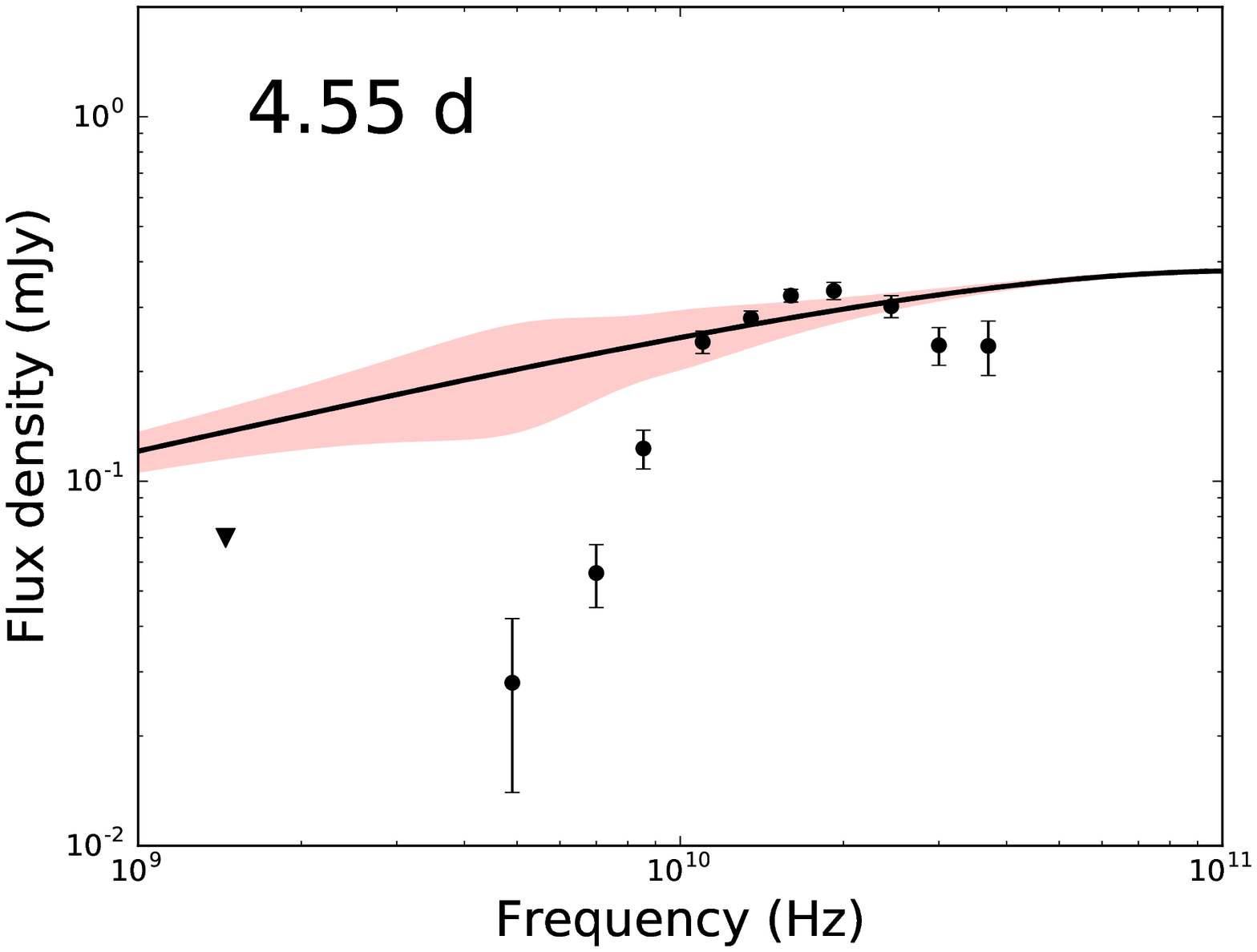} &
 \includegraphics[width=0.31\textwidth]{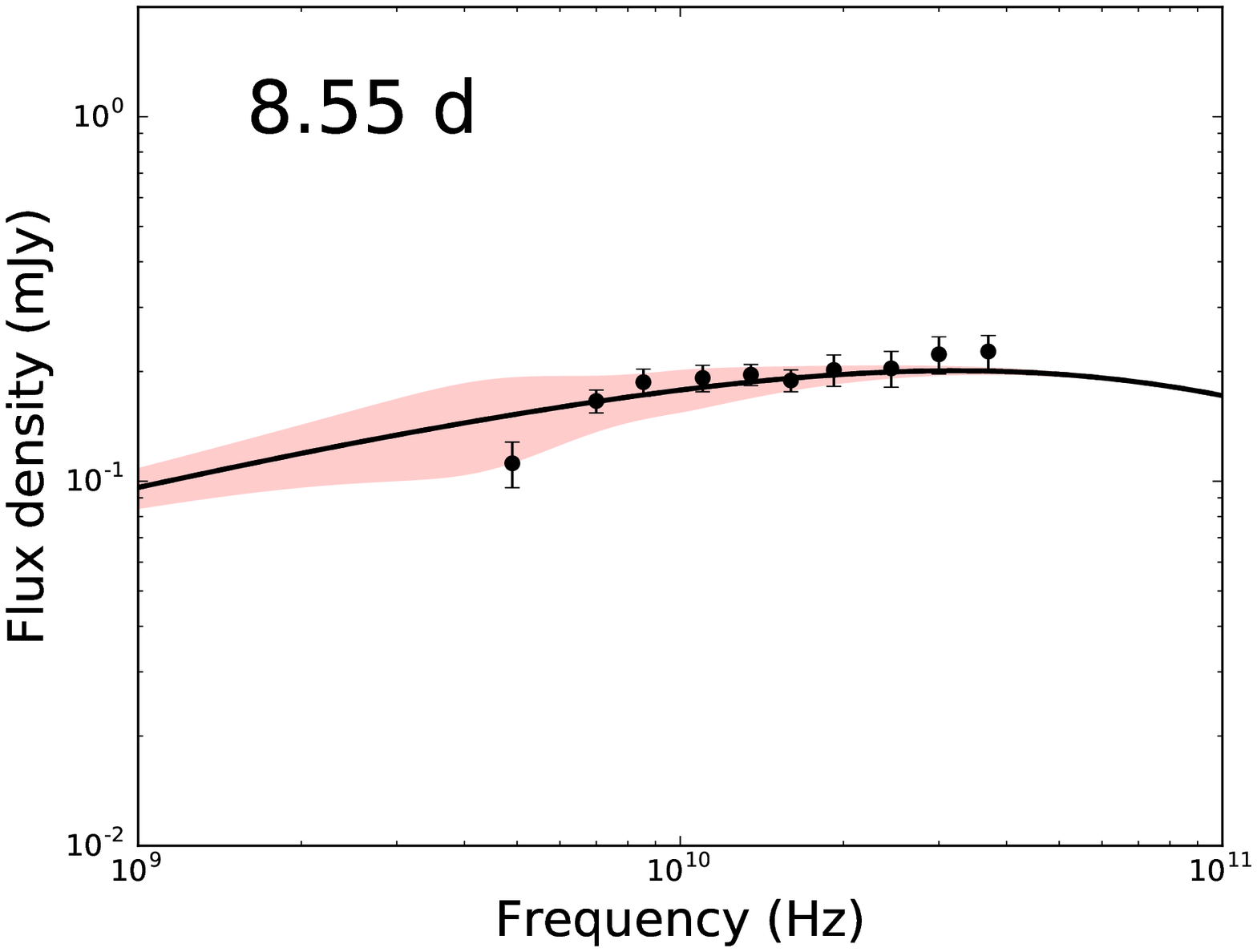} &
 \includegraphics[width=0.31\textwidth]{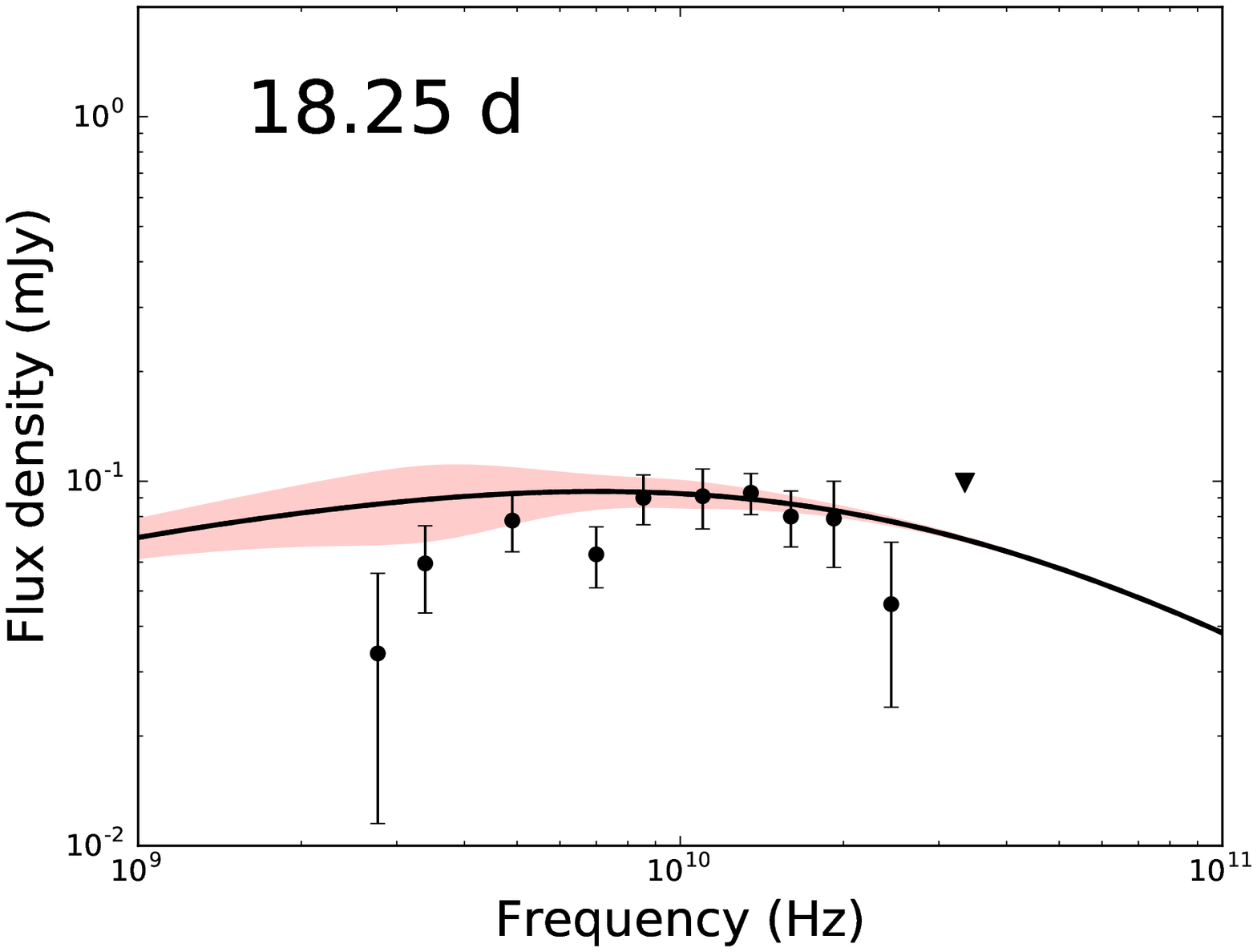} \\
 \includegraphics[width=0.31\textwidth]{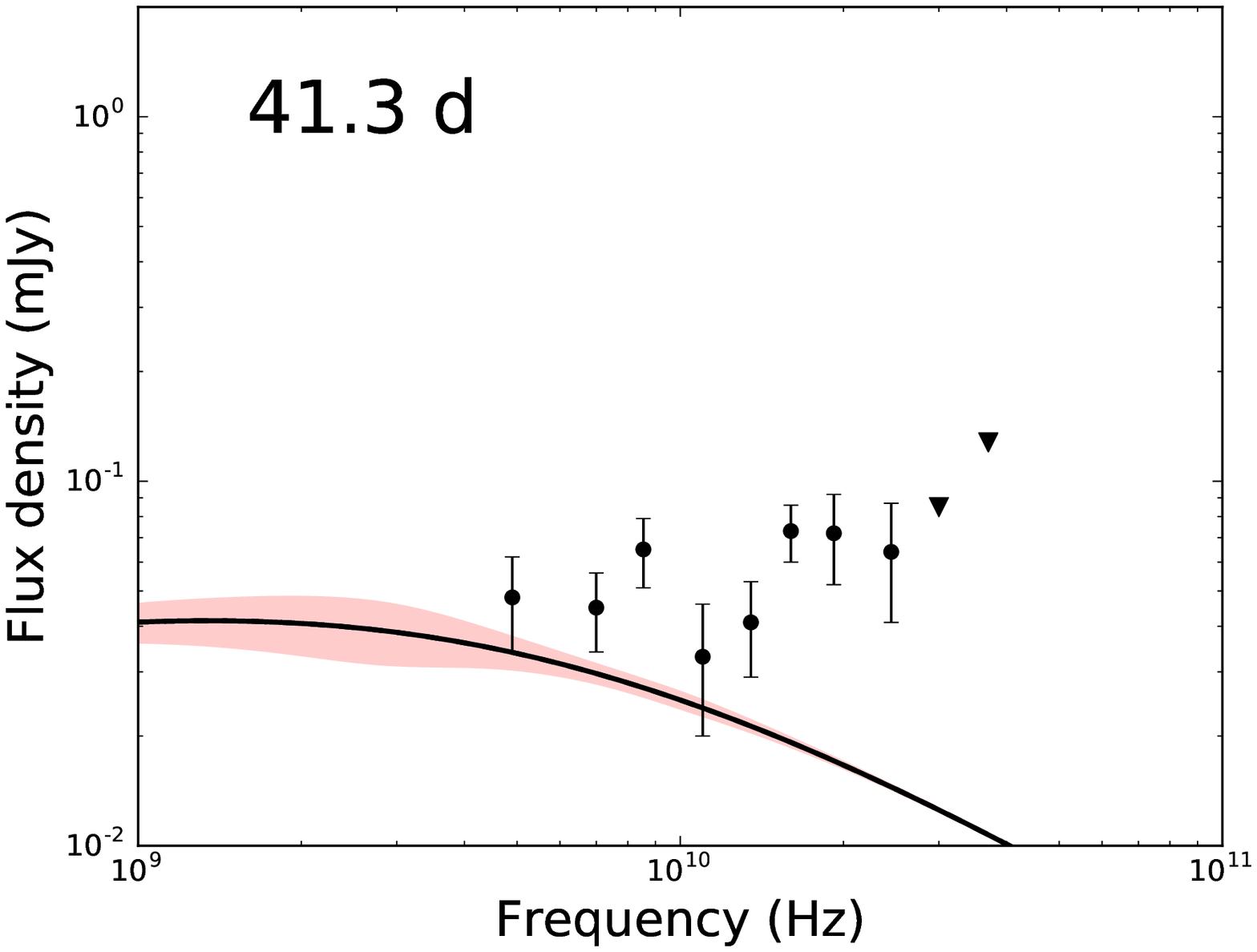} &
 \includegraphics[width=0.31\textwidth]{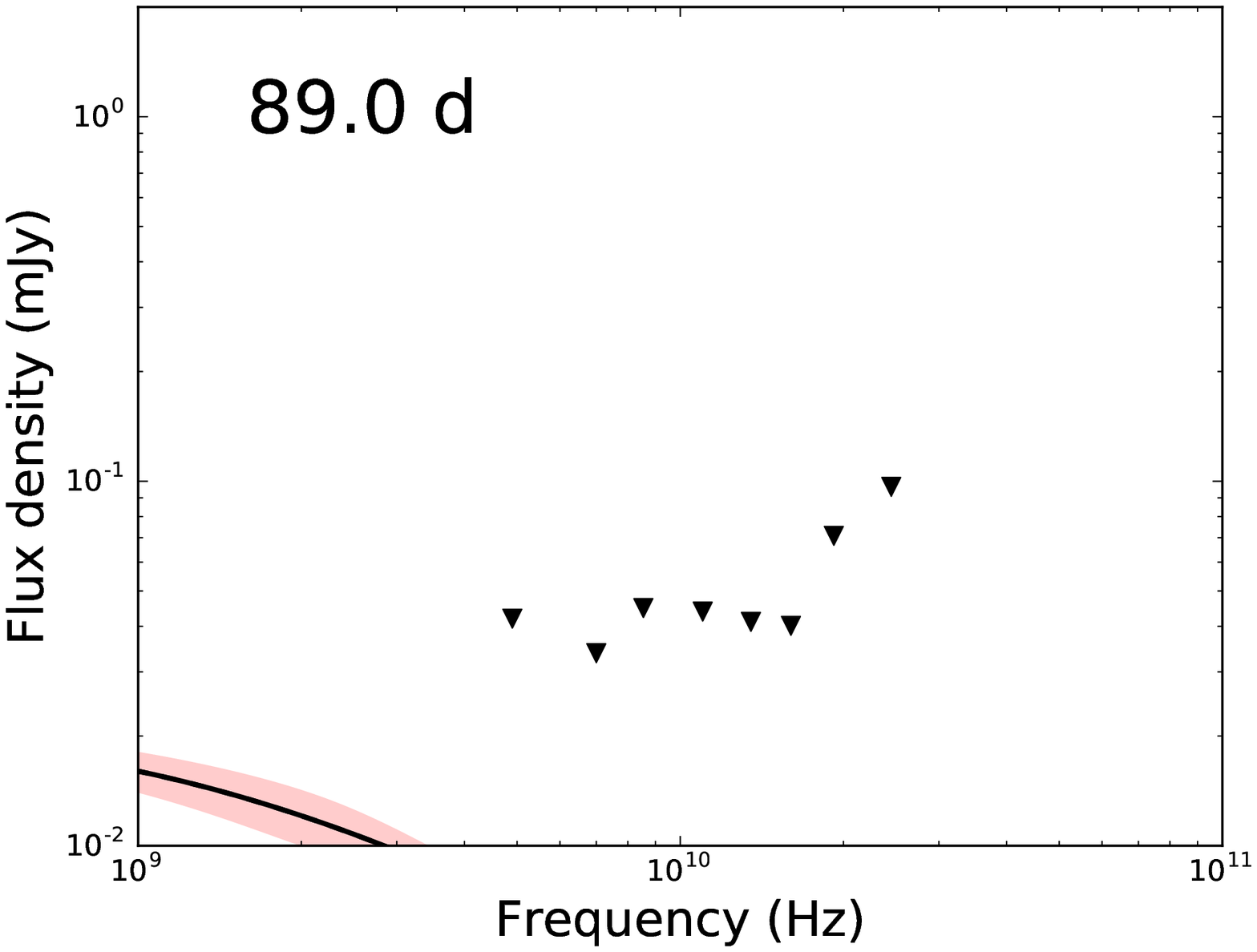} & 
\end{tabular}
\caption{Radio spectral energy distributions of the afterglow of \me\ at multiple epochs starting 
at 0.45~d, together with the same FS ISM model with $\nux<\nuc$ as in Figure 
\ref{fig:modellc_ISM14}. The red shaded regions represent the expected variability due to 
scintillation.
}
\label{fig:modelsed_ISM14}
\end{figure*}

If $\numax < \nunir < \nux < \nuc$, we require $p\approx2.9$ to explain the 
optical/NIR and X-ray light curves (Section \ref{text:wind_fs}). 
The resulting model fits the X-ray and NIR observations after 0.2\,d, and the radio SED at 8.5\,d 
and 18.25\,d. However, it over-predicts the 4.9\,GHz flux density and under-predicts the CARMA 
observation at 0.45\,d, results in an excess at $\approx15$\,GHz at 1.5\,d, severely 
over-predicts the 4.9 and 7.0\,GHz data at 2.5\,d, does not match the steep radio spectrum at 
4.5\,d, and marginally under-predicts the radio SED at 41.3\,d 
(Figures \ref{fig:modellc_ISM14} and \ref{fig:modelsed_ISM14}). 
In addition, the model requires\footnote{Since both $\nua$ and $\nuc$ are unconstrained by this 
model, the physical parameters can by scaled as 
$\epse = \epse^*(\nua/\nua^*)^{5/6}(\nuc/\nuc^*)^{1/4}$,
$\epsb = \epsb^*(\nua/\nua^*)^{-5/2}(\nuc/\nuc^*)^{-5/4}$,
$\dens = \dens^*(\nua/\nua^*)^{25/6}(\nuc/\nuc^*)^{3/4}$, and
$E_{\rm K,iso} = E_{\rm K,iso}^*(\nua/\nua^*)^{-5/6}(\nuc/\nuc^*)^{1/4}$ 
without modifying the spectrum (modulo inverse Compton corrections), where the parameters with 
asterisks refer to the values in Table \ref{tab:params_ISM}. In the limit $\epse\rightarrow1$, 
$E_{\rm K,iso}\rightarrow1.3\times10^{55}$\,erg remains high.}
a very high isotropic-equivalent kinetic energy, $E_{\rm K,iso}\approx3\times10^{56}$\,erg, a low 
density, and a very small jet opening angle, $\thetajet \approx 0.5$\,deg. 

\subsection{Low cooling frequency, $\nuc < \nunir$}
\label{text:lownuc}
\begin{figure*} 
 \begin{tabular}{cc}
  \includegraphics[width=0.47\textwidth]{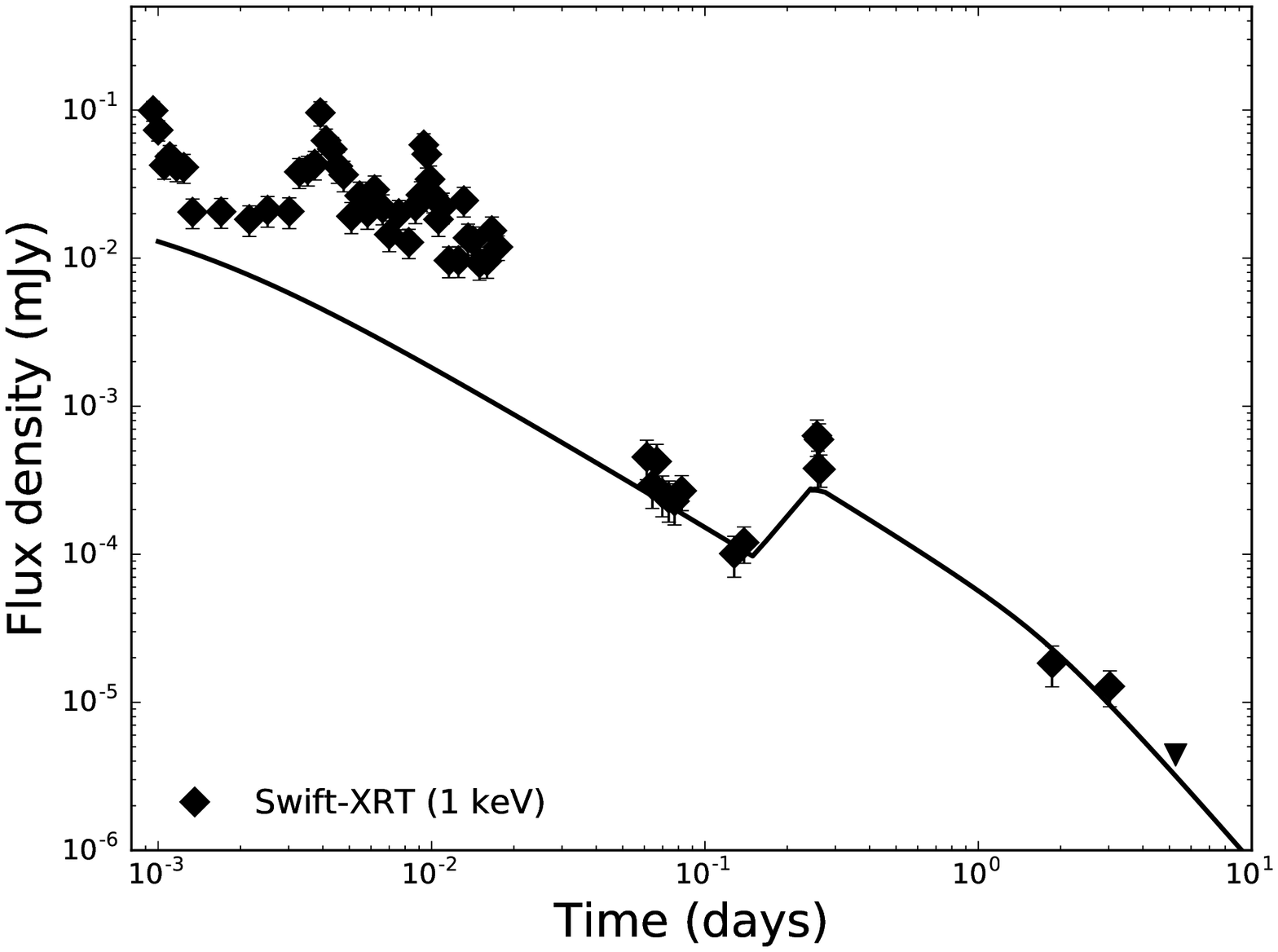} &
  \includegraphics[width=0.47\textwidth]{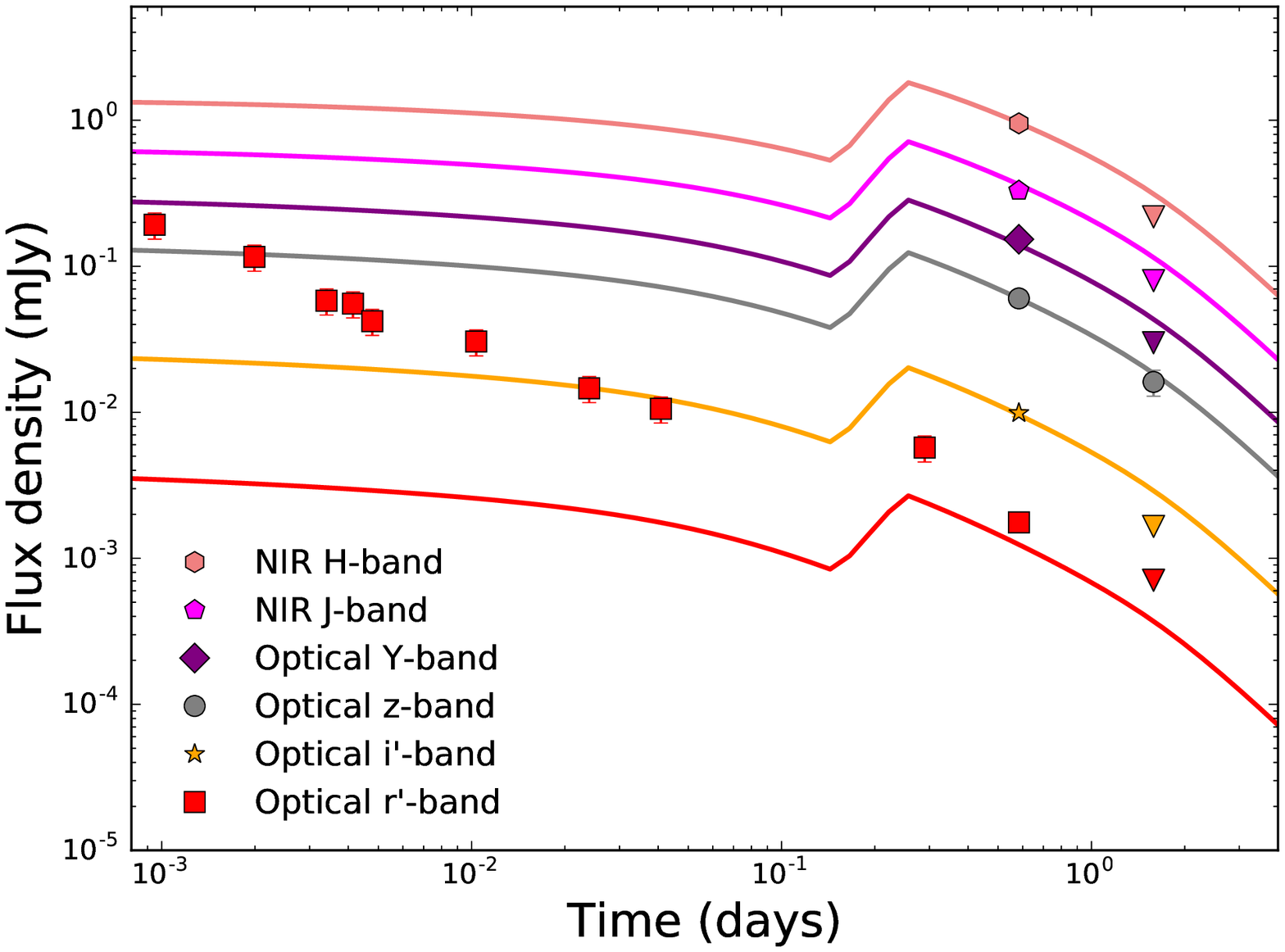} \\
  \includegraphics[width=0.47\textwidth]{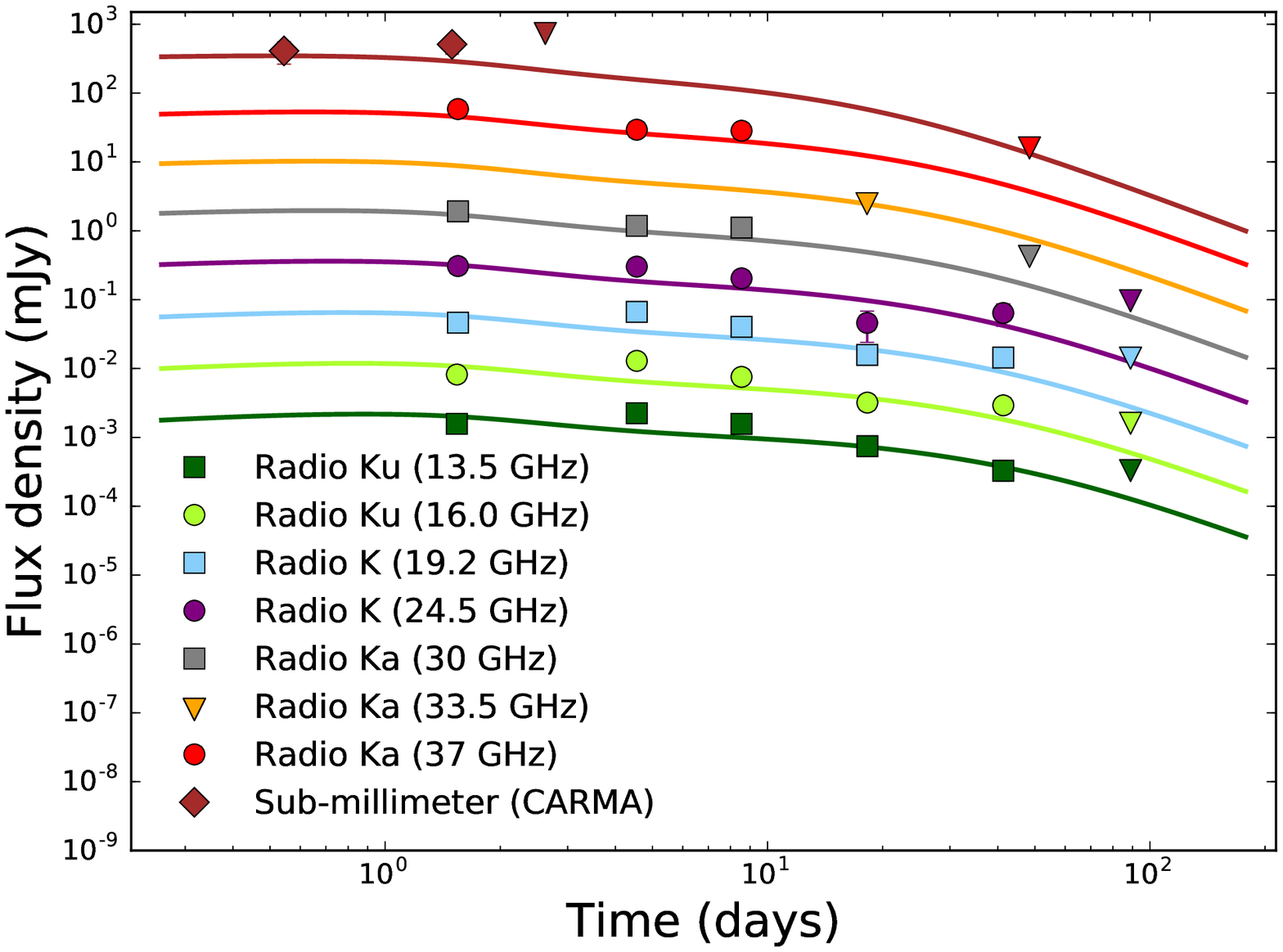} &
  \includegraphics[width=0.47\textwidth]{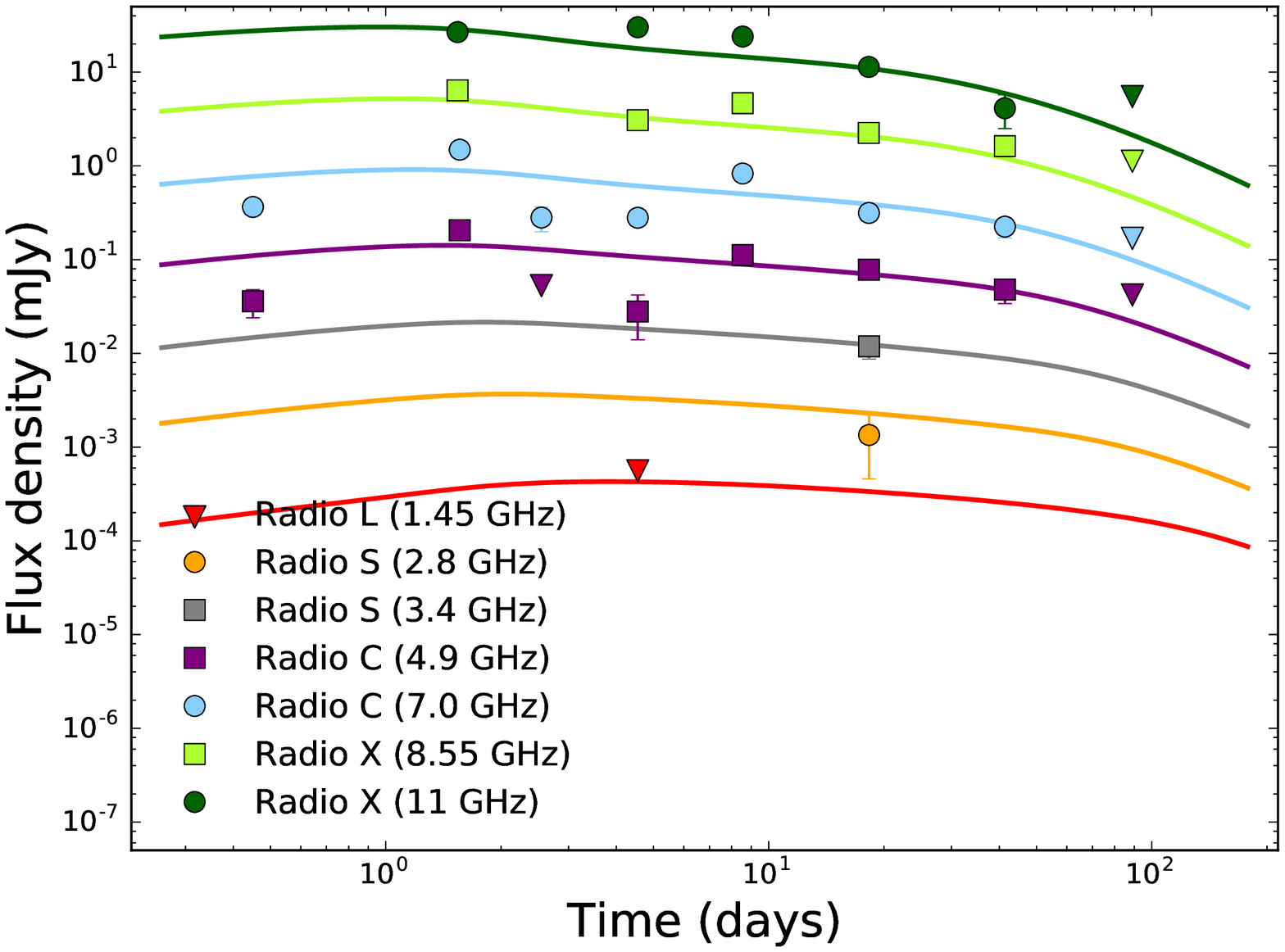} \\
 \end{tabular}
 \caption{X-ray (top left), optical/NIR (top right) and radio (bottom) light curves of the 
afterglow of GRB 140304A, together with an FS ISM model with $\nuc<\nunir$ including
energy injection between 0.15\,d and 0.26\,d (Section \ref{text:injection}). 
The model significantly under-predicts the optical light curve before $4\times10^{-2}$\,d,
and is therefore disfavored.}
\label{fig:modellc_ISM16}
\end{figure*}

\begin{figure*}
\begin{tabular}{ccc}
 \centering
 \includegraphics[width=0.31\textwidth]{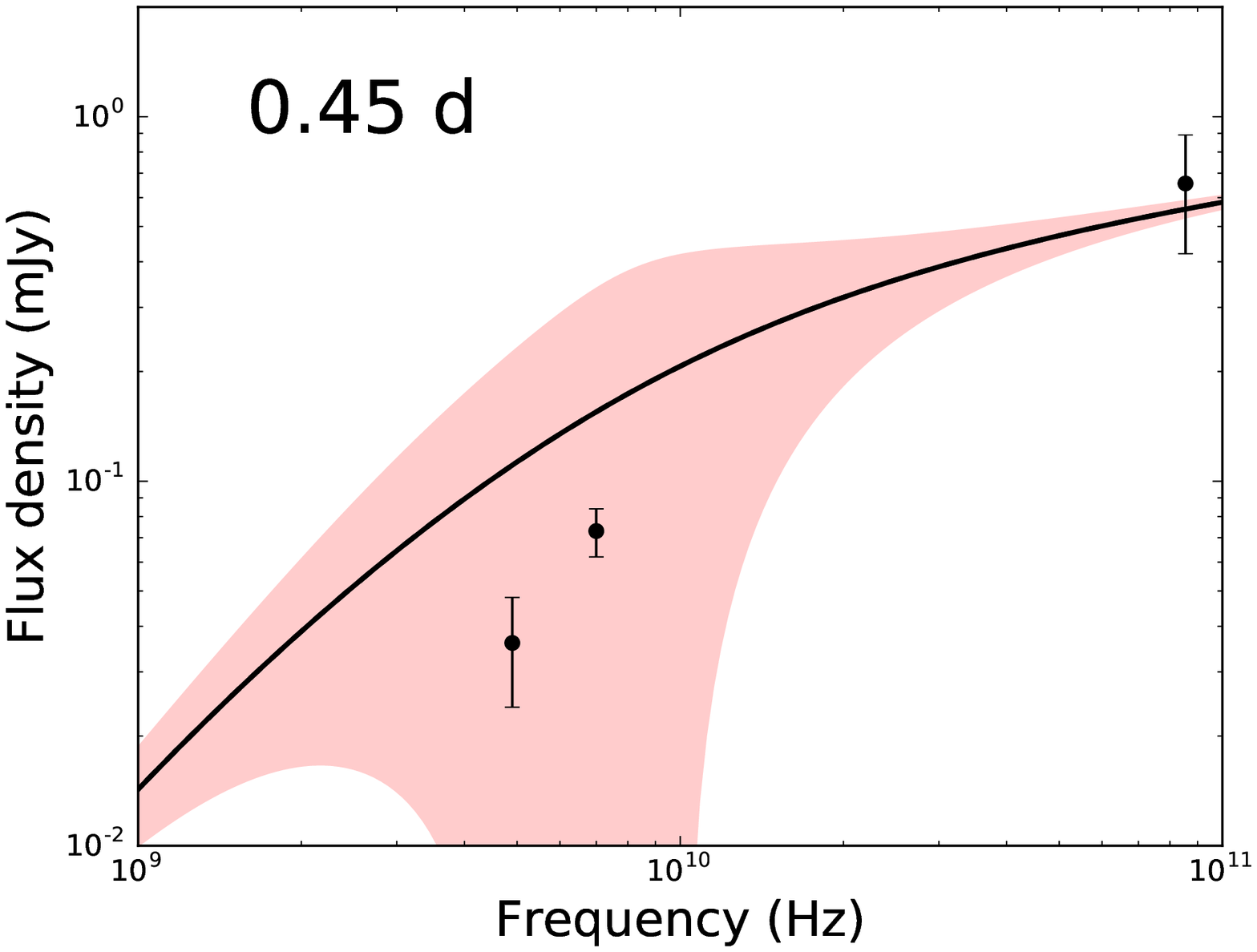} &
 \includegraphics[width=0.31\textwidth]{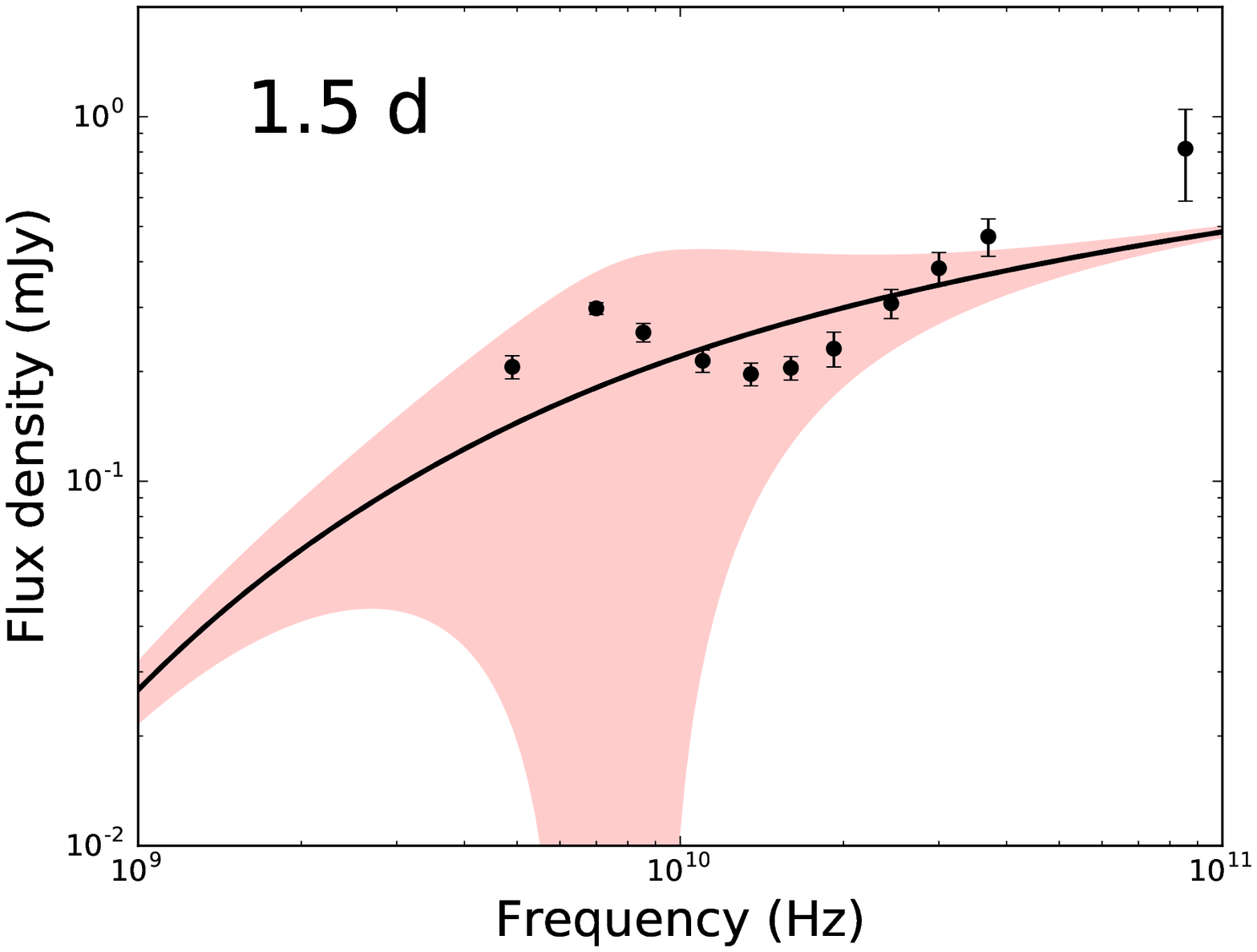} &
 \includegraphics[width=0.31\textwidth]{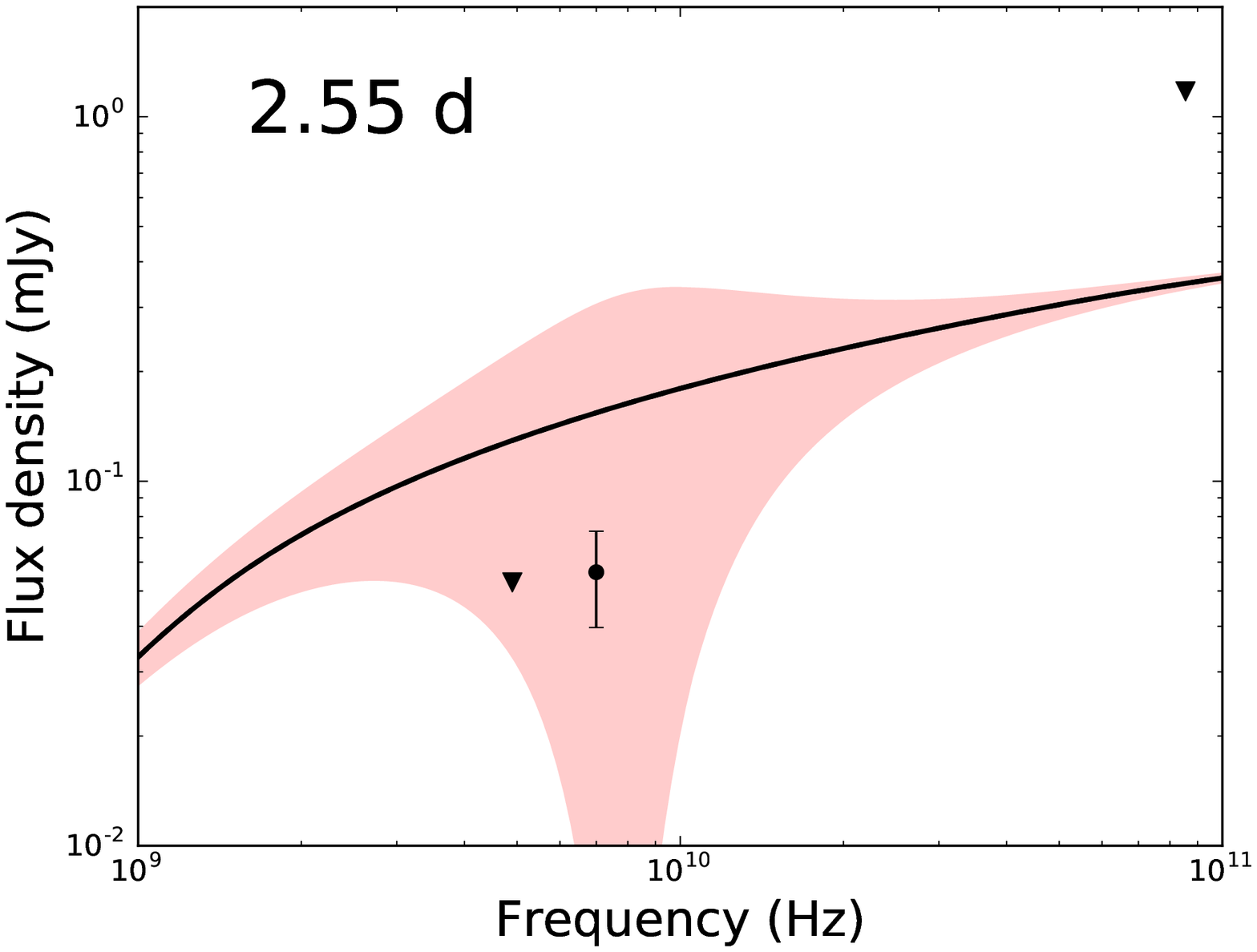} \\
 \includegraphics[width=0.31\textwidth]{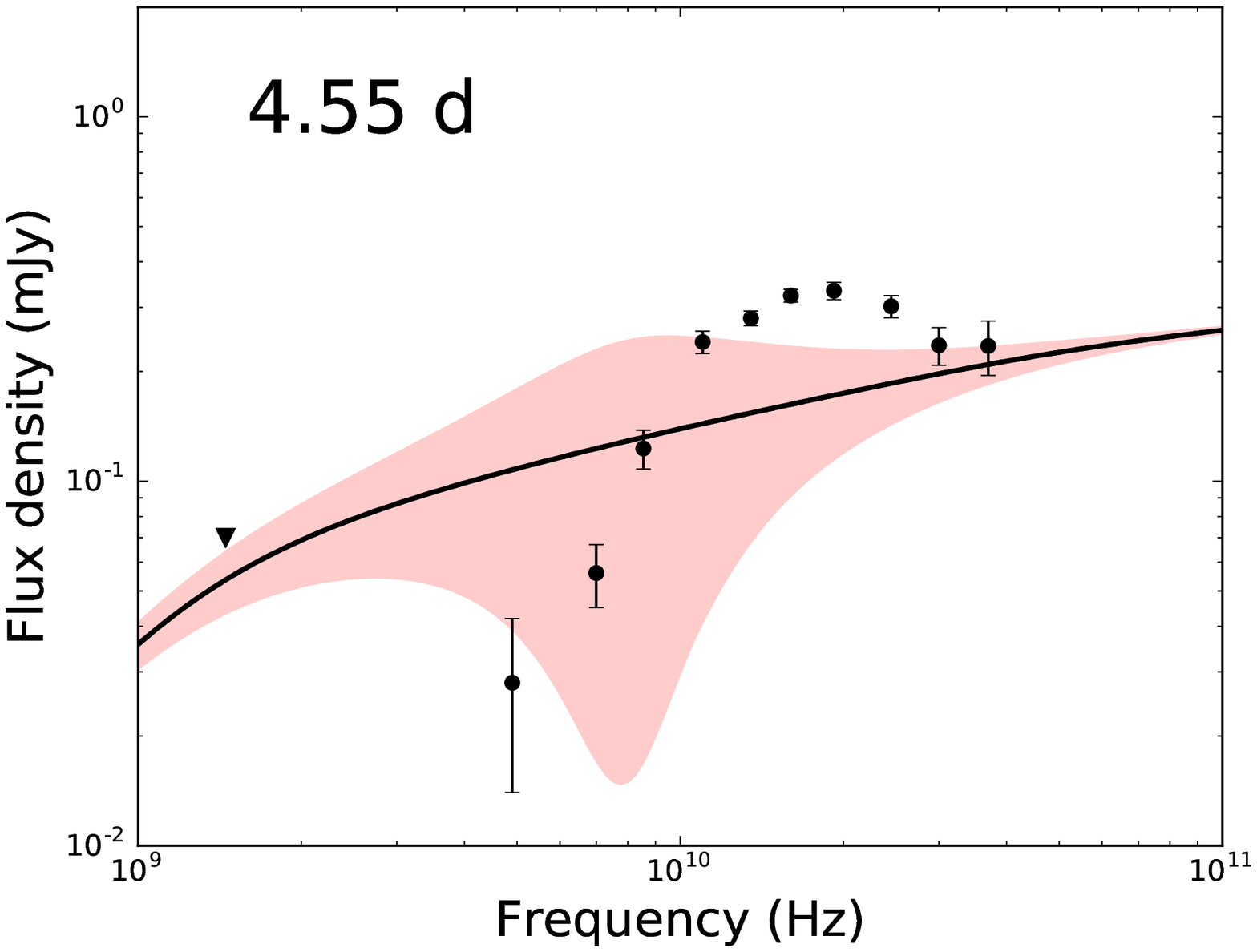} &
 \includegraphics[width=0.31\textwidth]{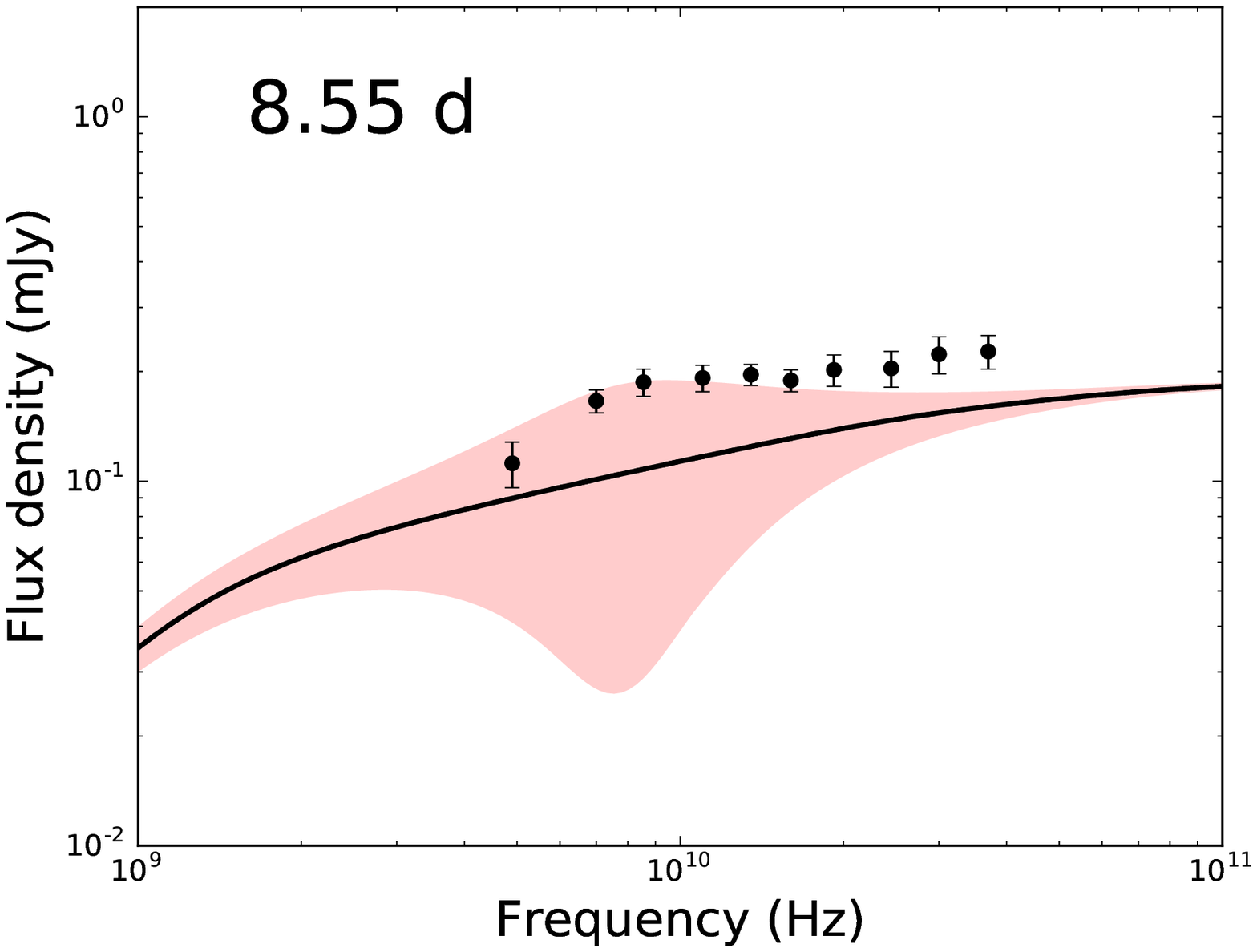} &
 \includegraphics[width=0.31\textwidth]{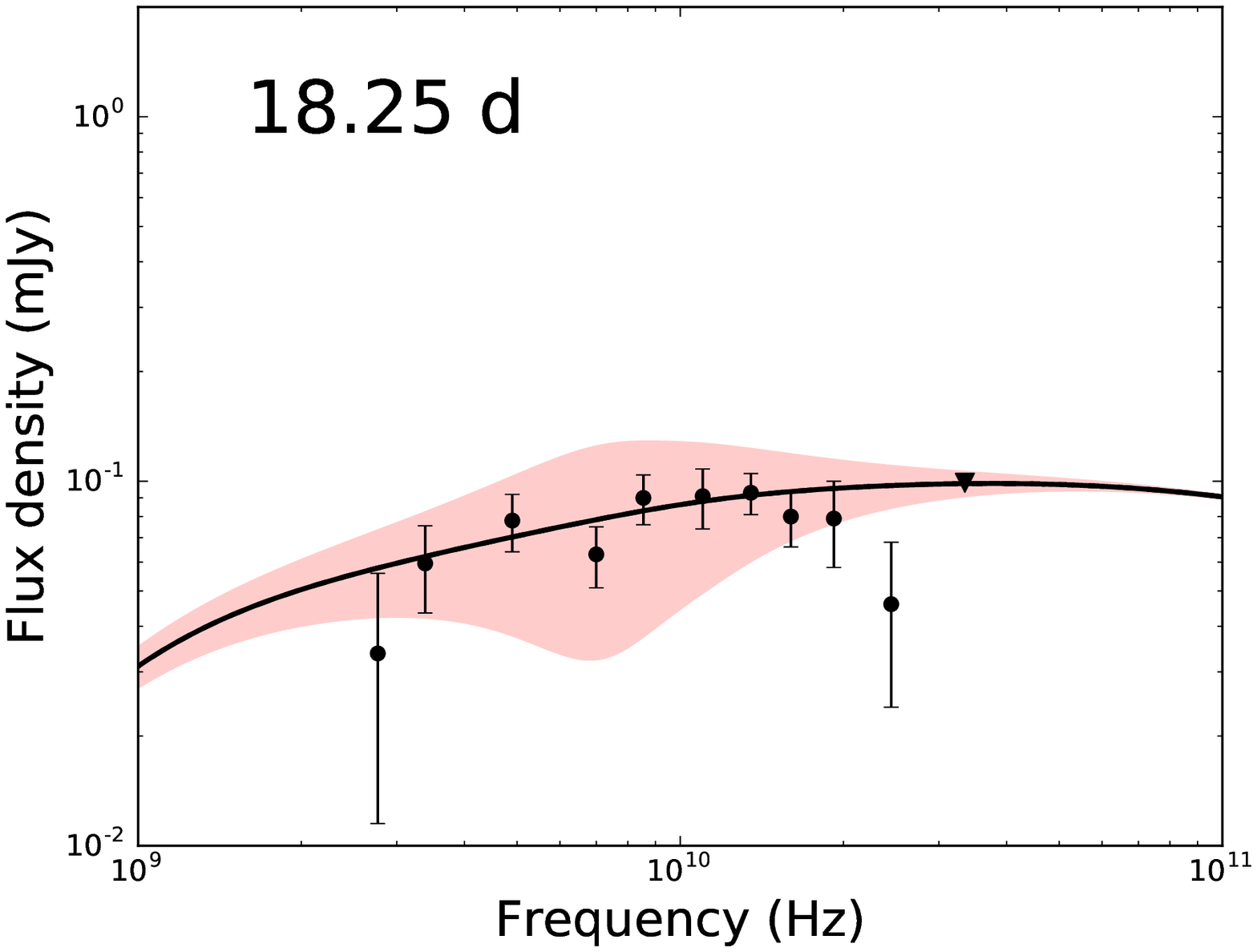} \\
 \includegraphics[width=0.31\textwidth]{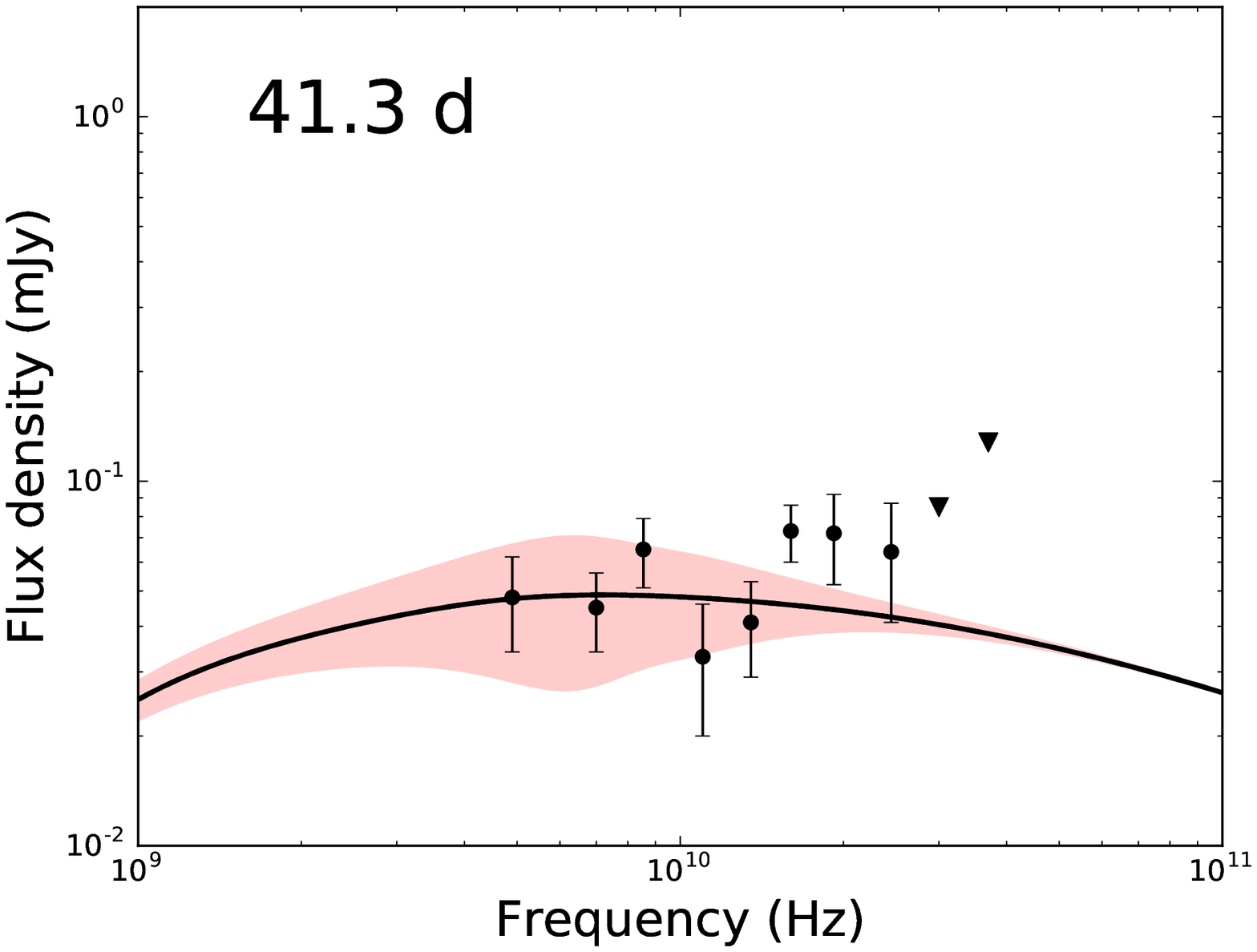} &
 \includegraphics[width=0.31\textwidth]{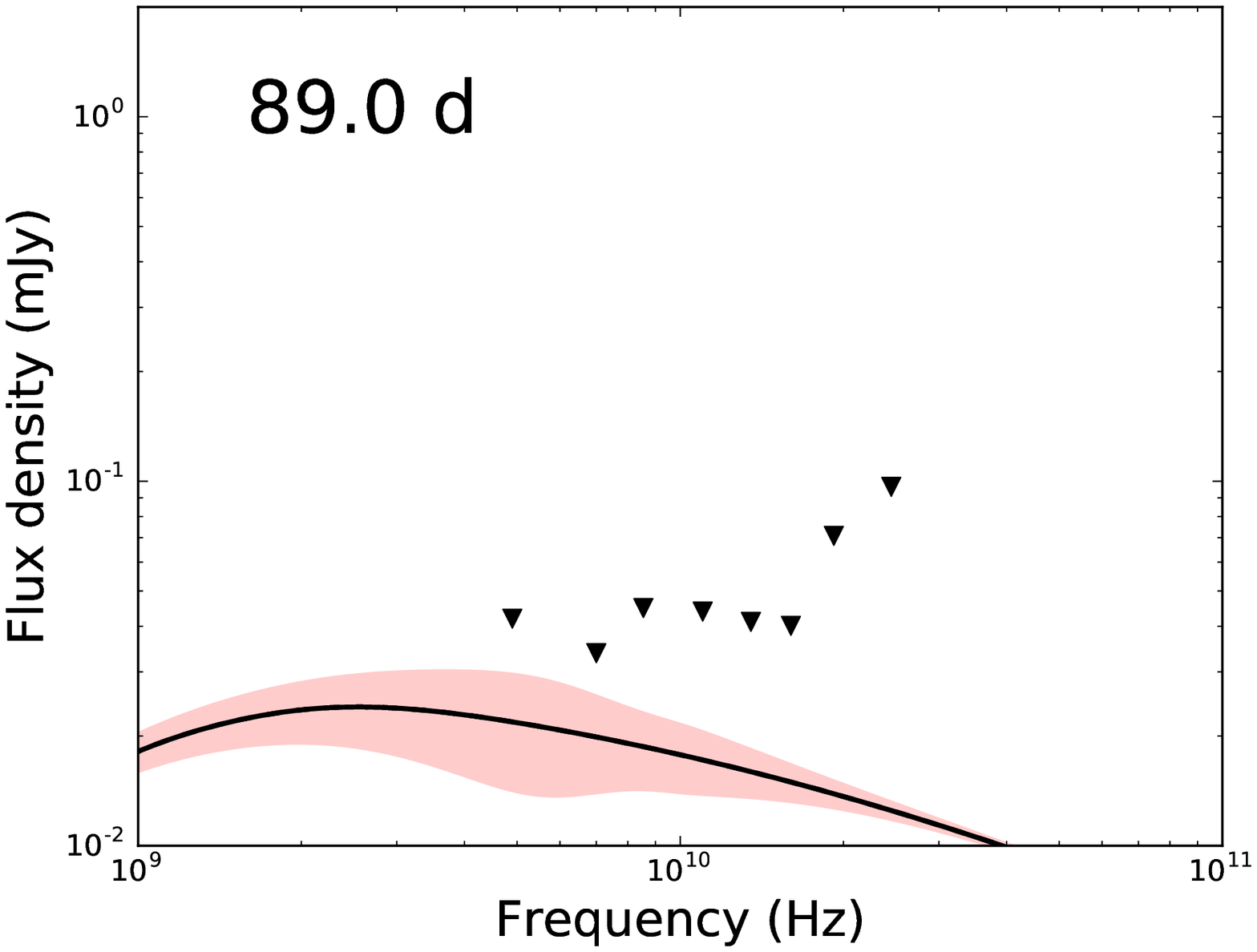} & 
\end{tabular}
\caption{Radio spectral energy distributions of the afterglow of \me\ at multiple epochs starting 
at 0.45~d, together with the same FS ISM model as in Figure \ref{fig:modellc_ISM16}. The 
red shaded regions represent the expected variability due to scintillation.
}
\label{fig:modelsed_ISM16}
\end{figure*}

If both \numax\ and \nuc\ are below \nunir, the optical/NIR and X-ray light curves do not 
distinguish 
between ISM and wind environments. In this scenario, the expected light curve decline rate is 
$\alpha \approx -1$. The observed decline rate is $\alpha_{\rm X}\approx-1.5$ and $\alpha_{\rm NIR} 
\lesssim -1.8$, which can be explained as a jet break between $\approx0.3$ and $\approx2$\,d.
This model matches the X-ray light curve after 0.2\,d, requiring a jet break at $\approx2.5$\,d. It 
also fits the 5--40\,GHz SED at 0.45\,d, 1.5\,d, and after 8.5\,d. However, it significantly 
over-predicts the 5--7\,GHz observations at 2.55\,d, does not match the steep spectrum below 
10\,GHz, and slightly over-predicts the optical/NIR limits at 1.58\,d 
(Figures \ref{fig:modellc_ISM16} and \ref{fig:modelsed_ISM16}).

\subsection{Intermediate cooling frequency, $\nunir < \nuc < \numax$}
\label{text:midnuc}
For the spectral ordering $\numax < \nunir < \nuc$ in the ISM 
environment, we expect $\alpha_{\rm NIR} = 3(1-p)/4$. Since $\alpha_{\rm NIR} \lesssim \alpha_{\rm 
z'} \approx -1.3$, this requires $p\gtrsim2.73$, which yields $\alpha_{\rm X} = (2-3p)/4 \lesssim 
-1.55$, consistent with observations. The $H$-band flux density at 0.58\,d corrected for Galactic 
extinction is $F_{\rm H} = (12.3\pm0.9)\times10^{-2}$\,mJy. Interpolating the X-ray light curve 
between 0.2\,d and 3\,d, the X-ray flux density at this time is $F_{\rm X} = 
(12.9\pm1.4)\times10^{-5}$\,mJy (Section \ref{text:basic_considerations}). For $p \approx 2.73$, 
the 
spectral index on either side of the cooling frequency is $\beta = (1-p)/2 \approx -0.87$ for $\nu 
< 
\nuc$ and $\beta = -p/2 \approx -1.36$ for $\nu > \nuc$. This allows us to locate $\nuc \approx 
6.6\times10^{16}$\,Hz at 0.58\,d. 

\begin{figure*} 
 \begin{tabular}{cc}
  \includegraphics[width=0.47\textwidth]{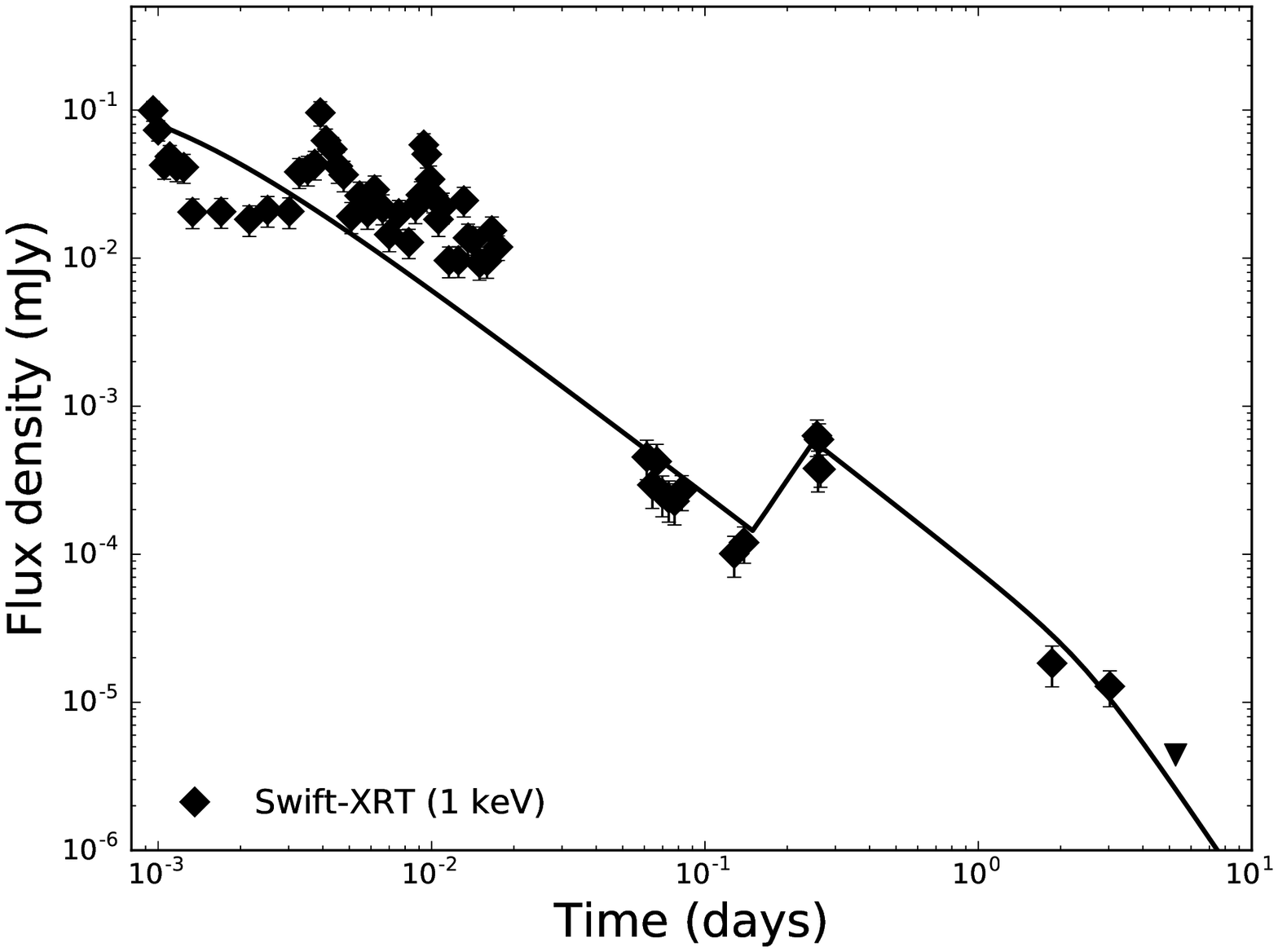} &
  \includegraphics[width=0.47\textwidth]{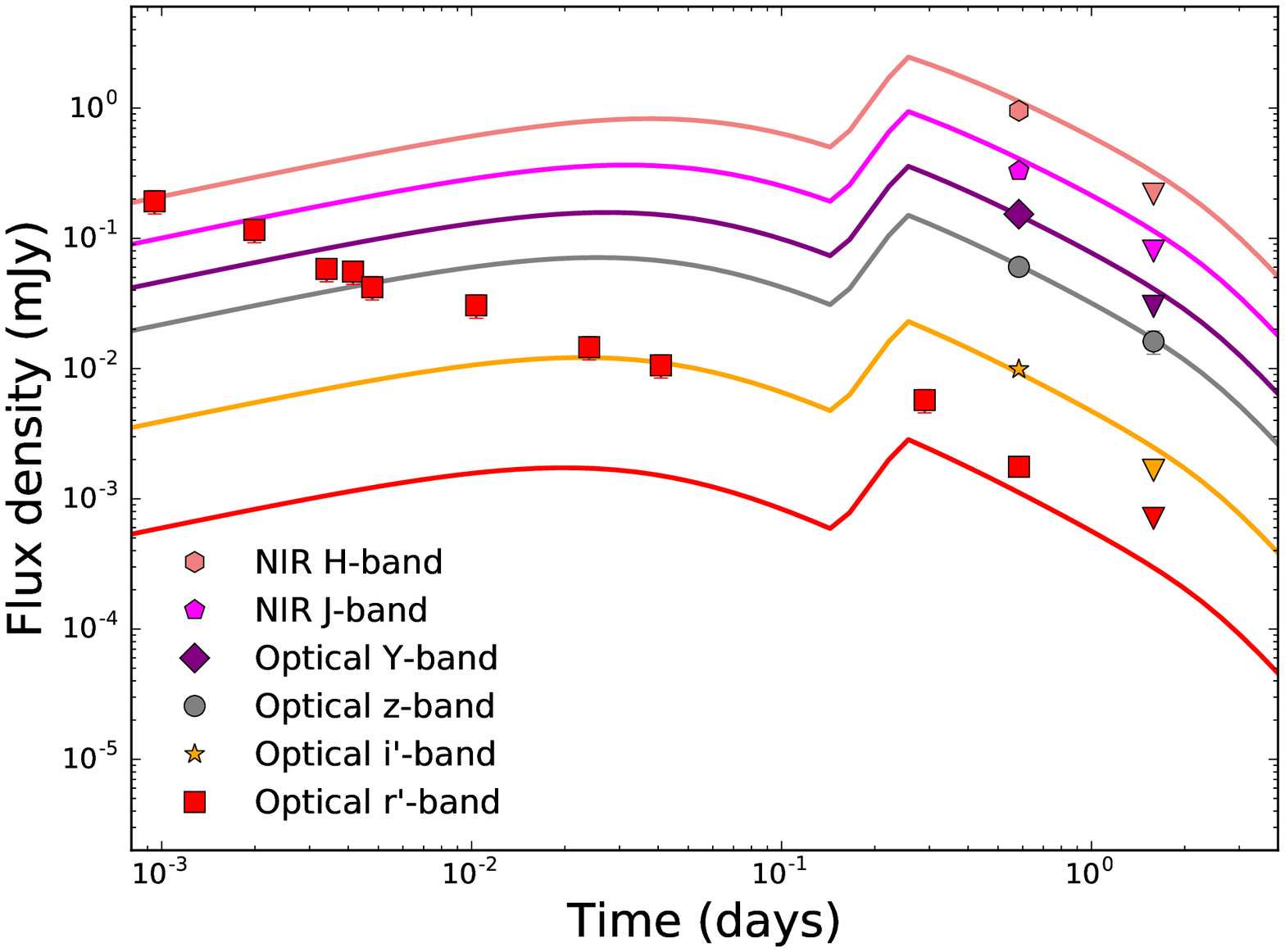} \\
  \includegraphics[width=0.47\textwidth]{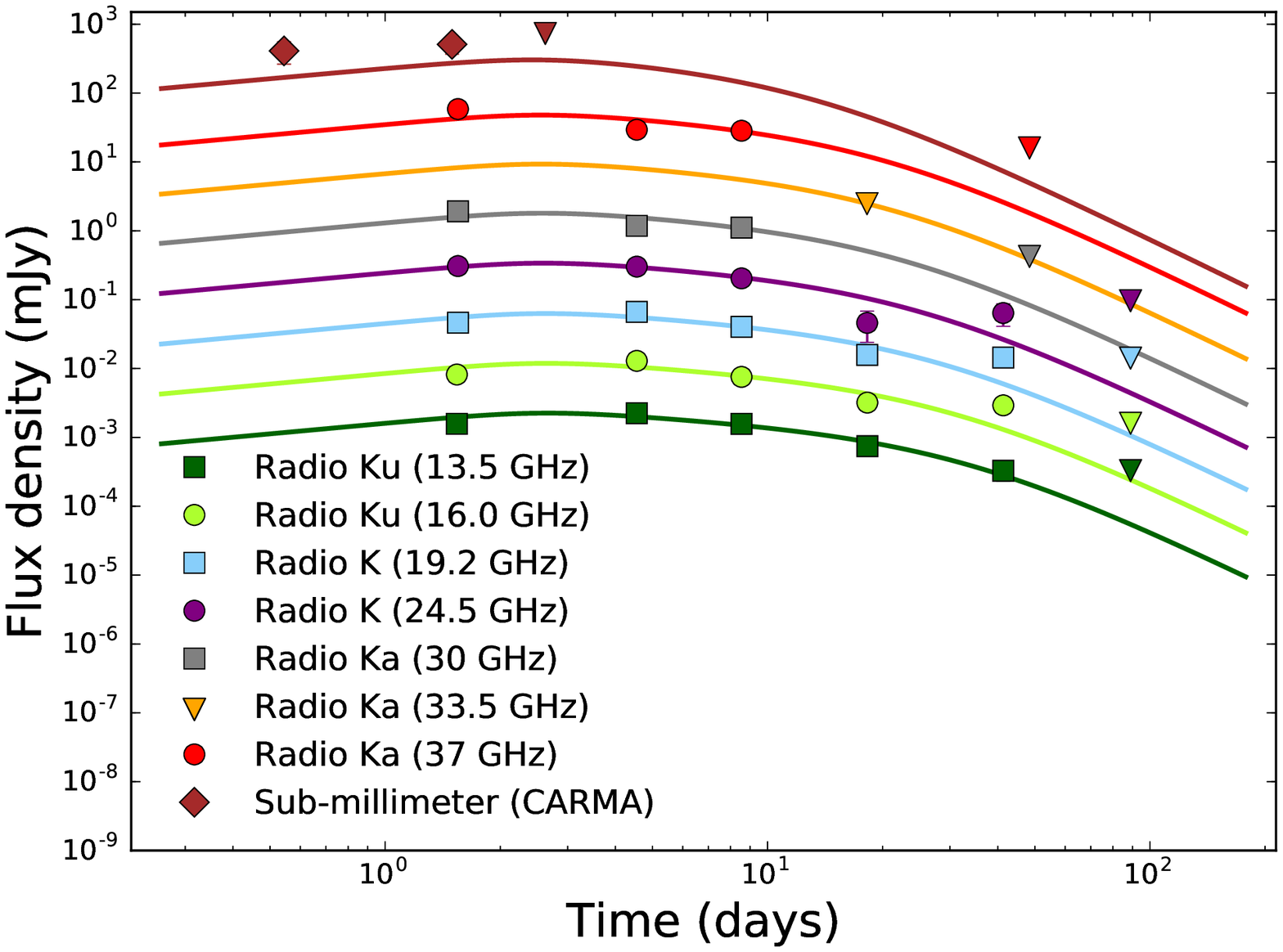} &
  \includegraphics[width=0.47\textwidth]{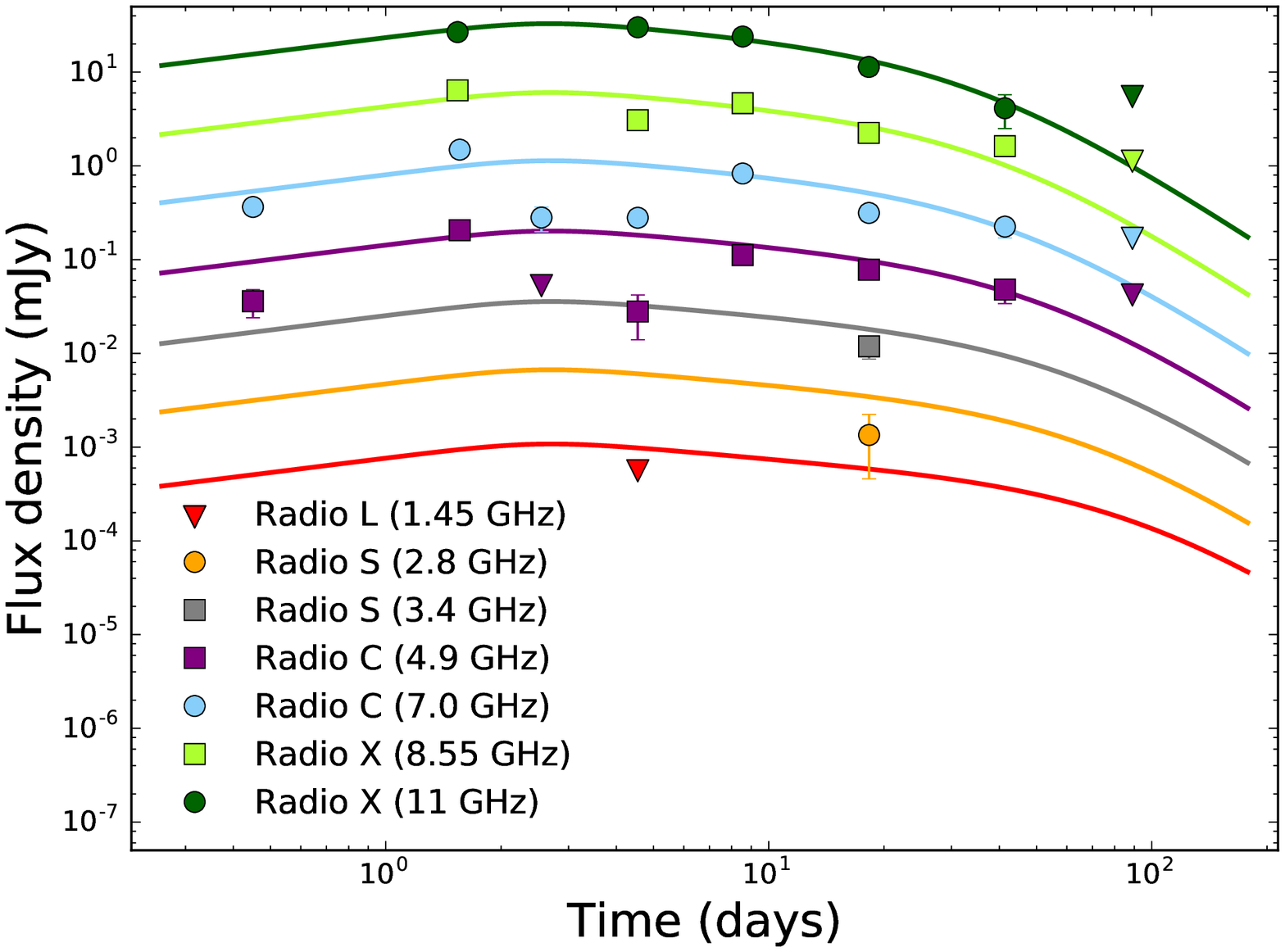} \\
 \end{tabular}
 \caption{X-ray (top left), optical/NIR (top right) and radio (bottom) light curves of the 
afterglow of GRB 140304A, together with an FS ISM model with $\nunir<\nuc<\nux$ including
energy injection between 0.15\,d and 0.26\,d (Section \ref{text:injection}). 
The different X-ray decay rate expected compared with the case of $\nuc>\nux$ is offset here by a 
slightly different value of $p$ (Table \ref{tab:params_ISM}).
The model significantly under-predicts the optical light curve before $4\times10^{-2}$\,d,
and is therefore disfavored.}
\label{fig:modellc_ISM15}
\end{figure*}

\begin{figure*}
\begin{tabular}{ccc}
 \centering
 \includegraphics[width=0.31\textwidth]{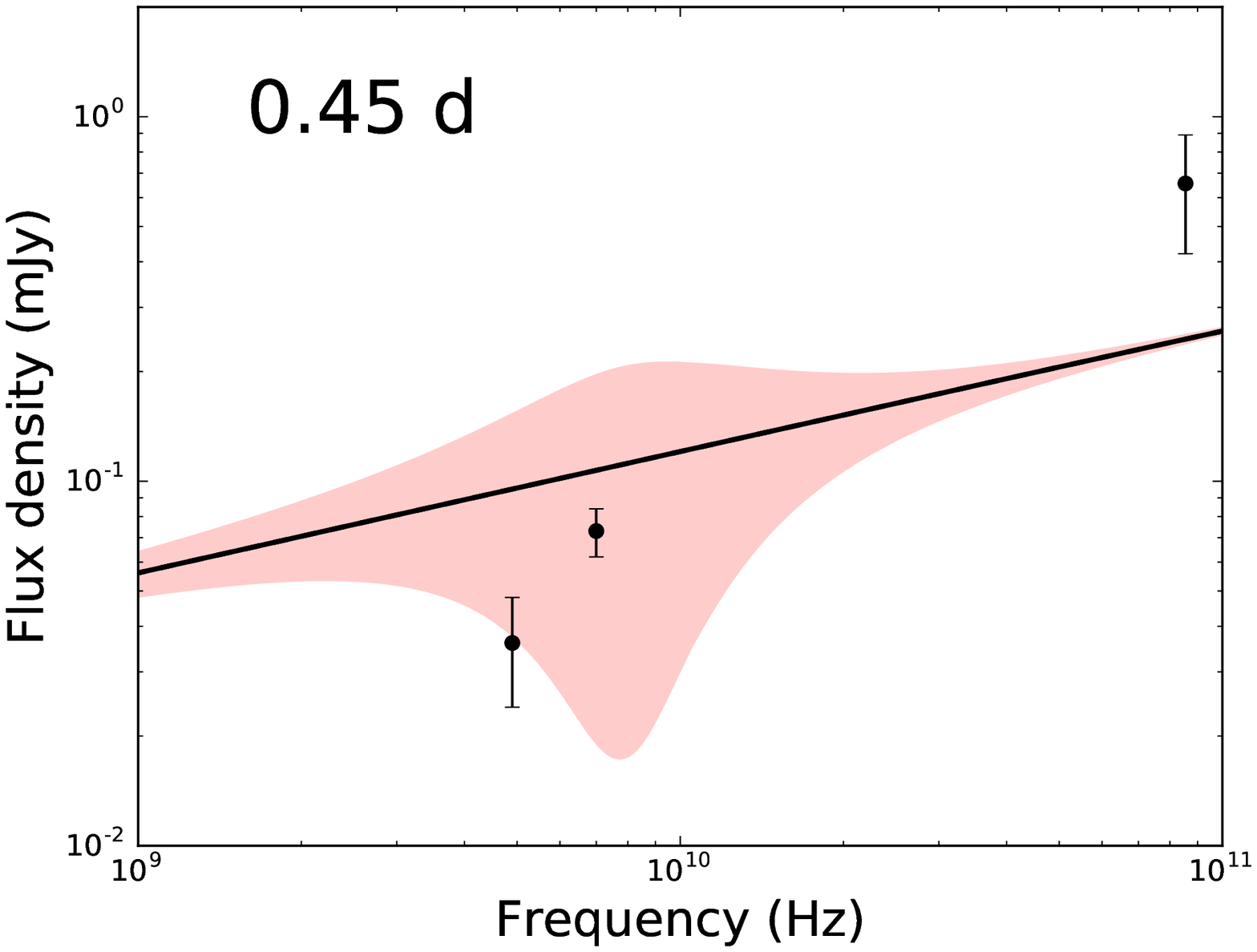} &
 \includegraphics[width=0.31\textwidth]{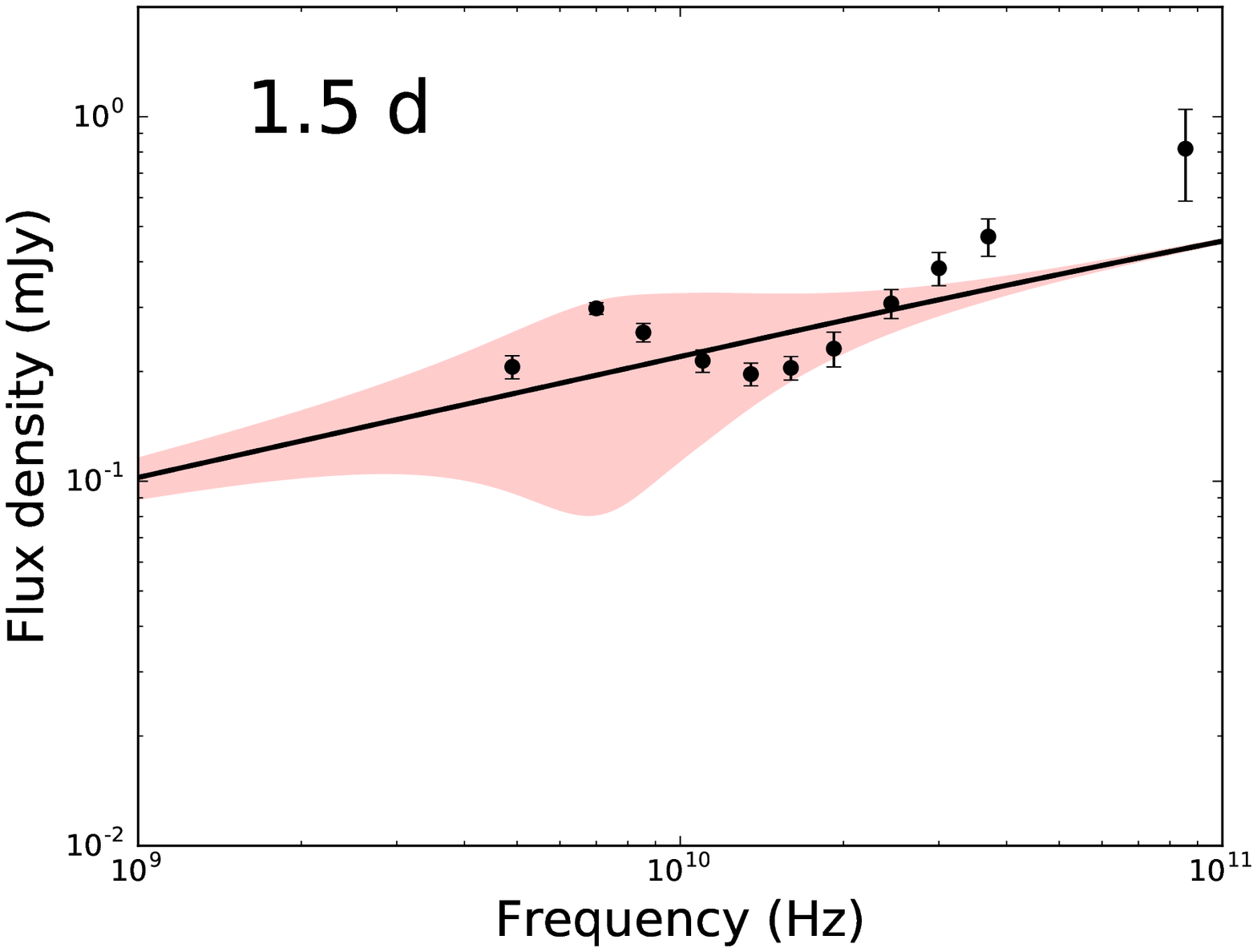} &
 \includegraphics[width=0.31\textwidth]{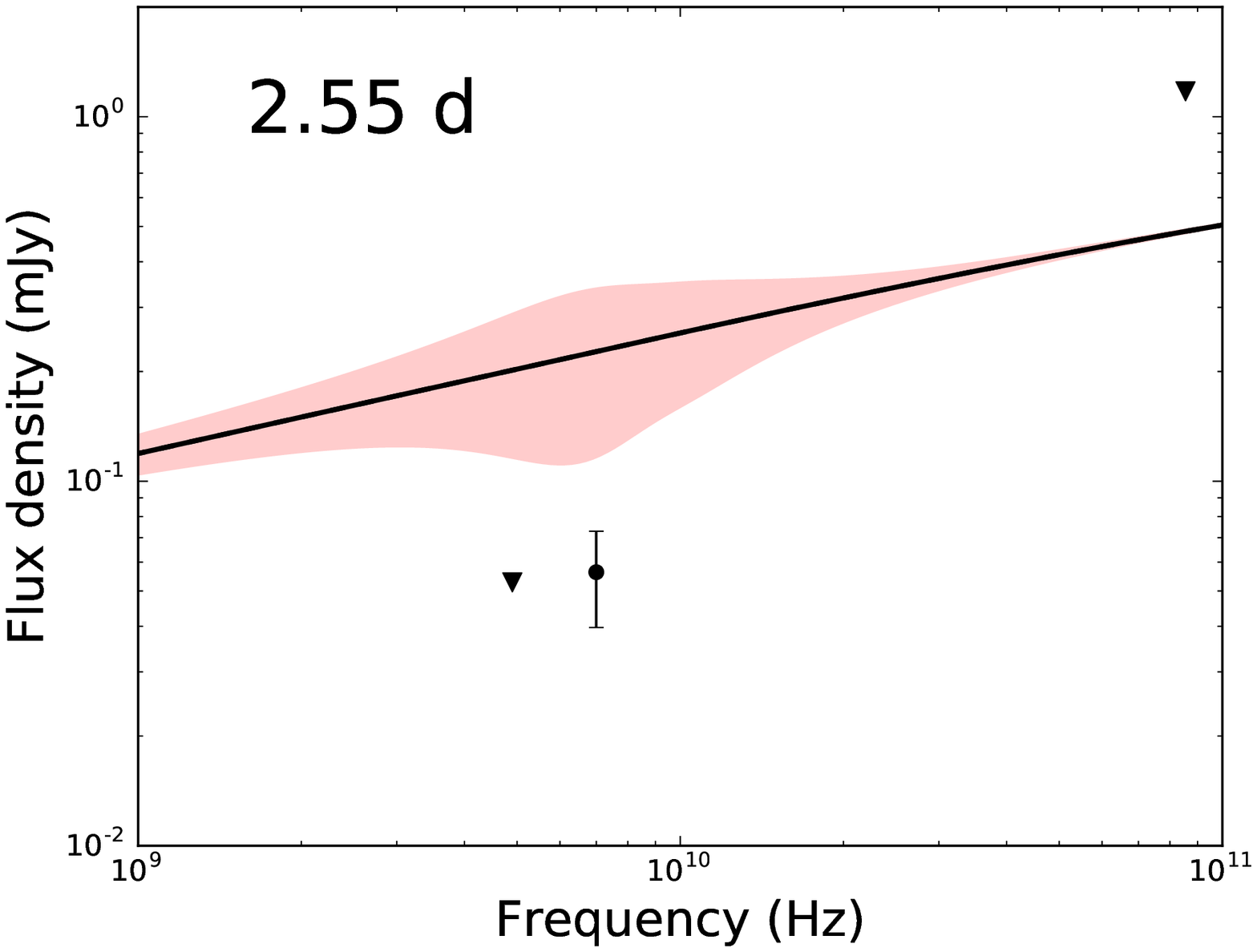} \\
 \includegraphics[width=0.31\textwidth]{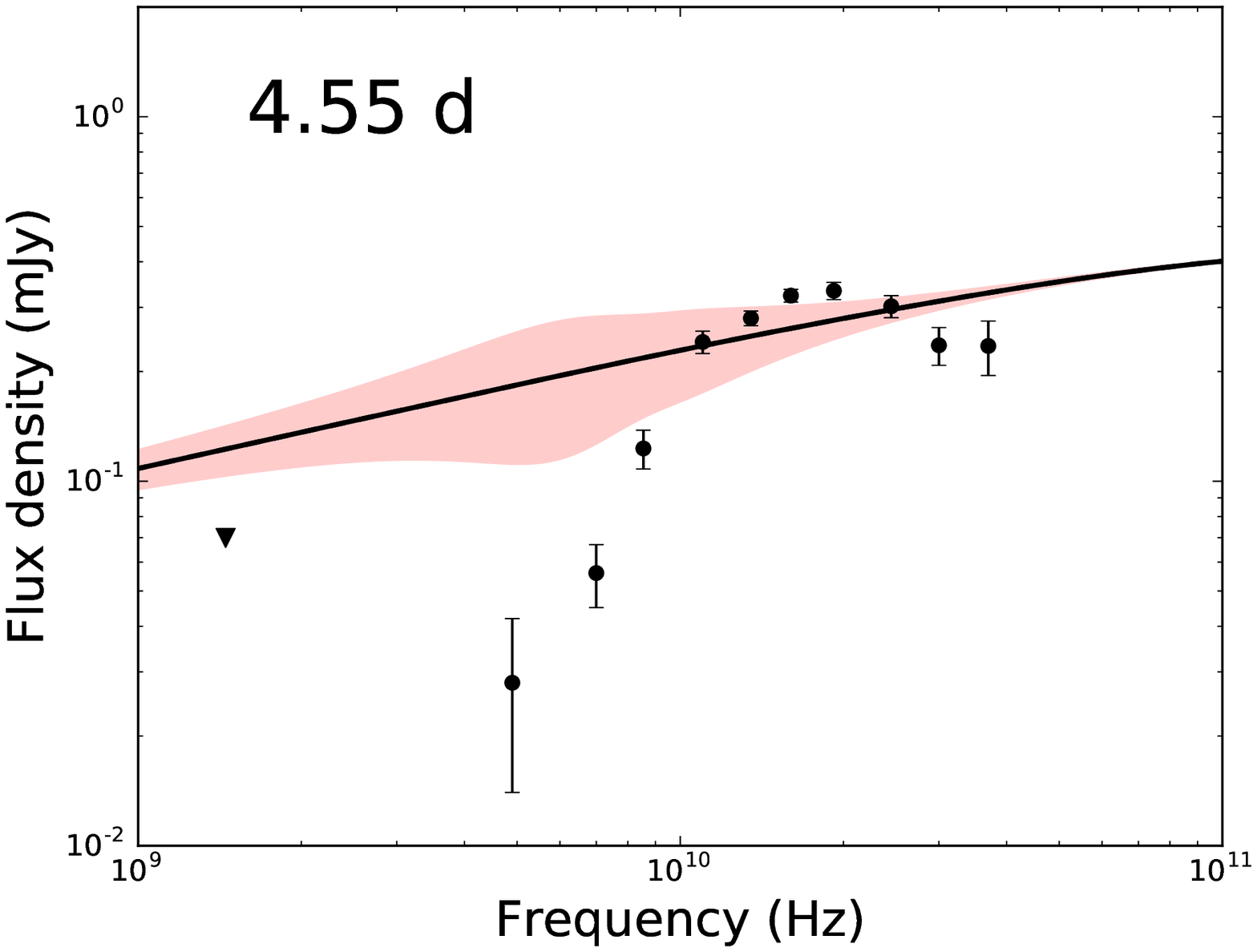} &
 \includegraphics[width=0.31\textwidth]{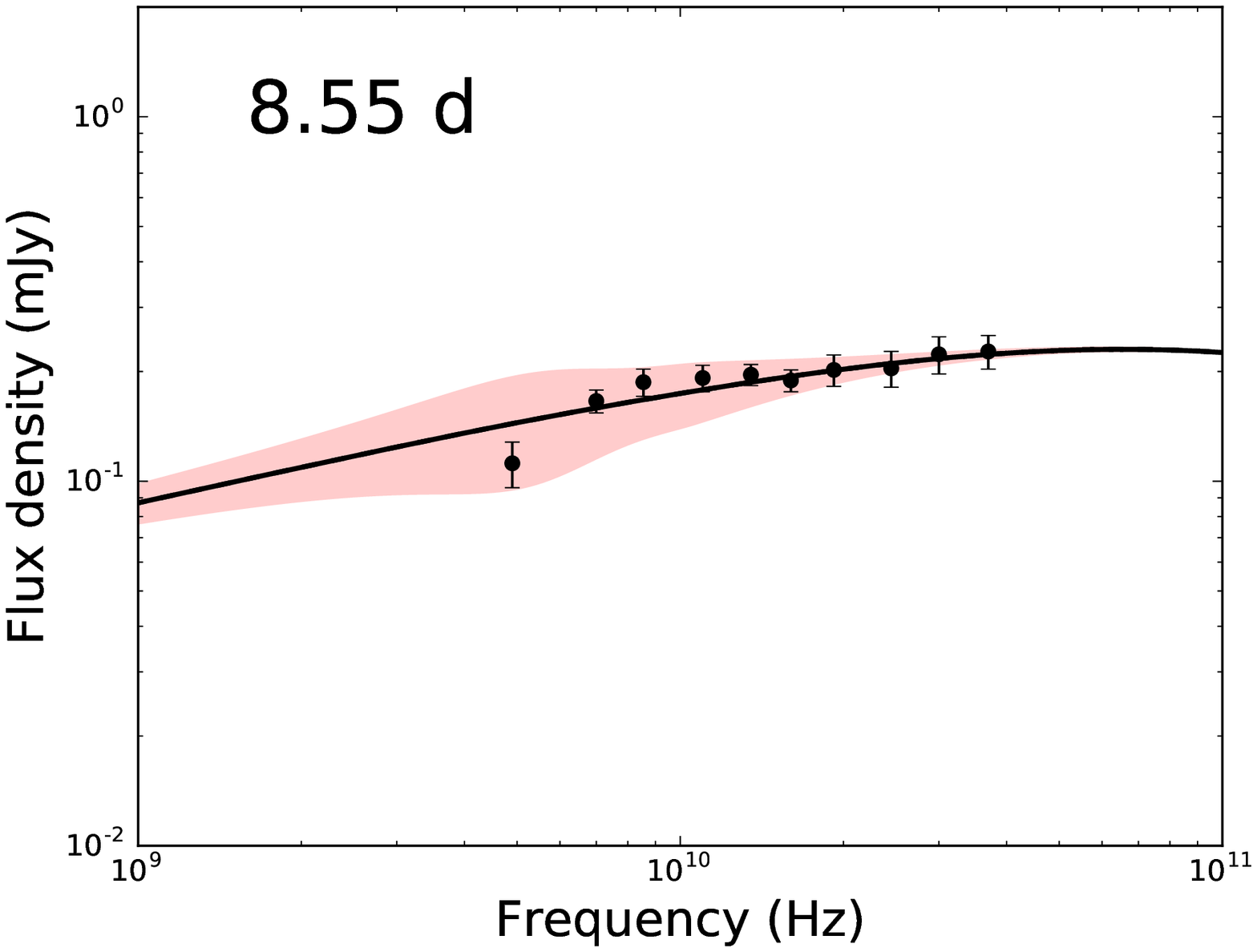} &
 \includegraphics[width=0.31\textwidth]{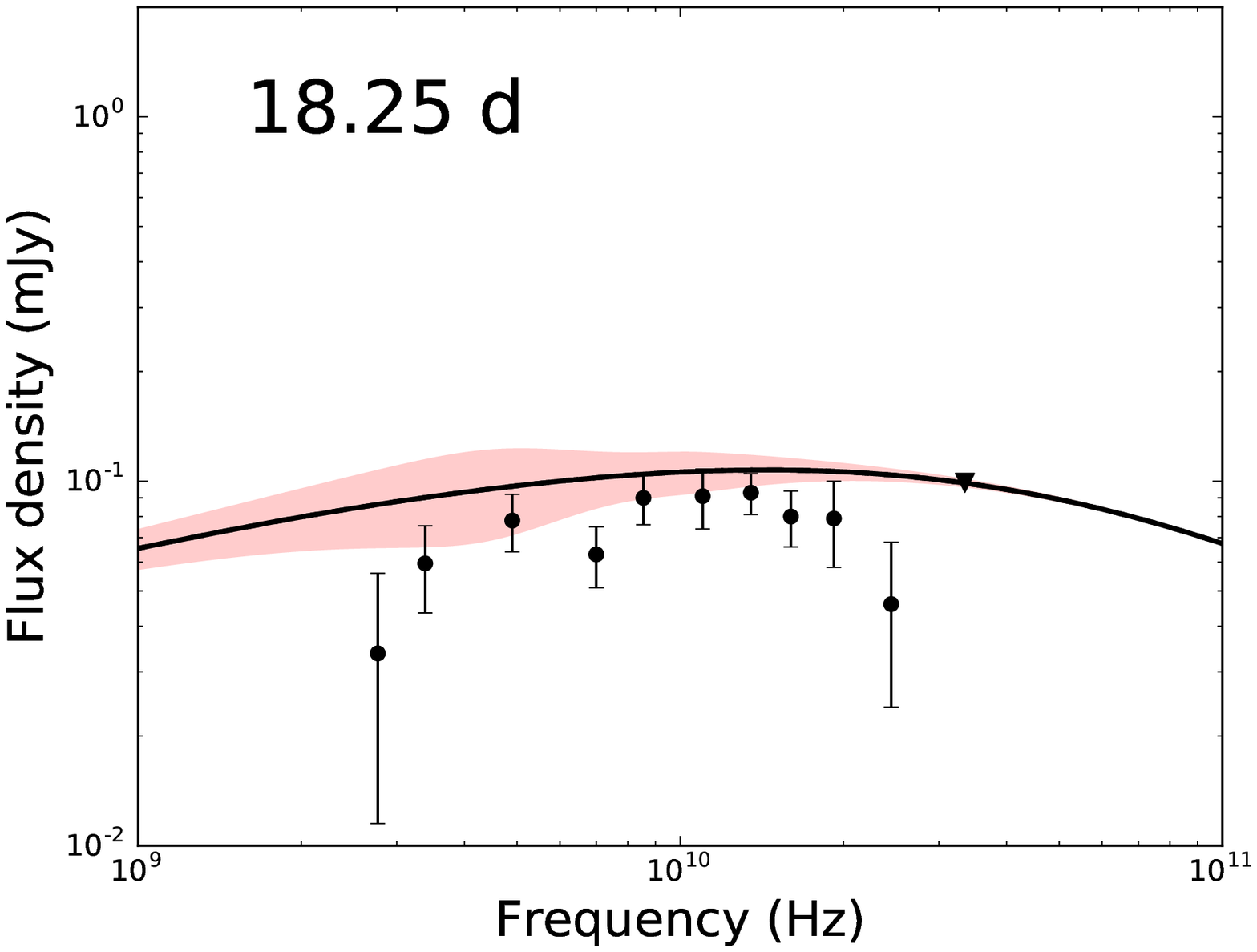} \\
 \includegraphics[width=0.31\textwidth]{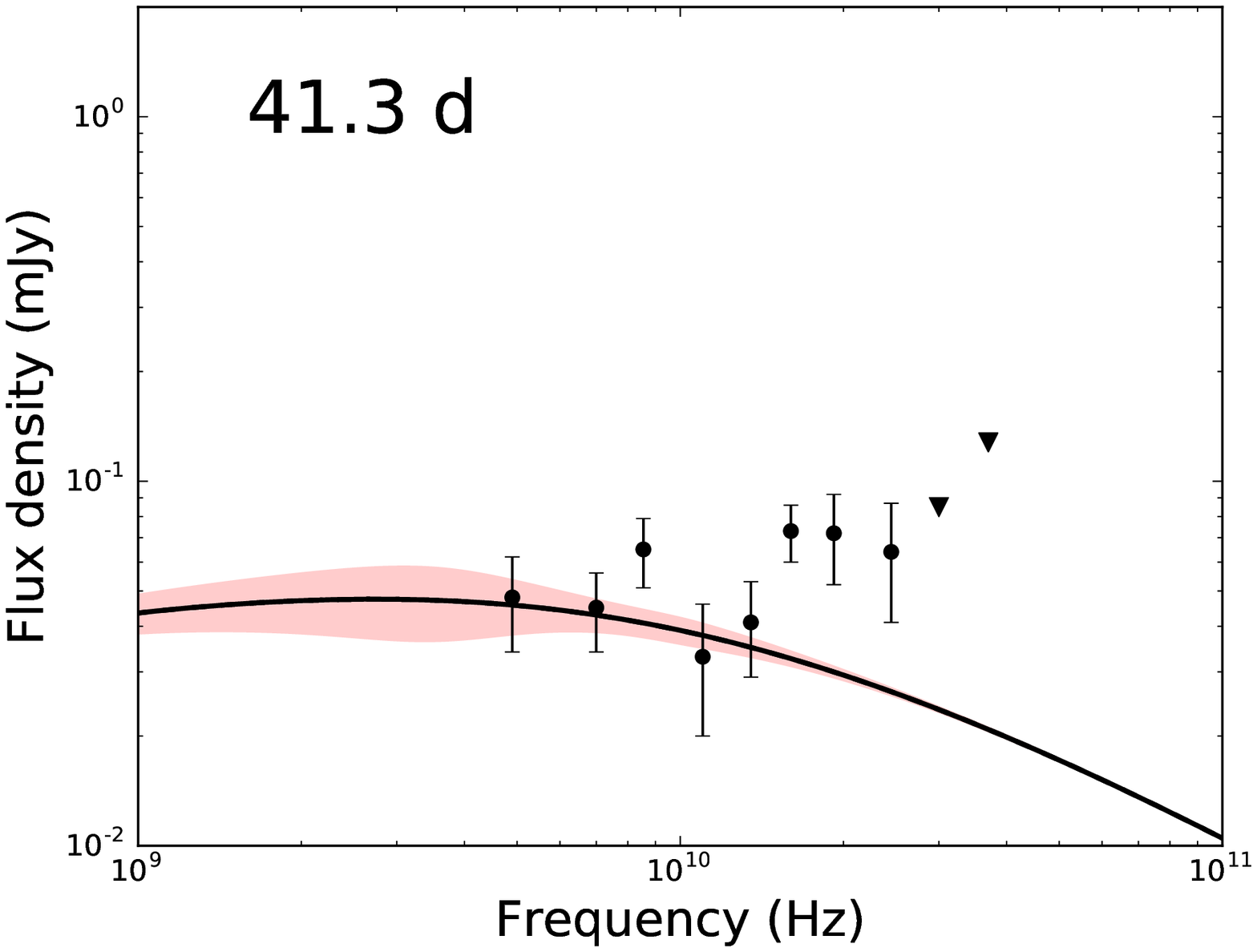} &
 \includegraphics[width=0.31\textwidth]{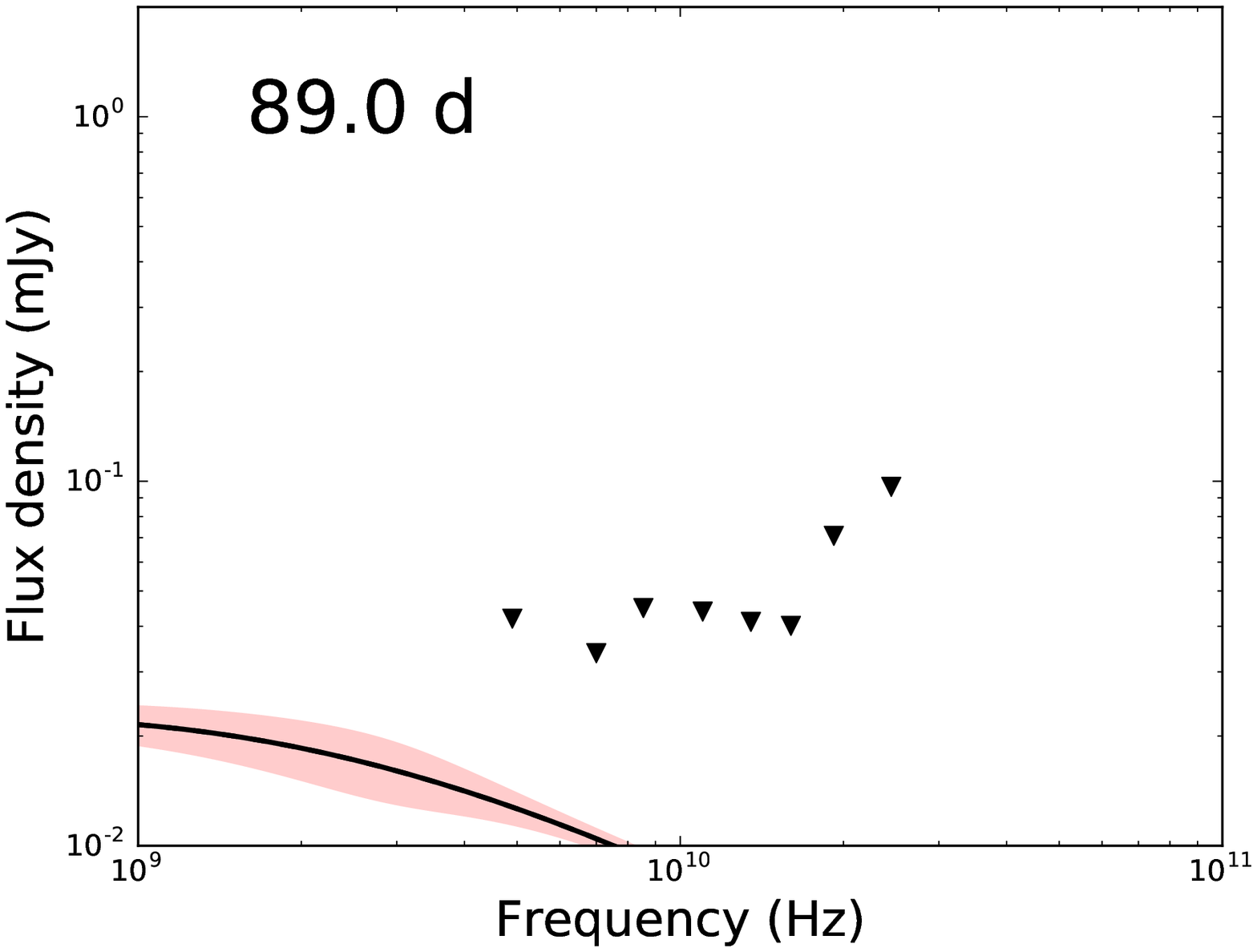} & 
\end{tabular}
\caption{Radio spectral energy distributions of the afterglow of \me\ at multiple epochs starting 
at 0.45~d, together with the same FS ISM model in Figure \ref{fig:modellc_ISM15}. The 
red shaded regions represent the expected variability due to scintillation.
}
\label{fig:modelsed_ISM15}
\end{figure*}

Fitting the gross features of the radio SED at 1.5\,d as a $\nu^{1/3}$ power law, the 
millimeter-band ($\nu_{\rm mm} = 85.5$\,GHz) flux density at 0.58\,d,
\begin{equation}
 F_{\rm mm}(0.58\,{\rm d}) = F_{\rm mm}(1.5\,{\rm d})\times(0.58/1.5)^{1/2} \approx 0.5\,{\rm mJy}.
\end{equation}
Requiring that this connect with the NIR SED at 0.58\,d, we have
\begin{equation}
 F_{\rm mm} = f_{\nu, \rm max}\left(\frac{\nu_{\rm mm}}{\numax}\right)^{\frac{1}{3}} 
            = F_{\rm H}\left(\frac{\numax}{\nu_{\rm H}}\right)^{\frac{1-p}{2}}
                       \left(\frac{\nu_{\rm mm}}{\numax}\right)^{\frac{1}{3}},
\end{equation}
which yields $\numax \approx 7\times10^{12}$\,Hz at 0.58\,d. This spectrum requires extreme 
parameters (low density and high energy), while \nua\ remains unconstrained (Table 
\ref{tab:params_ISM}). The resulting light curves and SEDs are presented in Figures 
\ref{fig:modellc_ISM15} and \ref{fig:modelsed_ISM15}. We note that the resulting parameters are 
very 
similar to the model described in Section \ref{text:highnuc}, the difference in the location of 
\nuc\ being offset by the slightly different value of $p$.

\section{The Early FS}
\label{appendix:earlyFS}
The energy of the FS increases before \tdec, while the RS is still crossing the ejecta. 
During this coasting period, the Lorentz factor, $\Gamma_0$, of the jet is approximately constant
if the RS is Newtonian \citep{glz+13}.
Thus, the observer time, $t_{\rm obs}\sim t/\Gamma_0^2\propto t$, where $t$ is the time in the frame
in which the circumburst medium is at rest. The FS radius, $r\propto t$, and the 
energy density of the shocked ISM, $\varepsilon \propto \Gamma_0^2\rho\propto r^{-k}$.
The energy of the FS then increases as 
\begin{equation}
\frac{dE}{dt} = \varepsilon 4\pi r^2 \frac{dr}{dt} \propto r^{2-k},
\end{equation}
and thus $E \propto r^{3-k}$. 
For a wind medium, this gives $E\propto r\propto t\propto t_{\rm obs}$,
and the blast wave energy increases linearly with observer time.

\section{Shell Collisions}
\label{appendix:shellcollision}
Consider a central engine that emits two shells of Lorentz factors, $\Gamma_1$, and $\Gamma_2$ at
times $t = 0$ and $t = \Delta t$, respectively, where $t$ is measured in the frame in which
the circumburst medium is at rest. After the first shell is decelerated by the environment,
its radius is given by the BM solution,
\begin{equation}
 R_1(t) = \left[1-\frac{1}{2(4-k)\Gamma_1(t)^2}\right]ct,
\end{equation}
The two shells collide at a time $t_1$, when the Lorentz factor
of the first shell is reduced to $\Gamma_1(t_1)$, also given by the BM solution,
\begin{equation}
 \Gamma_1(t) = \left[\frac{(17-4k)E}{8\pi A c^{-k} t^{3-k}}\right]^{1/2}
\end{equation}
The second shell moves with a constant Lorentz factor, and its radius is given by
\begin{equation}
 R_2(t) = c(t-\Delta t)\left(1-\frac{1}{2\Gamma_2^2}\right)
\end{equation}
We can take the time of collision between the shells to be when their radii are 
equal,
\begin{equation}
 R_1(t_1) = R_2(t_1),
\end{equation}
which yields
\begin{equation}
\label{eq:shellcollmaster}
 1-\frac{1}{2(4-k)\Gamma_1(t_1)^2} = \left(1-\frac{\Delta t}{t_1}\right)
\left[1-\frac{1}{2\Gamma_2(t_1)^2}\right].
\end{equation}
The observer time,
\begin{equation}
  t_{\rm z} = (1+z)(t-R/c) = (1+z)\frac{t}{2(4-k)\Gamma_1^2}
\end{equation}
Thus, the collision time in the observer frame,
\begin{equation}
 \tcol = (1+z) \frac{t_1}{2(4-k)\Gamma_1^2},
\end{equation}
while the shell launch delay in the observer frame,
\begin{equation}
 \Delta t_{\rm L} = (1+z)\Delta t.
\end{equation}
Equation \ref{eq:shellcollmaster} then reduces to
\begin{equation}
 \frac{\Delta t_{\rm L}}{\tcol} = 
 \frac{1-\frac{(4-k)\Gamma_1^2}{\Gamma_2^2}} {1-\frac{1}{2\Gamma_2^2}} 
 \approx 1-\frac{(4-k)\Gamma_1^2}{\Gamma_2^2},
\end{equation}
where we have ignored the second term in the denominator assuming
$\Gamma_2 \gg1$. For $k=2$, this gives
\begin{equation}
 \Gamma_2 = 2^{1/2}\Gamma_1\left(1-\frac{\Delta t_{\rm L}}{\tcol}\right)^{-1/2}.
\end{equation}

\bibliographystyle{apj}
\bibliography{grb_alpha,gcn}

\clearpage
\
\vfil
\hfil \textsc{On-line only material} \hfil
\vfil
\clearpage

\begin{deluxetable}{lccc}
 \tabletypesize{\footnotesize}
 \tablecolumns{4}
 \tablecaption{Parameters for best-fit ISM models}
 \tablehead{   
           \colhead{Parameter} &
           \colhead{Model 1} &
           \colhead{Model 2} &
           \colhead{Model 3}
   }
 \startdata   
   Ordering at 0.1\,d   & $\nuc>\nux$         & $\nuc<\nunir$         & $\nunir<\nuc<\nux$ \\   
   $p$                  & 2.86                & 2.11                  & 2.73               \\
   \epse                & $2.7\times10^{-2}$  & 0.38                  & $3.2\times10^{-2}$ \\
   \epsb                & $1.1\times10^{-4}$  & 0.13                  & $2.8\times10^{-3}$ \\
   \dens                & $1.0\times10^{-4}$  & 0.76                  & $4.2\times10^{-4}$ \\   
   $E_{\rm K, iso, 52}$ (erg)
                        & $2.6\times10^{4}$   & 14.1                  & $2.1\times10^{3}$  \\ 
   \tjet\ (d)           & 1.86                & 1.5                   & 2.5                \\
   \thetajet\ (deg)     & 0.53                & 3.9                   & 0.98               \\
   \AV\ (mag)           & 0.08                & 0.07                  & 0.15               \\
   \nusa (Hz)           & \ldots              & $4.7\times10^{8\dag}$ & \ldots             \\
   \nuac (Hz)           & \ldots              & $1.6\times10^{10}$    & \ldots             \\
   \nua\ (Hz)           & $2.6\times10^{7\dag}$ & \ldots              & $6.6\times10^{7\dag}$ \\ 
   \numax\ (Hz)         & $1.8\times10^{14}$  & $8.5\times10^{14}$    & $2.8\times10^{14}$ \\
   \nuc\ (Hz)           & $1.4\times10^{18}$  & $1.0\times10^{13}$    & $1.7\times10^{17}$ \\
   $F_{\nu, \rm max}$ (mJy) & 2.1             & 3.2                   & 1.7                \\
   $E_{\gamma}$ (erg)   & \\
   $E_{\rm K}$ (erg)    & $1.1\times10^{52}$  & $3\times10^{50}$      & $3.1\times10^{51}$ \\
   $E_{\rm tot}$ (erg)  & \\
   $\eta_{\rm rad}$     & 
 \enddata
 \tablecomments{All break frequencies are listed at 0.1\,d. \dag\ These break frequencies are not 
directly constrained by the data.}
\label{tab:params_ISM}
\end{deluxetable}

\end{document}